\documentclass[useAMS,usegraphicx,usenatbib]{mn2e}
\usepackage{times}   %for Times Roman font (used by MNRAS)
\usepackage{mathptm} %Type1 fonts for mathematical symbols
\usepackage{color}

\title[Boosting the accuracy of SPH]
{Boosting the accuracy of SPH techniques: \\
Newtonian and special-relativistic tests}
\author[Rosswog ]
{S. Rosswog\thanks{E-mail: s.rosswog@jacobs-university.de}$^{1,2,3}$\\
$^1$ The Oskar Klein Centre, Department of Astronomy, AlbaNova, Stockholm University, SE-106 91 Stockholm, Sweden\\
$^2$ School of Engineering and Science, Jacobs University Bremen, Campus Ring 1, 28759, Bremen, Germany\\
$^3$ TASC, Department of Astronomy and Astrophysics, University of California, Santa Cruz, CA 95064, USA}

\def\p{\partial}

\def\be{\begin{equation}}
\def\ee{\end{equation}}
\def\bi{\begin{itemize}}
\def\i{\item}
\def\ei{\end{itemize}}
\def\ben{\begin{enumerate}}
\def\een{\end{enumerate}}
\def\bea{\begin{eqnarray}}
\def\eea{\end{eqnarray}}
\def\bt{\begin{tabbing}}
\def\et{\end{tabbing}}

\def\edo{\end{document}}

\def\QIone{\mathcal{Q}_{\; \; 1}}
\def\QItwo{\mathcal{Q}_{\; \; 2}}
\def\QIthree{\mathcal{Q}_{\; \; 3}}
\def\QIfour{\mathcal{Q}_{\; \; 4}}

\newcommand{\enth}{\mathcal{E}}
\def\divv{\nabla \cdot \vec{v}}

\def\M4{M$_4$}
\def\M6{M$_6$}
\def\lcm6{LCM$_6$}
\def\qcm6{QCM$_6$}

\def\wh3{W$_{\rm H,3}$}
\def\wh4{W$_{\rm H,4}$}
\def\wh5{W$_{\rm H,5}$}
\def\wh6{W$_{\rm H,6}$}
\def\wh7{W$_{\rm H,7}$}

\begin{document}

\date{Draft version}

\pagerange{\pageref{firstpage}--\pageref{lastpage}} \pubyear{2013}

\maketitle

\label{firstpage}

\begin{abstract}
We study the impact of different discretization choices on the accuracy of SPH and we
explore them in a large number of Newtonian and special-relativistic benchmark tests. 
As a first improvement, we explore a gradient prescription that requires the (analytical) 
inversion of a small matrix. For a regular particle distribution this improves gradient 
accuracies by approximately ten orders of magnitude and the SPH formulations with this 
gradient outperform the standard approach in all benchmark tests.
Second, we demonstrate that a simple change of the kernel function can substantially increase
the accuracy of an SPH scheme. While the "standard" cubic spline kernel generally performs poorly, 
the best overall performance is found for a high-order Wendland kernel which allows for only 
very little velocity noise and enforces a very regular particle distribution, even in highly dynamical tests. 
Third, we explore new SPH volume elements that  enhance the treatment of fluid instabilities and,
last, but not least, we design new dissipation triggers. They switch on near shocks and in regions
where the flow --without dissipation-- starts to become noisy.
The resulting new SPH formulation yields excellent results even in challenging tests where
standard techniques fail completely.
\end{abstract}

\section{Introduction}
Smoothed Particle Hydrodynamics is a completely mesh-free, fully conservative hydrodynamics method
originally suggested by \cite{lucy77} and \cite{gingold77} in an astrophysical context. By now, SPH has 
spread far beyond its original scope and it has also found a multitude of applications in engineering. 
For detailed overviews over the various aspects of the method and its applications the interested reader 
is referred to recent SPH reviews 
\citep{monaghan05,rosswog09b,springel10a,price12a,monaghan12a,rosswog14c}.\\
SPH has long been appreciated for a number of properties that are highly desirable in an astrophysical 
simulation. One of them is SPH's natural adaptivity that comes without the  burden 
of additional infrastructure such as  an adaptive mesh. SPH is not constrained by any prescribed geometry
(as is usually the case in Eulerian approaches), the SPH particles simply follow the gas flow. 
SPH naturally has tendency to ``refine on density'' and therefore vacuum is modelled with ease: it is 
simply devoid of SPH particles and no computational resources are wasted for modelling it.
In Eulerian approaches, vacuum usually needs to be modelled as a low-density fluid and the interaction
of the ``real'' fluid with the background medium can introduce substantial artifacts such as spurious
friction or shocks which need to be disentangled from the physical results. \\
The probably most salient advantage of SPH, however, is that  the conservation of mass, energy, momentum
and angular momentum can be ``hardwired'' into discrete SPH formulations so that conservation is guaranteed 
independent of the numerical resolution. In practice, this conservation is only limited by, say, time integration
accuracy or the accuracy with which gravitational forces are calculated. These issues, however,
are fully controllable and can be adjusted to the desired accuracy. The exact conservation is 
usually enforced via symmetries in the SPH particle labels together with gradients that are antisymmetric 
with respect to the exchange of two particles. In SPH this is usually ensured by the use of radial kernels, 
$W(\vec{r})= W(|\vec{r}|)$, and the application of direct kernel gradients with the property $\nabla_a W(|\vec{r}_a - \vec{r}_b|)
= - \nabla_b W(|\vec{r}_a - \vec{r}_b|)$. See, for example, Sec. 2.4 in \cite{rosswog09b} for detailed discussion 
of conservation in SPH. Below we will also discuss an alternative gradient estimate that shares
the same antisymmetry properties.\\
Another, highly desirable property is Galilean invariance. For an Eulerian approach it is crucial to perform
a simulation in the best possible reference frame. For example,  a binary system modelled in a space-fixed 
frame may completely spuriously spiral in and merge while the same simulation performed in the frame of
 the binary system would have delivered the correct, stable orbital revolution \citep{new97}. For further examples 
related with Galilean invariance see \cite{springel10b}. Closely related is SPH's perfect advection property:
a property assigned to an SPH particle, say a nuclear composition, is simply (and exactly) carried along 
as the particle moves. This is highly desirable in a number of contexts, for example, when fluid trajectories 
need to be post-processed with nuclear reaction networks.\\
But as every numerical method, SPH has also properties where improvements would be welcome and this
is the topic of this paper. Most SPH codes use artificial viscosity to ensure the Rankine-Hugoniot relations
in shocks. Often, artificial viscosity is considered  a drawback, but a well-designed artificial dissipation 
scheme should perform similar to an approximate Riemann solver. The major challenge
in this respect is to design switches that apply dissipation only where needed and not elsewhere. We discuss 
such switches  in detail in Sec.~\ref{sec:AV}. Another drawback of standard SPH approaches that has received 
much attention in recent years is that, depending on the numerical setup, at contact discontinuities spurious 
surface tension forces can emerge  that can suppress subtle fluid instabilities \citep{agertz07,springel10a,read10}. 
The reason for these surface tension forces are a mis-match in the smoothness of the internal energy and the density. This observation also provides 
strategies to cure the problem, either by smoothing the internal energies via artificial conductivity \citep{price08a} 
or by reformulating SPH in terms of volume elements that differ from the usual choice $m/\rho$ \citep{saitoh13,hopkins13}.
Such approaches are discussed in detail and generalized to the special-relativistic case in 
Sec.~\ref{sec:volume_elements}.\\
SPH as derived from a Lagrangian has a built-in ``re-meshing'' mechanism: while the particles follow the fluid flow,
they try to arrange themselves into optimal relative positions that maximize the sum of the particle volumes. In 
other words, the SPH force is a sum of a regularization term and the approximation to the hydrodynamic force. For
optimally distributed particles the first term vanishes and the second one becomes identical to the Euler momentum
equation. If the particles are, however, arranged in a non-optimal way, they start to move to improve the local distribution and this
will appear as noise. Such motion is unavoidable, for example, in a multi-dimensional shock where the particles
have to transition from the pre- to the post-shock lattice. While this remeshing mechanism is highly desireable, the related
motions should be very subtle. This can be achieved by using very smooth kernels and some dissipation in regions
where such remeshing occurs. Two improvements for this ``noise issue''
are discussed below.  One is related to the choice of the smoothing kernel, see Sec.~\ref{sec:kernel_accuracy},
and the other one to a ``noise trigger'' for artificial dissipation, see Sec.~\ref{sec:AV}.\\
Last but not least, there is a drawback that is related to one of SPH's strengths, its refinement on density. If
low-density regions are geometrically close to high-density regions and they are the major interest of the investigation, 
SPH is seriously challenged: even when huge particle numbers are invested the low density region will always 
be poorly resolved. For such problems, SPH may not be the right method and one may possibly obtain good results 
with substantially less effort by resorting to adaptive mesh refinement methods.\\
In this article, we want to explore a number of technical improvements of SPH methods. Our main focus 
is multi-dimensional, special-relativistic hydrodynamics, but all discussed improvements can straight 
forwardly be applied also in purely Newtonian simulations. To demonstrate that the improvements also 
work in the Newtonian regime, we show a number of tests in this limit. Thereby we address in particular tests
where ``standard'' SPH approaches yield very poor results. As we will demonstrate below, the suggested new
SPH fromulations yield excellent results also in these challenging tests.
All the tests that are shown here --whether 
Newtonian or special-relativistic-- have been performed with a new, special-relativistic SPH code called SPHINCS\_SR.\\
The article is organized as follows. In Sec.~\ref{sec:kernel_estimates} we discuss a number of
kernel interpolation techniques as a basis for the following sections. In Sec.~\ref{sec:kernel_accuracy}
we discuss different kernel functions and assess their accuracy in estimating a constant density and the gradient
of a linear function for the case that particles are arranged as a perfect lattice. We subsequently generalize the
family of SPH volume elements that has recently been suggested by \cite{hopkins13} and an integral-based gradient
estimate \citep{garcia_senz12} to the special-relativistic case. These ingredients
are combined into a number of different SPH formulations in Sec.~\ref{sec:SPH} which are systematically
explored in Sec.~\ref{sec:tests}. By comparing the performance of the different SPH formulations one can 
gauge how important a particular measure is for the considered benchmark test. Our large set of benchmark tests
shows that the accuracy of SPH can be drastically improved with respect to commonly made choices
(high, constant dissipation; standard volume element; kernel gradients and the M$_4$ kernel).
Our results are finally summarized in Sec.~\ref{sec:summary}.

\section{Translation  between Newtonian and special-relativistic approaches}
The focus of this work is the {\em improvement of  SPH techniques}, and we demonstrate them
at Newtonian and --with future applications in mind-- special-relativistic hydrodynamic examples. 
In the following we mainly stay in the special-relativistic picture, but the notation can be straight 
forwardly translated to the Newtonian case:
\bi
   \i {\em Mass}\\ We use here the (fixed) baryon number per SPH particle $\nu$, this corresponds to
     (fixed) SPH particle mass $m$.
  \i {\em Density}\\ In the relativistic treatment we need to distinguish between
      densities measured in our computing frame ($N$) and those measured in the local fluid 
      rest frame ($n$; they differ by a Lorentz factor, see Eq.~\ref{eq:N_vs_n}). In the
      Newtonian limit ($\gamma=1$), of course, both densities coincide. The prescription
      how to calculate $N$, Eq.~(\ref{eq:std_density_SPH}), corresponds to the usual Newtonian density summation for $\rho$.
  \i {\em Velocity}\\ The derivation from a relativistic Lagrangian suggests to use the canonical momentum
     per baryon, $\vec{S}$, as momentum variable. This is beneficial, since the form of the equations becomes very
     similar to the Newtonian case and since it avoids numerical complications (such as time derivatives
     of Lorentz factors) which would otherwise appear \citep{laguna93a}. Inspection of Eq.~(\ref{eq:S_a})
    ($\gamma \rightarrow 1$, $\nu \rightarrow m$, inserting speed of light) shows that the corresponding 
     Newtonian quantity is the velocity.
  \i {\em Energy}\\ Again, the Lagrangian derivation suggests to use the canonical energy per baryon. The resulting
     energy equation looks very similar to the Newtonian equation for the thermokinetic energy, $u + v^2/2$,
     see, for example, Sec. 2.3.2 in \cite{rosswog09b}. But the simplicity of the evolution equations comes at the 
     price of recovering the primitive variables by an iteration at every time step. This topic is discussed
     in Sec.~\ref{sec:recovery}.
  \i {\em Suggested improvements}\\ 
      The suggestions such as the gradient estimate, see Sec.~\ref{sec:kernel_estimates}, kernel choice, 
      see Sec.~\ref{sec:kernel_accuracy}, volume element, see Sec.~\ref{sec:volume_elements},
      and dissipation triggers, see Sec.~\ref{sec:AV}, then carry over straight forwardly. We have not performed 
      extensive tests, though, to double-check whether our parameters can be blindly applied in Newtonian formulations.
\ei

\section{Interpolation and  gradients}
\label{sec:kernel_estimates}

As a basis for the later  SPH discretizations in Sec. \ref{sec:SPH} we briefly discuss 
here some basic properties of discrete kernel interpolations. For now we keep the 
SPH volume element, $V_b$, unspecified, we discuss particular choices in 
Sec.~\ref{sec:volume_elements}. Nothing in this section is specific to the
special-relativistic case. 
In the following, we use the convention that upper indices refer to spatial directions while
lower indices refer to SPH particle identities. Usually, we will use ``a'' for the 
particle of interest and ``b'' for a neighbor particle.
We will follow the convention of summing from 
1 to the number of spatial dimensions $D$ over spatial indices that appear twice. 
In some cases, though, we explicitly write out the summation for clarity.

\subsection{SPH kernel interpolation}
\label{sec:SPH_interpolation}
At the heart of SPH is the smooth representation of a quantity $f$ known at discrete 
("particle") positions $\vec{r}_b$ by
\be
\langle f \rangle (\vec{r})= \sum_b V_b f_b W(|\vec{r}-\vec{r}_b|,h),
\label{eq:SPH_interpolant}
\ee
where $W$ is a smoothing kernel whose width is set by the smoothing length $h$. 
Particular choices of the smoothing kernel are discussed in Sec.~\ref{sec:kernel_accuracy}.
To assess the accuracy of the approximation Eq.~(\ref{eq:SPH_interpolant}) one can 
Taylor-expand $f_b$ around $\vec{r}$ 
\be
f_b\approx f(\vec{r})+ (\vec{r}_b - \vec{r}) \cdot \nabla f(\vec{r}) +  ... 
\ee
and insert it into Eq.~(\ref{eq:SPH_interpolant})
\bea
\langle f \rangle(\vec{r})                                     
\simeq  f(\vec{r})  \sum _b V_b W(|\vec{r}-\vec{r}_b|,h)
+ \nabla f \cdot \sum _b V_b (\vec{r}_b - \vec{r})W(|\vec{r}-\vec{r}_b|,h).    
\label{eq:quality_SPH_interpolation_0}             
\eea
Requiring that $\langle f \rangle(\vec{r})$ be a close approximation to $f(\vec{r})$ then
provides us with the "quality indicators" for the discrete kernel interpolation
\be
\QIone: \quad \sum _b V_b \; W(|\vec{r}-\vec{r}_b|,h) \simeq 1 
\label{eq:quality_SPH_interpolation1}
\ee
\be
\QItwo : \quad \sum _b 
V_b \; (\vec{r}_b - \vec{r}) \; W(|\vec{r}-\vec{r}_b|,h) \simeq 0
\label{eq:quality_SPH_interpolation2}
\ee
which --for a good particle distribution-- should be fulfilled to high accuracy. $\QIone$
simply states that the particles should provide a good partition of unity.

\subsection{Standard SPH gradient}
\label{sec:SPH_gradient}
A standard SPH procedure is to take directly the gradient of the interpolant Eq.~(\ref{eq:SPH_interpolant})
\be
\left(\nabla f \right)_{SPH} (\vec{r})= \sum_b V_b f_b \nabla W(|\vec{r}-\vec{r}_b|,h).
\label{eq:grad_SPH_interpolant}
\ee
Although this estimate is known to be of moderate accuracy only, it is advantageous
because the involved kernel gradient has the desired antisymmetry property, 
$\nabla_r W(|\vec{r}-\vec{r}_b|,h)= - \nabla_{r_b} W(|\vec{r}-\vec{r}_b|,h)$, which makes it
easy to obtain a fully conservative set of SPH equations, see e.g. Sec. 2.4 in \cite{rosswog09b}
for an explicit discussion of numerical conservation in SPH.
One can again proceed as above and insert a Taylor expansion into Eq.~(\ref{eq:grad_SPH_interpolant}) to obtain
\bea
\langle \nabla f \rangle (\vec{r}) = \sum_b V_b \left\{ f(\vec{r}) + (\vec{r}_b - \vec{r}) 
\cdot \nabla f + ... \right\}\nabla W(|\vec{r}-\vec{r}_b|,h).
\label{eq:grad_SPH_interpolant2}
\eea
Requiring $\langle \nabla f \rangle (\vec{r})$ be a close approximation to $\nabla f (\vec{r})$
then delivers the quality indicators of this gradient estimate:
\be
\QIthree: \quad \sum_b V_b \nabla W(|\vec{r}-\vec{r}_b|,h)\simeq 0, 
\label{eq:quality_SPH_gradient1}
\ee
\be
\QIfour: \quad  \sum_b V_b  
(\vec{r}_b - \vec{r})^i \nabla^j W(|\vec{r}-\vec{r}_b|,h)\simeq \delta^{ij}.
\label{eq:quality_SPH_gradient2}
\ee
$\QIthree$ is simply the gradient of $\QIone$ and therefore again an expression
of the partition of unity requirement. 

\subsection{Constant-exact gradient}
\label{sec:LE_gradients}
It is obvious that the gradient estimate Eq.~(\ref{eq:grad_SPH_interpolant}) does not necessarily
vanish  for constant function values $f_b= f_0$, which is sometimes referred to as lack of zeroth
order consistency. The gradient only vanishes in the case when $\QIthree$ is fulfilled exactly. 
This property, however,  can be enforced by simply subtracting the leading error term, see 
Eq.~(\ref{eq:grad_SPH_interpolant2}),
\be
\left(\nabla f \right)_{CE} (\vec{r})= \sum_b V_b (f_b - f(\vec{r})) \nabla W(|\vec{r}-\vec{r}_b|,h),
\label{eq:grad_CE}
\ee
so that now a constant function is reproduced exactly.
For a non-regular particle distribution this substantially improves the gradient estimate, see Sec.~\ref{sec:gradient_accuracy},
but it comes at the price that it does not have the desired antisymmetry and therefore makes it much harder 
to obtain exact conservation.

\subsection{Linear-exact gradient}
\label{sec:LE_gradients}
A linear-exact gradient estimate can be constructed \citep{price04c} starting from 
Eqs.~(\ref{eq:grad_SPH_interpolant}) and (\ref{eq:grad_SPH_interpolant2}) specified 
to particle position $a$
\be
\sum_b V_b f_b \nabla_a W_{ab}= \sum_b V_b \left\{ f_a + (\vec{r}_b - \vec{r}_a) 
\cdot \nabla_a f + ... \right\}\nabla_a W_{ab},
\ee
which can be rearranged into
\be
\sum_b V_b (f_b-f_a) \nabla_a^k W_{ab}= \nabla_a^i f \sum_b 
V_b (\vec{r}_b - \vec{r}_a)^i \nabla_a^k W_{ab}.
\ee
By matrix inversion one obtains a linearly exact gradient estimate
\be
\left(\nabla_a f\right)^i_{\rm LE}= M^{ik} J_f^k,
\label{eq:LEgradient}
\ee
where
\be
M^{ik}= \left(\sum_b V_b (\vec{r}_b - \vec{r}_a)^i \nabla_a^k W_{ab}\right)^{-1}
\label{eq:MIK}
\ee
and
\be
J_f^k= \sum_b V_b (f_b-f_a) \nabla_a^k W_{ab}.
\label{eq:JFK}
\ee
$M^{ik}$ contains information about the local particle distribution while $J_f^k$ 
contains the function values at the neighboring particles. Obviously, the calculation of $M^{ik}$ 
requires the inversion of a $D \times D$-matrix, but this can be done analytically and 
does not represent a major computational burden.

\subsection{Integral-based gradient}
\label{sec:IA_gradients}
More than five decades ago it was realized that derivatives can also be estimated by 
actually performing an integration \citep{lanczos56}. The resulting generalized 
derivative has a number of interesting properties. Among them is its existence 
even where conventional derivatives are not defined and the property that its 
value is the average of the left- and right-hand side limit of the derivative. 
As an example, the Lanczos derivative of $|x|$ at $x=0$ is $D_L(|x|)= 0$. From a 
numerical perspective, this derivative has the desirable property that it is
rather insensitive to noise in the data from which the derivative is to be 
estimated.\\
In an SPH context, integral-based estimates for second derivatives have been 
applied frequently, mainly because they are substantially less noise-prone than 
those resulting from directly taking second derivatives of kernel approximations 
\citep{brookshaw85,monaghan05}. For first order derivatives, however, 
such integral approximations have only been explored very recently 
\citep{garcia_senz12,cabezon12a,jiang14}.
We will assess the accuracy of the integral-based gradient estimates under idealized
conditions in Sec.~\ref{sec:gradient_accuracy}, and, in more practical, dynamical tests
in Sec.~\ref{sec:tests}.\\
The function $f(\vec{r}')$ in the expression 
\be
\tilde{\vec{I}}_f (\vec{r}) \equiv \int [f(\vec{r}') - f(\vec{r})]  \; 
(\vec{r}' - \vec{r}) \; W(|\vec{r} - \vec{r}'|, h) \; dV'
\label{eq:tilde_I}
\ee
can be Taylor-expanded around $\vec{r}$, so that one finds 
\be
\tilde{I}^i_f (\vec{r}) =  \int [ (\nabla f)^k _{\vert_{\vec{r}}} (\vec{r}' - \vec{r})^k]  
\; (\vec{r}' - \vec{r})^i \; W(|\vec{r} - \vec{r}'|, h) \; dV' + O(f'').
\ee
Therefore the gradient component representation, which is exact for linear functions, is given by
\be
(\nabla f)^k (\vec{r})= \tilde{C}^{ki}(\vec{r}) \; \tilde{I}^i_f (\vec{r}),
\label{eq:integral_derivative}
\ee
where the matrix $\tilde{C}$ is the inverse of the symmetric matrix $\tilde{\tau}$ whose components
read
\be
\tilde{\tau}^{ki}(\vec{r}) = \int (\vec{r}' - \vec{r})^k \; 
(\vec{r}' - \vec{r})^i \; W(|\vec{r} - \vec{r}'|, h) \; dV'. 
\label{eq:tilde_tau}
\ee 
$\tilde{\tau}^{ki}$ contains only position information while $\vec{I}$ also contains the function to be
differentiated. 
In the following we will approximate the integrals in Eqs.~(\ref{eq:tilde_I}) and 
(\ref{eq:tilde_tau}) by conventional SPH summations over particles (the resulting summation approximations
have no tilde), which yields
\be
\tau^{ki}(\vec{r})= \sum_b V_b (\vec{r}_b - \vec{r})^k \; (\vec{r}_b - \vec{r})^i \; W(|\vec{r} - \vec{r}_b|, h) 
\ee
and 
\be
\left(\vec{I}_f (\vec{r})\right)_{\rm fIA} =  \sum_b V_b  [f_b - f(\vec{r})]  \; (\vec{r}_b - \vec{r}) \; W(|\vec{r} - \vec{r}_b|, h). \label{eq:fIA}
\ee 
Whenever we use this expression in a gradient estimate, we refer to it as the ``full integral approximation'', or fIA for short.\\
It is worth mentioning that for a radial kernel the gradient can be written as
\be
\nabla_a W_{ab}(h_a)= - \frac{\p W}{\p u} \frac{\vec{r}_b - \vec{r}_a}{h_a |\vec{r}_a - \vec{r}_b|}= (\vec{r}_b - \vec{r}_a) Y_{ab}(h_a),
\label{eq:kernel_gradient}
\ee
where $u= |\vec{r}_a - \vec{r}_b|/h_a$ and $Y$ is also a valid, positively definite and compactly supported kernel function. Therefore,
if Eq.~(\ref{eq:kernel_gradient}) is inserted in Eqs.~(\ref{eq:MIK}) and (\ref{eq:JFK}), one recovers the fIA-gradient formula, i.e.
the LE- and fIA-gradient approach are actually equivalent.\\
If we now assume that the quality indicator $\QItwo$, Eq.~(\ref{eq:quality_SPH_interpolation2}),
is fulfilled to good accuracy we can drop the term containing $f(\vec{r})$ to obtain
\be
\left(\vec{I}_f (\vec{r})\right)_{\rm IA}= \sum_b V_b  f_b   \; (\vec{r}_b - \vec{r}) \; W(|\vec{r} - \vec{r}_b|, h). 
\label{eq:IA}
\ee
We refer to Eq.~(\ref{eq:integral_derivative}) with $\left(\vec{I}_f (\vec{r})\right)_{\rm IA}$ as 
``integral approximation'' or IA for short.
How good this approximation is in practice depends on the regularity of the particle distribution,
see also Sec.~\ref{sec:gradient_accuracy}. As pointed out by \cite{garcia_senz12} this last
approximation breaks the exactness of the gradient of linear functions, but, on the other hand, it rewards 
us with a gradient estimate that is antisymmetric with respect to the exchange of $\vec{r}_b$ and $\vec{r}$.
This is crucial to ensure that the strongest property of SPH, the exact conservation, remains 
preserved. From Eq.~(\ref{eq:IA}) it is obvious that the gradient only vanishes exactly in the case of 
constant $f_b$ if $\QItwo$, Eq.~(\ref{eq:quality_SPH_interpolation2}), is fulfilled exactly.
So rather than having to fulfill several quality criteria for the function interpolant and its 
gradient,  Eqs.~(\ref{eq:quality_SPH_interpolation1}) to (\ref{eq:quality_SPH_gradient2}), we only need 
to ensure the interpolation quality in the form of Eq.~(\ref{eq:quality_SPH_interpolation2}) in order to also 
have accurate gradient estimates.\\
So our gradient estimate in integral approximation reads explicitly
\bea
(\nabla f)_{\rm IA}^k (\vec{r})&=& C^{kd}(\vec{r}) \; \left(I^d_f (\vec{r})\right)_{\rm IA}\nonumber\\
                                   &=& \sum_b V_b f_b \sum_{d=1}^{D} C^{kd}(\vec{r},h) 
(\vec{r}_b - \vec{r})^d W(|\vec{r}-\vec{r}_b|,h).
\label{eq:integral_gradient}
\eea
From the comparison with Eq.~(\ref{eq:grad_SPH_interpolant}) it is obvious that the second
sum takes over the role that is usually played by the kernel gradient:
\be
\nabla^k W(|\vec{r} - \vec{r}_b|,h) \rightarrow 
\sum_{d=1}^{D} C^{kd}(\vec{r},h) (\vec{r}_b - \vec{r})^d W(|\vec{r}-\vec{r}_b|,h).
\label{eq:form_replacement}
\ee
We will make use of this replacement in Sec.~\ref{sec:IA_SPH} to obtain an alternative SPH formulation
with integral-based derivative estimates.

\subsection{Assessment of the gradient accuracy}
\label{sec:gradient_accuracy}
%-----------------------------------------------------------------------
\begin{figure*} 
   \centering
   \centerline{\includegraphics[width=10cm,angle=0]{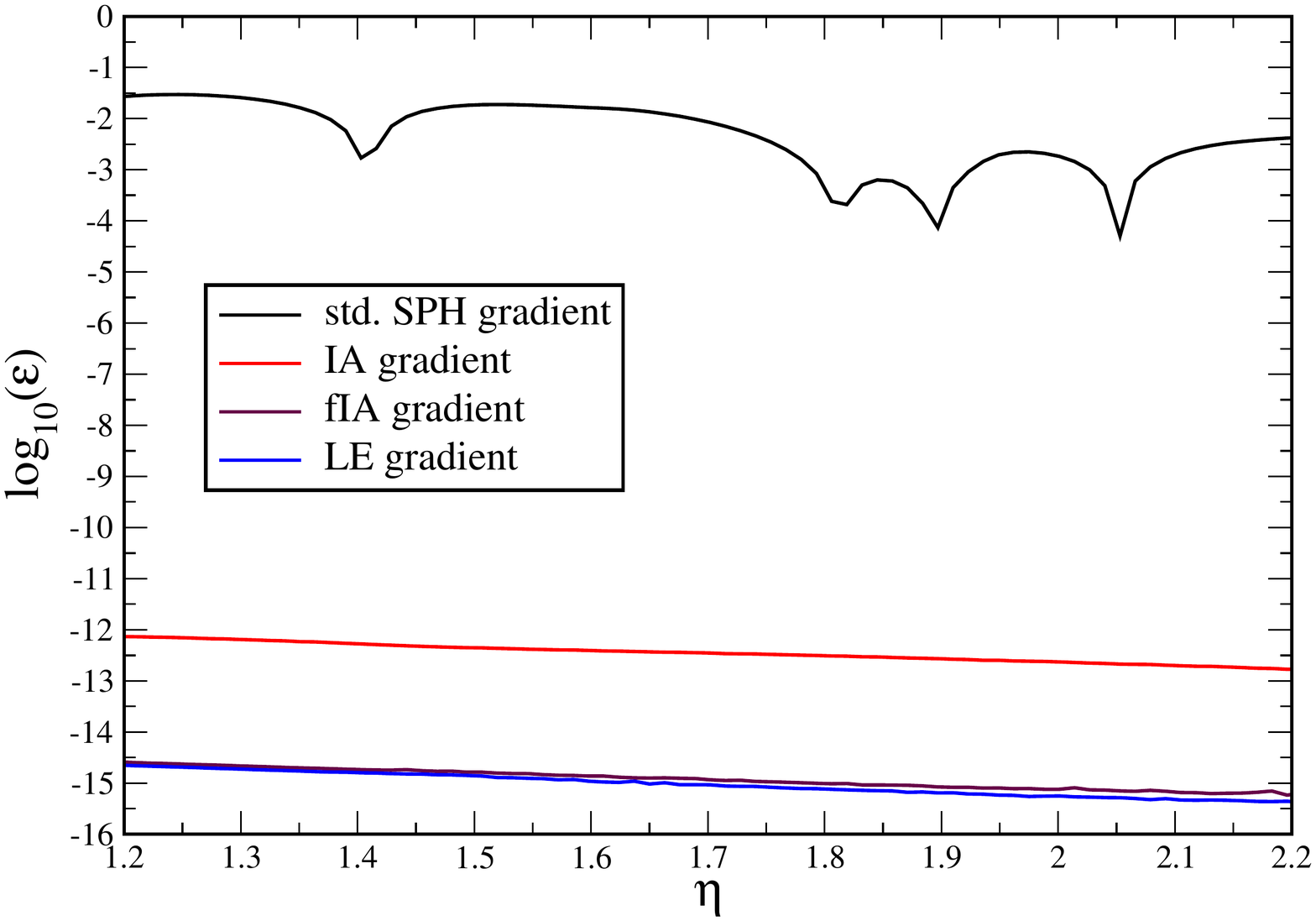}\hspace*{-0.5cm}
                     \includegraphics[width=10cm,angle=0]{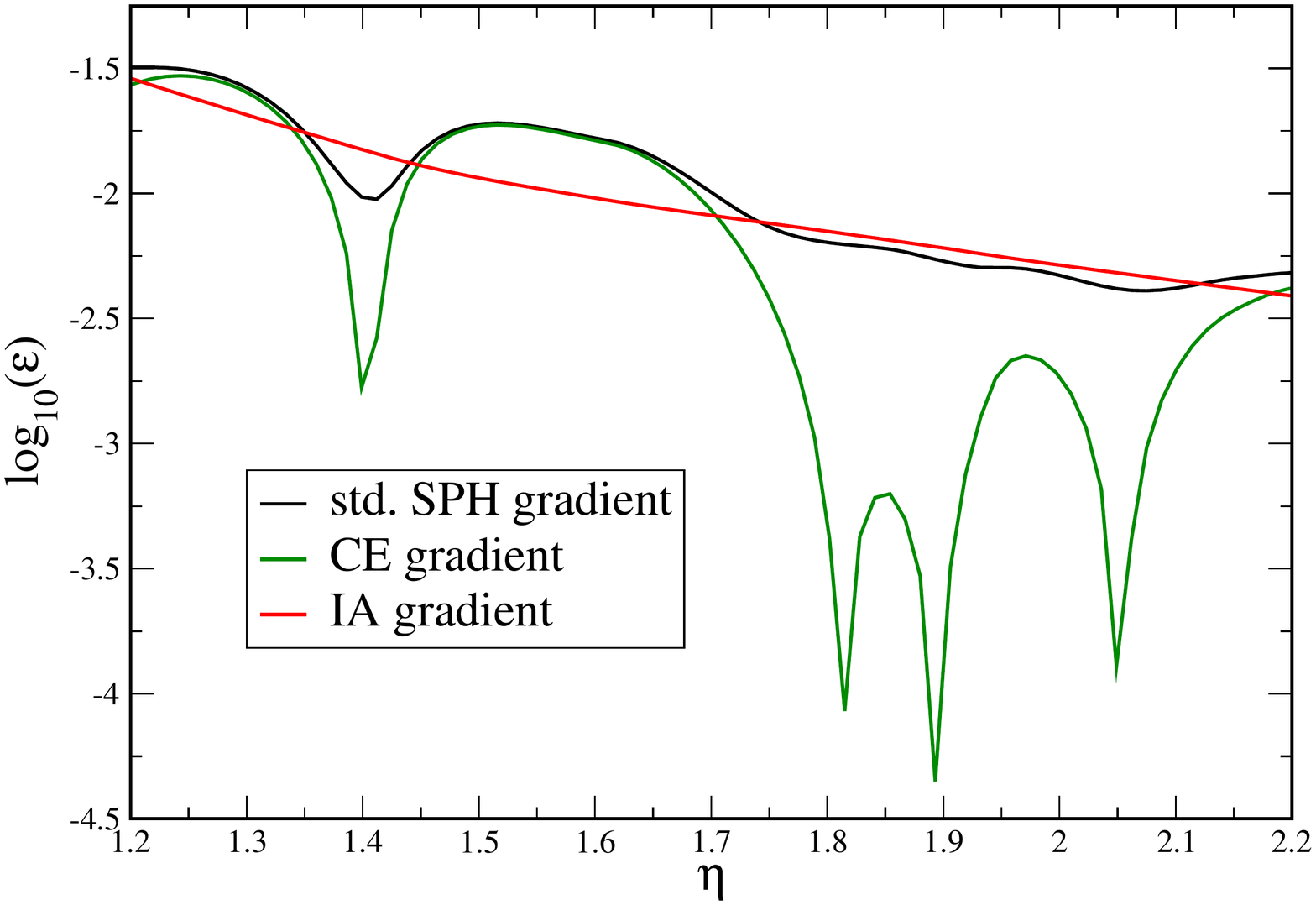}}
    \vspace*{-0.4cm}
    \caption{Sensitivity of different gradient prescriptions to the regularity of the particle distribution. The left panel shows results
                  for a perfect 2D hexagonal lattice corresponding to the closest packing of spheres of radius $r_s$.
                  The right panel shows results for a slightly perturbed hexagonal lattice that was obtained by 
                  displacing each particle in a hexagonal lattice in a random direction by a distance $\Delta r_a$ 
                  that has been randomly chosen from $[0,10^{-3} r_s]$.
                  The parameter $\eta$ determines the smoothing length via $h_a= \eta V_a^{1/D}$, where $V_a$ is the particle
                  volume.
                  ``Std. SPH gradient'' refers to  the direct gradient of the SPH interpolant 
                  Eq.~(\ref{eq:grad_SPH_interpolant}),  ``CE gradient'' stands for ``constant exact gradient'' and is
                  calculated according to Eq.~(\ref{eq:grad_CE}), ``IA gradient'' is calculated from
                  Eq.~(\ref{eq:integral_gradient}),  the ``fIA gradient'' from Eq.~(\ref{eq:fIA})
                  and the ``LE  gradient'' from Eq.~(\ref{eq:LEgradient}).  For a regular particle distribution the gradient
                  estimate can be improved by about ten orders of magnitude by using the IA-prescription (left; the CE
                  gradient coincides with the standard SPH estimate and is therefore not shown). But even
                  a small perturbation of the lattice degrades the gradient quality to an accuracy similar to the
                  standard SPH estimate (right panel). The LE- and the fIA gradient are hardly affected and therefore not shown in the right panel.
                  See text for more details.}
   \label{fig:acc_different_gradients}
\end{figure*}
%-----------------------------------------------------------------------
We briefly want to  assess the accuracy of the different gradient estimates 
Eqs.~(\ref{eq:grad_SPH_interpolant}), (\ref{eq:grad_CE}), (\ref{eq:LEgradient}), 
(\ref{eq:fIA})  and (\ref{eq:integral_gradient}) (with the standard cubic spline kernel) 
in a numerical experiment. 
Our experiment is similar to the one in \cite{rosswog10b}: we set up particles on a 2D 
hexagonal lattice in $[-1,1] \times  [-1,1]$, corresponding to a close-packed distribution 
of spheres with radius $r_s$. The particles are assigned the same baryon number/masses
and pressures that rise linearly with the x-coordinate so that the slope is 
$\p_x P=1$. The numerical gradient estimate, $(\p_x P)_{\rm num}$, is calculated via 
Eq.~(\ref{eq:grad_SPH_interpolant}) (``SPH-gradient''), the linear-exact 
gradient (``LE-gradient''), Eq.~(\ref{eq:LEgradient}), Eq.~(\ref{eq:fIA}) (full integral 
approximation, ``fIA-gradient'') and Eq.~(\ref{eq:IA}) (``IA-gradient''). In 
Fig.~\ref{fig:acc_different_gradients} we display the error 
$\epsilon= |(\p_x P)_{\rm num} - \p_x P|/|\p_x P|$ as a
function of the parameter $\eta$ by which we set the smoothing length
\be
h_a= \eta V_a^{1/D}
\label{eq:h_vol}
\ee 
based on the particle volume $V_b$. For this perfectly regular particle distribution the quality indicators, 
Eq.~(\ref{eq:quality_SPH_interpolation1}) and (\ref{eq:quality_SPH_interpolation2}), are fulfilled 
to high accuracy and therefore the constant exact (CE) gradient is practically 
identical to the standard  SPH estimate and therefore it is not shown in the left panel of 
Fig.~\ref{fig:acc_different_gradients}. For the same reason the IA-approximation, 
Eq.~(\ref{eq:IA}), is very accurate and yields a gradient estimate (red) that is {\em roughly ten orders 
of magnitude better than the standard SPH gradient estimate} (black). The full 
integral approximation and the linear-exact prescription reproduce the exact result to 
within machine precision, but, as noted above, they lack the desirable antisymmetry property 
that facilitates exact numerical conservation.\\ 
As expected from the term that was neglected to obtain Eq.~(\ref{eq:integral_gradient}), the quality
of antisymmetric gradient estimate is sensitive to the particle distribution. To illustrate this,
we perform a variant of the previous experiment in which we slightly perturb the perfect
hexagonal lattice. To each particle position $\vec{r}_a$ we add a randomly oriented vector $\Delta \vec{r}_a$ 
whose modulus is chosen randomly from the interval $[0, 10^{-3} r_s]$. Even this very subtle 
perturbation substantially degrades the accuracy of those gradients that do not account
for the actual particle distribution, i.e. for the standard SPH- and the IA-gradient prescriptions.
The CE-, LE- and fIA-gradient estimates make use of the information on the local particle distribution
and are therefore hardly deteriorated. For the perturbed distribution,
the IA-gradient has no more obvious advantage with respect to the standard SPH
gradient, both now show comparable errors. Therefore,  further, dynamical, tests are required to 
see whether a regular enough particle distribution can be maintained during dynamical simulation
and to judge whether the IA-gradient prescription really improves the accuracy in practice. As will 
be shown below, however, and consistent with the findings of \cite{garcia_senz12}, the integral-based, 
antisymmetric IA-approach by far outperforms the traditional  direct kernel gradients in all dynamical tests.

\section{Kernel choice}
\label{sec:kernel_accuracy}

\subsection{Kernels}
In Sect. \ref{sec:kernel_estimates} we have briefly collected some kernel interpolation basics.
Traditionally, ``bell-shaped'' kernels with vanishing derivatives at the origin have been preferred
as SPH kernels
because they are rather insensitive to the exact positions of nearby particles and therefore they are good 
density estimators \citep{monaghan92}. For most bell-shaped kernels, however, a ``pairing instability'' sets 
in once the smoothing length exceeds a critical value. When it sets in, particles start to form pairs and,
in the worst case, two particles are effectively replaced by one. This has no dramatic effect other than halfing the 
effective particle number, though, still at the original computational expense. Recently, there have 
been several studies that (re-)proposed peaked kernels \citep{read10,valcke10} as a possible cure
for the pairing instability. Such kernels have the benefit of producing very regular particle distributions, 
but as the experiments below show, they require the summation over many neighboring particles 
for an acceptable density estimate. It has recently been pointed out by \cite{dehnen12}, however, that, contrary
to what was previously thought, the pairing instability is {\em not} caused by the vanishing central
derivative, but instead a necessary condition for stability against pairing is a non-negative Fourier transform
of the kernels. For example, the Wendland kernel family (out of which we explore one member below),
has a vanishing central derivative, but does not fall pray to the pairing instability. This is, of course,
advantageous for convergence studies since  the kernel support size is not restricted.\\
In the following, we collect a number of kernels whose properties are explored below. We give the kernels
in the form in which they are usually presented in the literature, but to ensure a fair comparison in tests,
we scale them to a support of $2h$, as the most commonly used SPH kernel, M$_4$, see below. 
So if a kernel has a normalization $\sigma_{lh}$ for a support of $l h$,  it has, in $D$ spatial 
dimensions, normalization $\sigma_{kh}= (l/k)^D \sigma_{lh}$ when it is stretched to a support of $k h$. 
%-----------------------------------------------------------------------
\begin{figure*}
   \centering
   \centerline{\includegraphics[width=10cm,angle=0]{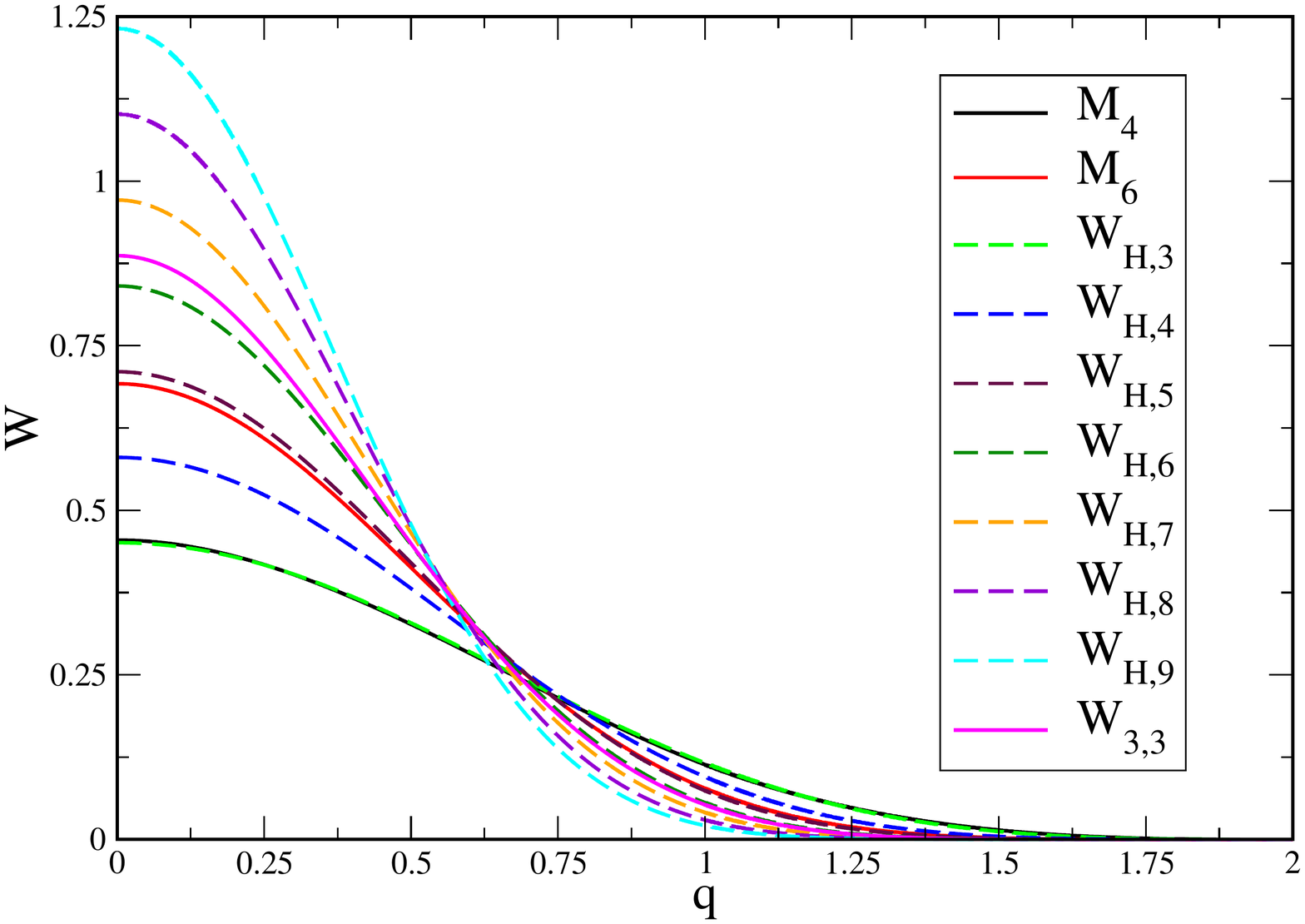} \hspace*{-0.5cm}
                     \includegraphics[width=10cm,angle=0]{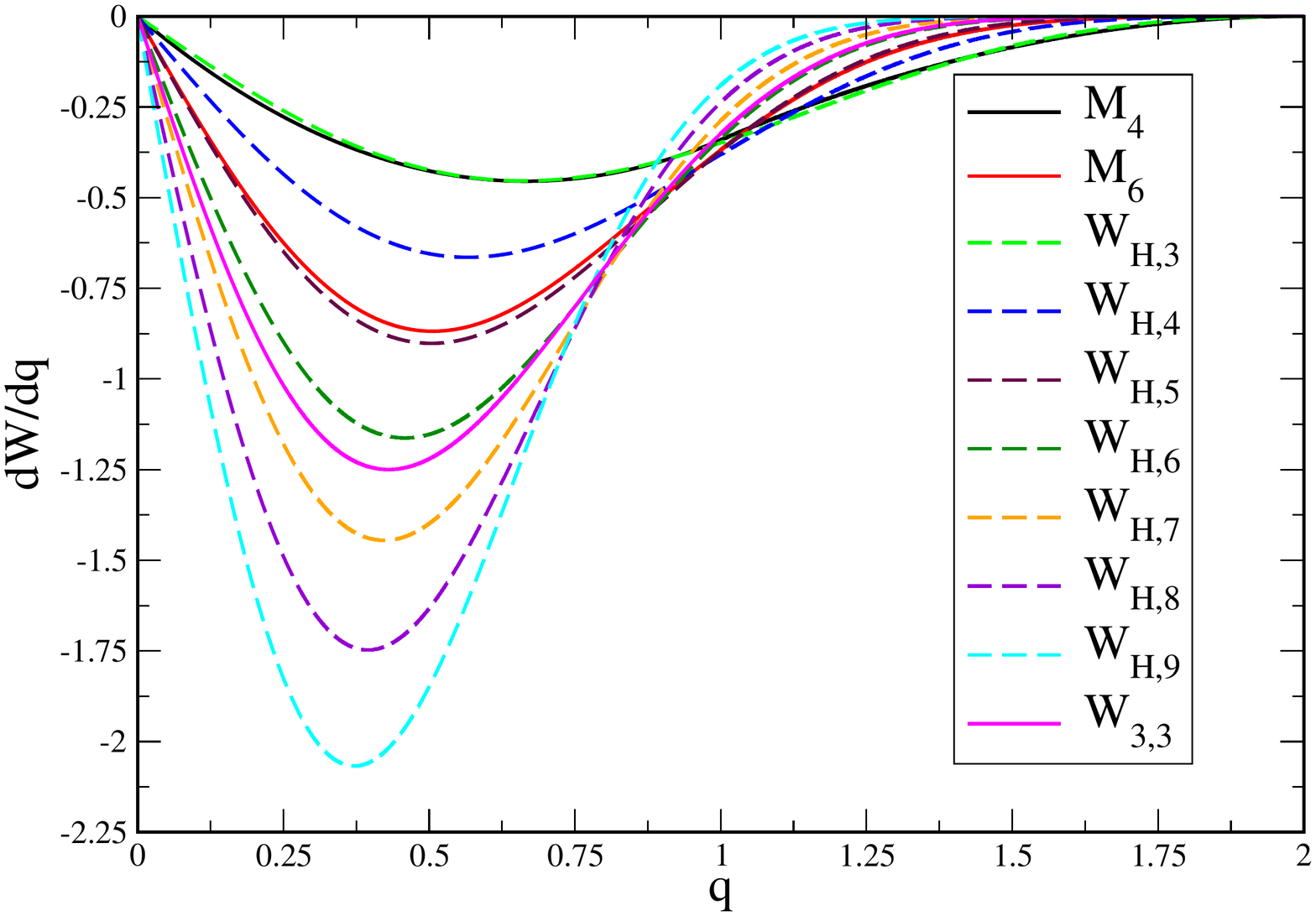}}
   \caption{Comparison of the standard spline kernels M$_4$ and M$_6$  with the kernel family 
                  W$_{H,n}$ (kernels left, their derivatives right). Note that for ease of comparison the
                  M$_6$ kernel has been stretched to a support of $R_k= 2$. W$_{\rm H,3}$ is a close approximation 
                  of M$_4$ and, if scaled to the same support,  W$_{\rm H,5}$ is similar to M$_6$. Also shown 
                  is the Wendland kernel $W_{3,3}$.}
   \label{fig:smooth_kernels}
\end{figure*}
%-----------------------------------------------------------------------
%-----------------------------------------------------------------------
\begin{figure*} 
   \centering
   \centerline{\includegraphics[width=10cm,angle=0]{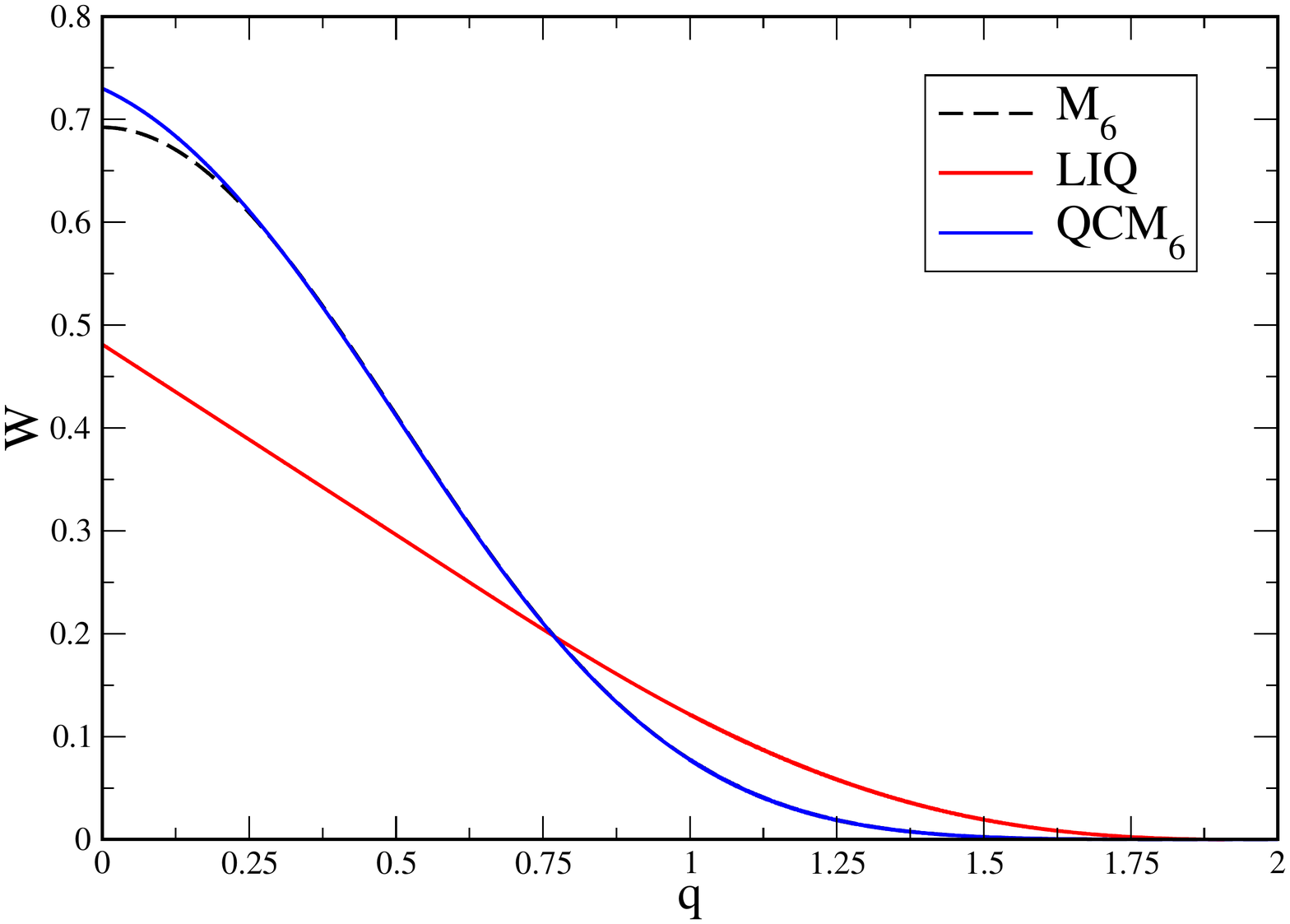} \hspace*{0.5cm}
                     \includegraphics[width=10cm,angle=0]{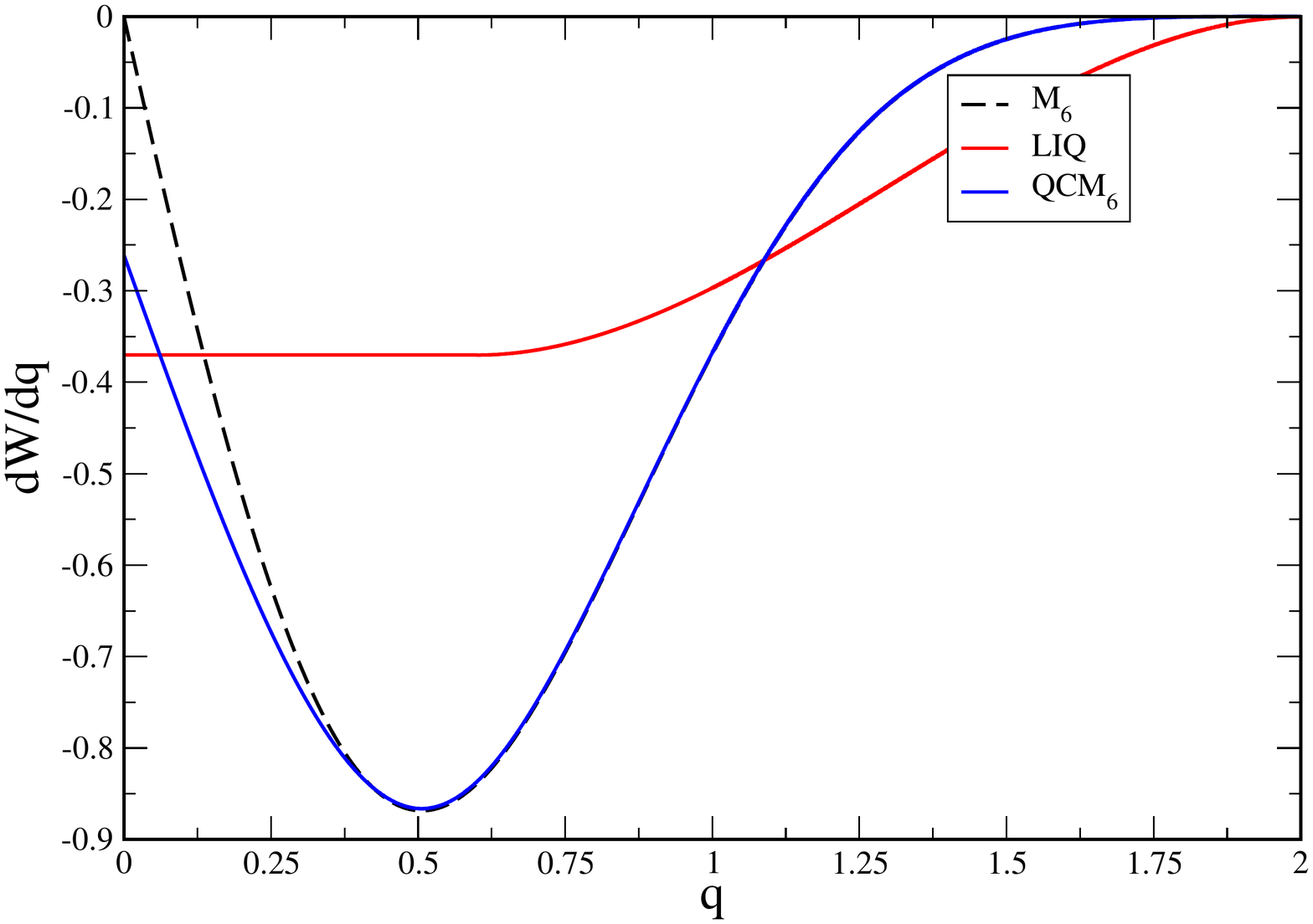}}
    \caption{The peaked kernels LIQ and QCM$_6$: kernel $W$ on the left and its derivative 
                  $dW/dq$ on the right. The QCM$_6$ kernel has been constructed so that it 
                  hardly deviates from M$_6$, but has a non-vanishing central derivative. 
                  For comparison also the M$_6$ kernel plotted (dashed lines).}
   \label{fig:peaked_kernels}
\end{figure*}
%-----------------------------------------------------------------------
\subsubsection{Kernels with vanishing central derivatives}
{\bf B-spline functions: M$_4$ and M$_6$ kernels}\\
The most commonly used SPH kernels are the so-called B-spline 
functions \citep{schoenberg46}, $M_n$, which are generated as Fourier transforms:
\be
M_n(x,h)= \frac{1}{2 \pi} \int_{-\infty}^\infty \left[ \frac{\sin(kh/2)}{kh/2}\right]^n \cos(kx) dk.
\ee
The smoothness of the $M_n$ functions increases with $n$ and they are
continuous up to the $(n-2)$-th derivative. Since SPH requires at the very least the continuity in the 
first and second derivative, the cubic spline kernel M$_4$ 
\be
w_4(q)=  \left\{\begin{array}{ll}  \frac{1}{4}(2-q)^3 - (1-q)^3
& 0 \le q < 1\\
         \frac{1}{4}(2-q)^3 & 1 \le q < 2\\
         0      & {\rm else}
         \end{array}\right.
\ee
is the lowest-order member of this kernel family that is a viable option. It is often considered the ``standard choice'' in SPH.
The normalized SPH kernel then has the form
\be
W(|\vec{r}-\vec{r}'|,h)= \frac{\sigma_n}{h^D} w_n(q), \label{eq:normalized_SPH_kernel}
\ee
where\footnote{We use the convention that $W$ refers to the full normalized kernel while
$w$ is the un-normalized shape of the kernel.} $q= |\vec{r}-\vec{r}'|/h$. The normalizations are obtained from
\be
\sigma^{-1}= \left\{\begin{array}{ll}  
                        2 \int_0^{Q} w(q) dq  & {\rm in \; 1D}\\\\
                          \int_0^{Q} w(q) \; 2 \pi \; q\; dq & {\rm in \; 2D}\\\\
                          \int_0^{Q} w(q) \; 4 \pi \; q^2 \; dq & {\rm in \; 3D},
                        \end{array}\right. \\
\label{eq:normalization}
\ee
where $Q$ is the kernel support which is equal to 2 for M$_4$. This yields values of 
$[2/3,10/(7 \pi),1/\pi]$ in 1, 2 and 3 dimensions.\\
The M$_6$ kernel (truncated at support $Q= 3$) is given as
\be
w_6(q)= \left\{\begin{array}{ll}  
         (3-q)^5 - 6(2-q)^5 + 15(1-q)^5 & 0 \le q < 1\\
         (3-q)^5 - 6(2-q)^5                       & 1 \le q < 2\\
         (3-q)^5                                          & 2 \le q < 3\\
                           0      & {\rm else}
         \end{array}\right. 
\ee
with normalizations of $[1/120, 7/(478 \pi), 1/(120 \pi)]$ in 1, 2 and 3 dimensions. Note
that for a fair comparison in all plots  the M$_6$ kernel is rescaled to a support on $2h$. \\

\noindent{\bf A parameterized family of Kernels}\\
More recently, a one parameter family of kernels has been suggested \citep{cabezon08}
\be
W_{{\rm H},n}= \frac{\sigma_{{{\rm H},n}}}{h^D} \left\{\begin{array}{ll}  
         1 & q = 0\\
         \left( \frac{\sin [\frac{\pi}{2} q]}{\frac{\pi}{2} q}\right)^n    & 0 < q \le 2\\
         0      & {\rm else,}
         \end{array}\right. 
\ee
where $n$ determines the smoothness and the shape of the kernel, see Fig.~\ref{fig:smooth_kernels}.
The normalization of this kernel family can be expressed as a fifth order polynomial whose coefficients can
be found in Tab. 2 of \cite{cabezon08}. In this form, $n$ is allowed to vary continuously between 2 and 7. 
We use here exactly the form described in their paper, in particular all kernels have a support of $R_k=2$.
The $W_{{\rm H},3}$ kernel is a very close approximation to the  M$_4$, see Fig.~ \ref{fig:smooth_kernels},
while $W_{{\rm H},5}$ is very similar to M$_6$, provided they are stretched to have the same support. For practical
calculations we use the $W_{{\rm H},n}$ kernels for $n$-values from 3 to 9, the corresponding normalizations are
given in Table~\ref{tab:WHn_normal}.\\
\begin{table*}
\centerline{\bf Normalization $\sigma_{H,n}$ of the W$_{H,n}$ kernels}

\vspace*{0.2cm}

\begin{tabular}{ l | l l l l l l l}
               & n= 3    & n= 4             & n= 5            & n= 6            & n= 7             & n= 8             & n= 9 \\
\hline \\
1D  & 0.66020338 & 0.75221501 & 0.83435371 & 0.90920480 & 0.97840221 & 1.04305235  & 1.10394401\\
2D  & 0.45073324 & 0.58031218 & 0.71037946 & 0.84070999 & 0.97119717 & 1.10178466  & 1.23244006\\
3D  & 0.31787809 & 0.45891752 & 0.61701265 & 0.79044959 & 0.97794935 & 1.17851074  & 1.39132215\\
\end{tabular}
\caption{Normalization constant $\sigma_{H,n}$ of the W$_{H,n}$ kernels in one, two and three dimensions.}
\label{tab:WHn_normal}
\end{table*}

\noindent{\bf Wendland kernels}\\
An interesting class of kernels with compact support and positive Fourier transforms
are the so-called Wendland functions \citep{wendland95}. In various areas of applied mathematics
they have long been appreciated for their good interpolation properties, but they 
have not received much attention as SPH kernels. Recently, they have been discussed
in some detail in \cite{dehnen12}, where it was in particular found that these kernels
are not prone to the pairing instability, despite having a vanishing central 
derivative. We only experiment here with one particular example, the $C^6$ smooth
\be
W_{3,3}= \frac{\sigma_W}{h^3} \left(1 - q \right)_{+}^8 \left(  32 q^3 + 25 q^2 +  8 q + 1 \right)
\label{eq:wend33},
\ee
see e.g. \cite{schaback06}, where the symbol $(.)_+$ denotes the cutoff function $\rm{max}(.,0)$. 
The normalization $\sigma_W$ is 78/(7$\pi$) and 1365/(64 $\pi$) in  2 and 3 dimensions. 
As we will demonstrate in the below benchmark tests, this kernel has some very interesting properties,
in particular it maintains a highly ordered particle distribution in dynamical simulations and it
does not fall prey to the pairing instability. \\

\subsubsection{Kernels with  non-vanishing central derivatives}
Here, we  briefly discuss two kernel functions with non-vanishing central derivatives. The first example
is the ``linear quartic'' (LIQ) kernel that has been suggested \citep{valcke10} to achieve a regular particle distribution and to
improve SPH's performance in Kelvin-Helmholtz tests. The second example is shown mainly for pedagogical
reasons: it illustrates how a very subtle change to the core of the well-appreciated M$_6$ kernel seriously compromises
its accuracy.\\

\noindent{\bf Linear quartic kernel, LIQ}\\
The centrally peaked "linear quartic" (LIQ) kernel \citep{valcke10} reads
\be
W_{\rm LIQ}(q)= \frac{\sigma_{\rm LIQ}}{h^D}   \left\{
  \begin{array}{ l l l}
     F - u   \hspace*{2.8cm} {\rm for \; } u \le x_s\\
     A u^4 + B u^3 + C u^2 + D u + E  \;\; {\rm for \; }  x_s < u \le 1\\
     0   \hspace*{3.4cm} {\rm else } 
       \end{array} \right.\ee
with $x_s=0.3, A=-1.458, B=3.790, C=-2.624, D=-0.2915, E=0.5831$ and $F= 0.6500$ and $u= q/2$. The 
normalization constant $\sigma_{\rm LIQ}$ is 2.962 in 2D and 3.947 in 3D.\\ 

\noindent{\bf Quartic core M$_6$ kernel, QCM$_6$}\\
We briefly explore a modification of the M$_6$ kernel so that it remains very smooth, but
has a non-vanishing derivative in the center. This {\em quartic core $M_6$ kernel (QCM$_6$)} 
is constructed by replacing the second derivative of the $M_6$ kernel for $q<q_c$ by a 
parabola whose parameters have been chosen so that it fits smoothly and differentially the 
$M_6$ kernel at the transition radius $q_c=0.759298$ which is defined by the condition 
$d^2 w_6/dq^2(q_c)= 0$. The QCM$_6$-kernel then reads:
\be
W_{QCM_6}(q)= \frac{\sigma_{QCM_6}}{h^D}  \left\{\begin{array}{ll}  
         A q^4 + B q^2 + C q + D& 0 \le q < q_c\\
         (3-q)^5 - 6(2-q)^5 + 15(1-q)^5& q_c \le q < 1\\
         (3-q)^5 - 6(2-q)^5            & 1 \le q < 2\\
         (3-q)^5                       & 2 \le q < 3\\
                           0      & {\rm else .}
         \end{array}\right. 
\ee
The coefficients $A, B, C$ and $D$ are determined from the conditions $w_{\rm QCM_6}(q_c)= w_6(q_c)$,
$w'_{\rm QCM_6}(q_c)= w'_6(q_c)$, $w''_{\rm QCM_6}(q_c)= w''_6(q_c)$ and $w'''_{\rm QCM_6}(q_c)= w'''_6(q_c)$, 
where the primes indicate the derivatives with respect to $q$. The resulting numerical values are 
given in Table \ref{tab:params_QCM6}. Note that $QCM_6$ is continuous everywhere up to the third derivative.
The peaked kernels LIQ and QCM$_6$ are compared in Fig.~\ref{fig:peaked_kernels}.
\begin{table}
\begin{tabular}{ l l }
  parameter  & numerical value  \\
  \hline
  $A$                           &   11.017537\\
  $B$                           &   -38.111922\\
  $C$                           &  -16.619585 \\
  $D$                           &   69.785768 \\  
  $\sigma_{\rm 1D}$ &     8.245880  E-3 \\
  $\sigma_{\rm 2D}$ &     4.649647  E-3\\
  $\sigma_{\rm 3D}$ &     2.650839  E-3\\\\
 \end{tabular}
 \caption{Parameters of the Quartic-Core-$M_6$ kernel (QCM$_6$).}
 \label{tab:params_QCM6}
\end{table}

\subsection{Accuracy assessment}
\subsubsection{Kernel support size}
We give in Tab.~\ref{tab:eta_crit} the values $\eta$ that are used in the numerical experiments. They are 
chosen to be very large, but small enough to avoid pairing. Also given is the value $q_c$, where $|dW/dq|$
has its maximum, $\eta_c= 1/q_c$ and the corresponding neighbor number for a hexagonal lattice, $N_c$. 

\subsubsection{Density estimates}
%-----------------------------------------------------------------------
\begin{figure*} 
\centerline{
   \includegraphics[width=18cm,angle=0]{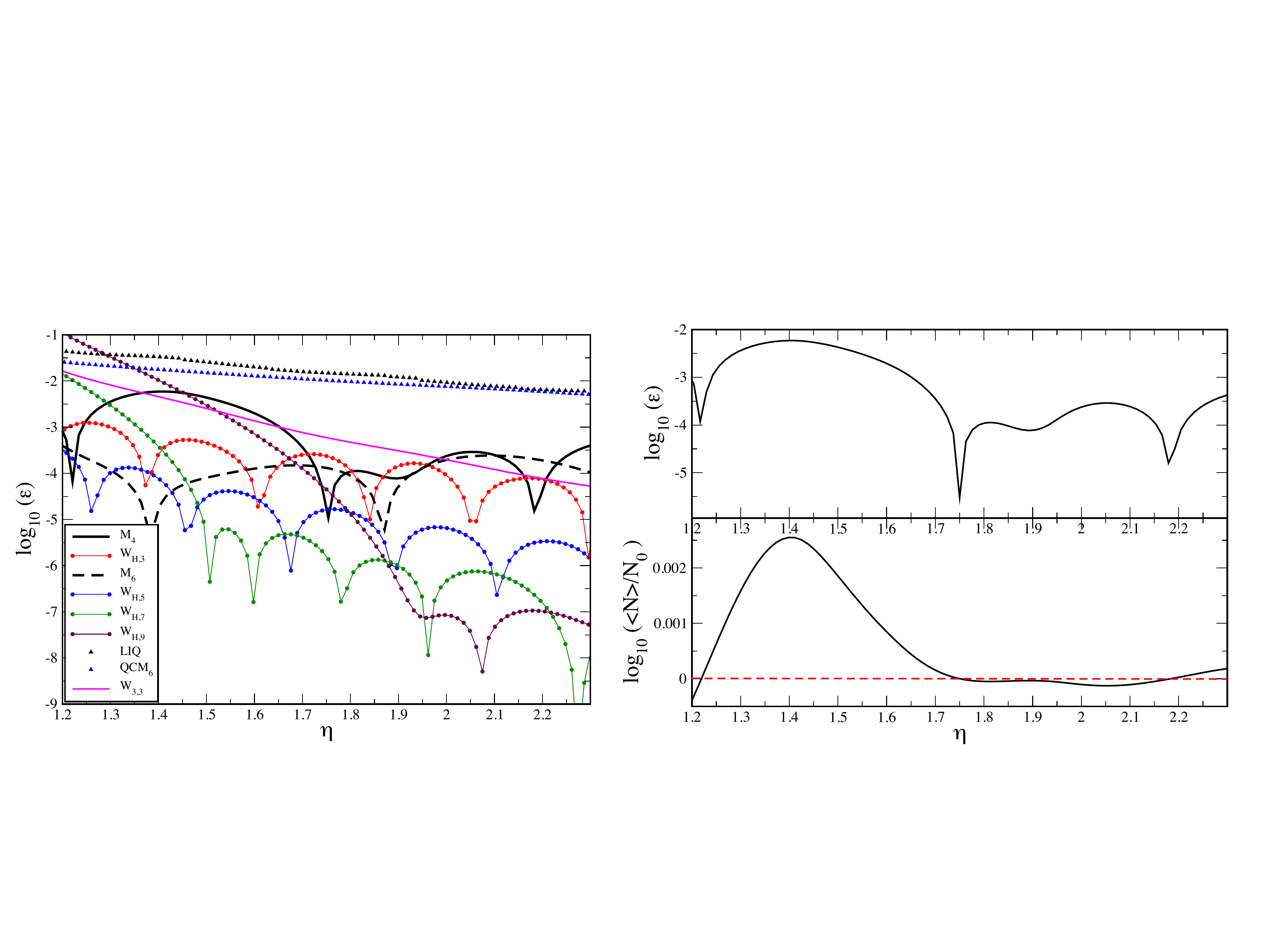} }
   \vspace*{-3cm}
   \caption{Accuracy of density estimates for different kernels.
                 Particles are placed on a hexagonal lattice so that the density is uniform
                 and the SPH density estimate is compared with the theoretical value to calculate
                 the relative density error, see left panel. The parameter $\eta$ determines the smoothing length, 
                 $h_a= \eta V_a^{1/D}$;  LIQ: linear-quartic,  QCM$_6$: quartic core M$_6$). To clarify the nature
                 of the ``dips'' in the left panel, we plot in the right panel the logarithm of the relative density
                 error (upper right) and $\log(\langle N \rangle/N_0)$ for the case of the cubic spline kernel M$_4$.
                 The dips occur when the density error changes sign (exact value is indicated by the dashed red line).
                 }
   \label{fig:dens_accuracy}
\end{figure*}
%-----------------------------------------------------------------------

To assess the density estimation accuracy, we perform a simple experiment:
we place the SPH particles on a hexagonal lattice and assign all of them a constant value for the baryon 
number\footnote{This is equivalent to assigning a constant particle mass for the Newtonian case.},
$\nu_0$. Since 
the effective area belonging to each particle with radius $r_s$ in a close packed configuration is $A_{\rm eff}= 2 \sqrt{3} r_s^2$, 
the theoretical density value is $N_0= \nu_0/A_{\rm eff}$.
We now measure the average relative deviation between the estimated and the 
theoretical density value, $\epsilon_{N}= \sum_{a=1}^{N_{\rm part}} |N_a - N_0|/(N_0 \; 
N_{\rm part})$, the results are shown in the left panel of Fig.~\ref{fig:dens_accuracy}
as a function of the smoothing length parameter $\eta$, see Eq.~(\ref{eq:h_vol}). 
For clarity, we only show the odd members of the W$_{\rm H,n}$ family. Note that we 
show the results up to large values of $\eta$ where, in a dynamical simulation, some of the kernels would
already have become pairing unstable. The ``dips'' in the left panel occur where the density error changes
sign. The right panel of Fig.~\ref{fig:dens_accuracy} illustrates this for the case of the M$_4$ kernel. \\
The ``standard'', cubic spline kernel M$_4$ does not perform particularly well and simply replacing it by, say,
the M$_6$ kernel increases the density estimate already by roughly two orders of magnitude. If larger smoothing
length can be afforded, however, the density estimate can be further substantially improved. For example, the
W$_{\rm H,9}$ kernel for $\eta > 2$ achieves in this test approximately four orders of magnitude lower errors
than what the M$_4$ can achieve for $\eta=1.2$ (for larger values it becomes pairing unstable). The Wendland
kernel achieves a continuous improvement in the density estimate with increasing $\eta$ and, as \cite{dehnen12}
argue, this protects the kernel against becoming pairing unstable. Although the $W_{3,3}$ results
in this test are less accurate than those of the higher-order W$_{H,n}$ kernels, we gave preference to the Wendland
kernel as our standard choice, since $W_{3,3}$ allows only for very little noise, see below.\\
Both peaked kernels perform very poorly in this test, even for a very large support.  It is particularly interesting to
observe how the subtle change of the core of the M$_6$ kernel (compare dashed black and the blue curve in 
Fig.~\ref{fig:peaked_kernels}) seriously compromises the density estimation ability (compare the dashed black curve
with the blue triangles in Fig.~\ref{fig:dens_accuracy}).

\begin{table}
\begin{tabular}{ l l l l l}
  kernel        & $q_c$ & $\eta_c$ & $N_c$ & $\eta$\\
  \hline
  M$_4$           & 0.667 &   1.500  & 28   & 1.2\\
  M$_6$           & 0.506 &   1.976  & 49  & 1.6\\
  $W_{{\rm H},3}$ & 0.661 &   1.512  &  28  & 1.2\\  
  $W_{{\rm H},4}$ & 0.567 &   1.765  & 39   & 1.5\\  
  $W_{{\rm H},5}$ & 0.504 &   1.984  & 49   & 1.6\\
  $W_{{\rm H},6}$ & 0.458 &   2.183  & 59   & 1.7\\
  $W_{{\rm H},7}$ & 0.423 &   2.364  & 70   & 1.8\\
  $W_{{\rm H},8}$ & 0.395 &   2.531  & 80   & 1.9\\
  $W_{{\rm H},9}$ & 0.372 &   2.690  & 90   & 2.2\\
  $W_{3,3}$        & 0.430  &  2.323  & n.a.  & 2.2\\ 
 LIQ                 &  n.a.    &   n.a.     & n.a. & 2.2\\         
 QCM$_6$        & 0.506 &   1.976  & 49   & 2.2
\\
 \end{tabular}
 \caption{Values $q_c$ where the kernel derivative $|dW(q)/dq|$ has the maximum, $\eta_c= 1/q_c$.
               $N_c$ is the neighbor number for a hexagonal lattice that corresponds to $q_c$
               and $\eta$ is the value used in our experiments. The latter has been chosen so that it is
               as large as possible without becoming pairing unstable, or, in cases where no pairing occurs,
               as large as computational sources allowed.} 
 \label{tab:eta_crit}
\end{table}

%-----------------------------------------------------------------------
\begin{figure} 
\centerline{
   \includegraphics[width=10cm,angle=0]{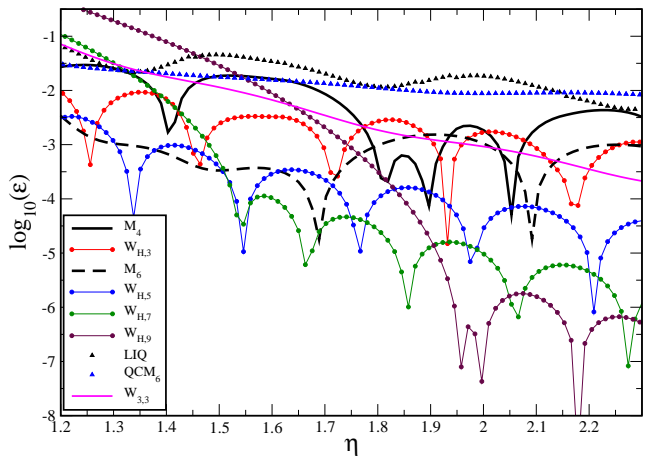}}
   \caption{Accuracy of gradient estimates (calculated via 
            Eq.~(\ref{eq:grad_SPH_interpolant})) for different kernels.
            Particles are placed on a hexagonal lattice so that the density is uniform and
            the pressure linearly increasing.
            The parameter $\eta$ determines the smoothing length, $h_a= \eta V_a^{1/D}$;  
            LIQ: linear-quartic,  QCM$_6$: quartic core M$_6$.}
   \label{fig:grad_accuracy}
\end{figure}
%-----------------------------------------------------------------------

\subsubsection{Gradient estimates}
We repeat the experiment described in Sec.~\ref{sec:gradient_accuracy} with the ``standard'' SPH
gradient, Eq.~(\ref{eq:grad_SPH_interpolant}), for a number of different kernels.
We set up particles on a 2D hexagonal lattice in $[-1,1] \times  [-1,1]$, corresponding to a 
close-packed distribution of spheres with radius $r_s$. All particles 
possess the same baryon numbers $\nu_b$ and are assigned pressures that rise linearly
with the x-coordinate so that the slope is $(\p_x P)=1$. 
In Fig.~\ref{fig:grad_accuracy}  we show the impact of the kernel choice
on the gradient accuracy (again, the M$_6$ is stretched to a support of 2h). The M$_6$ kernel (dashed black) yields, 
roughly speaking, a two orders of magnitude improvement over the standard SPH kernel M$_4$ (solid black).
Provided one is willing to apply larger neighbor numbers, the accuracy can be further improved by
using smoother kernels. Once more, the high-order members of the $W_{{\rm H},n}$ kernel family 
perform very well for large $\eta$ values. And, again, the Wendland kernel continuously improves
the gradient estimate with increasing $\eta$. None of peaked kernels reaches an accuracy 
substantially below $10^{-2}$, not even for very large $\eta$-values.

\section{Generalized volume elements}
\label{sec:volume_elements}
Here we discuss the volume elements that are needed for the kernel techniques described
in Sec.~\ref{sec:kernel_estimates}. In the following discussion we use the baryon number $\nu$ 
and the computing frame baryon number density $N$, but every relation straight forwardly translates
into Newtonian SPH by simply replacing $\nu$ with the particle mass $m$ and $N$ with the 
mass density $\rho$.\\
At contact discontinuities the pressure is continuous, but density and internal energy suffer
a discontinuity.
For a polytropic EOS, $P= (\Gamma-1) u n$, where $n$ is the baryon number density in the local 
rest frame\footnote{As will be described in more detail below, we measure energies here in 
units of the baryon rest mass energy $m_0 c^2$. The Newtonian correspondence of
the expression is, of course, $P= (\Gamma-1) u \rho $.}, the product of density and internal energy must be the
same on both sides of the discontinuity to ensure a single value of $P$ at the discontinuity, 
i.e. $n_1 u_1= n_2 u_2$. Here the subscripts label the two sides of the discontinuity. If this 
constraint is violated on a numerical level, say, because density and internal energy
have a different smoothness across the discontinuity, spurious forces occur that 
act like a surface tension. This can occur in "standard" SPH formulations since the density 
is estimated by a kernel-weighted sum over neighboring particles and therefore is smooth, 
while the internal energy is a property assigned to each particle that enters "as is" (i.e. un-smoothed) 
in the SPH evolution equations. 
Such spurious surface tension forces can compromise the ability to accurately handle subtle
fluid instabilities, see for example  \cite{agertz07,springel10a,read10}. One may, however, question
whether an unresolvably sharp transition in $u$ is a viable initial condition in the first place.
Note that Godunov-type SPH formulations \citep{inutsuka02,cha03,cha10,murante11,puri14} do 
not seem to suffer from such surface tension problems.\\
The problem can be alleviated if also the internal energy is smoothed, for example by applying 
some artificial thermal conductivity. This approach has been shown to work well in the
case of Kelvin-Helmholtz instabilities \citep{price08a,valdarnini12}. But it is actually 
a non-trivial problem to design appropriate  triggers that supply conductivity in the right amounts
exclusively where needed, but not elsewhere. Artificial conductivity applied where it is undesired
can have catastrophic consequences, say by spuriously removing pressure gradients 
that are needed to maintain a hydrostatic equilibrium.\\ 
An alternative and probably more robust cure comes from using different volume elements in
the SPH discretization process. \cite{saitoh13} pointed out that SPH formulations that do not 
include density explicitly in the equations of motion avoid the pressure becoming multi-valued 
at contact discontinuities. Since the density usually enters the equation of motion via the 
choice of the volume element $\nu_a/N_a$ (or $m_a/\rho_a$, respectively, in the Newtonian case), 
a different choice can possibly avoid the problem altogether. This observation is consistent with the findings of
\cite{hess10} who used a Voronoi tessellation to calculate particle volumes. In their approach 
no spurious surface tension effects have been observed. Closer to the original SPH spirit 
is the class of kernel-based particle volume estimates that have recently been suggested by
\cite{hopkins13} as a generalization of the \cite{saitoh13} approach. In the following we will make
use of these ideas for our relativistic SPH formulations.\\
We explore different ways to calculate kernel-based particle volume estimates $V_a$
from which the densities follow as
\be
N_a= \frac{\nu_a}{V_a} \label{eq:CF_density}.
\ee  
An obvious possibility for a volume element is the inverse of the 
local SPH-particle number density (calculated in the computing frame), estimated by a kernel sum
\be
V^{(1)}_a= \left(\sum_b W_{ab}(h_a)\right)^{-1}.
\label{eq:vol_1}
\ee
While this is a natural choice, one is in principle free to generalize this estimate by weighting each 
kernel with an additional quantity $X$
\be
V^{(X)}_a= \frac{X_a}{\sum_b X_b W_{ab}(h_a)} \equiv \frac{X_a}{\kappa_{X, a}}.
\label{eq:gen_vol_element}
\ee
There is a lot of freedom in the choice of $X$ and we will here only explore a small set of 
(Lorentz invariant) weights and assess their suitability in numerical experiments.
%
%--weight X=1
%
If the weight $X=1$ is chosen, one obviously recovers the volume element of Eq.~(\ref{eq:vol_1}) and
the baryon number density is simply given by  the number density estimate weighted with the
particle's own baryon number
\be
N^{(1)}_a= \nu_a \sum_b W_{ab}(h_a).
\label{eq:N_1}
\ee
Since only the baryon number of the particle itself enters, this form in principle
allows for sharp density transitions (say, via a uniform particle distribution with 
discontinuous $\nu$-behavior). As confirmed by the experiments in Sec.~\ref{sec:surf_tension}, 
this removes spurious surface tension effects.\\
%
%--weight X=nu
%
If instead $X= \nu$ is chosen, one recovers the standard SPH density estimate
\be
N^{(\nu)}_a= \sum_b \nu_b W_{ab}(h_a).
\label{eq:std_density_SPH}
\ee
%
%--weight X= P^k
%
Another choice is $X= P^k$, which yields 
\be
N^{(P^k)}_a= \nu_a\sum_b \left(\frac{P_b}{P_a} \right)^k  W_{ab}(h_a).
\ee
One may wonder whether the pressure  $P_a$ in the denominator may not give an inappropriately 
large weight in case of substantial pressure differences between neighboring particles, say in a shock.
Indeed, for the relativistic Sod-type shock, Sec.~\ref{sec:riemann_I},  with a pressure ratio of order 
$10^7$, and the choice $k=1$ we have observed a small density ``precursor'' spike within $\sim$ one 
smoothing length of the shock. To avoid such artifacts, we choose a small value, $k=0.05$, for which 
no anomalies have been observed. Obviously, for $k=0$ one recovers the previous case of $X=1$. 
If the pressure is used as a weight in the volume/density estimate an iteration is required
for self-consistent values of $N$ and $P$. This is explained in detail in Sec.~\ref{sec:dens_it}.
\\
To illustrate the impact that the choice of the volume element has on the resulting pressure across
a contact discontinuity we perform the following experiment. We place particles on a uniform hexagonal
lattice, assign baryon numbers so that the densities for $x<0$ have value $N_1=1$ and  $N_2=2$ for $x>0$ and 
internal energies as to reproduce a constant pressure $P_0=1$ on both sides, $u_i= P_0/((\Gamma-1) n_i)$.
Here $i$ labels the side and $\Gamma=5/3$ is the polytropic exponent. Once set up, we measure the densities
according to Eq.~(\ref{eq:CF_density}) and since here $n= N$ we calculate the pressure distribution 
across the discontinuity, once for each choice of $X$.
%-----------------------------------------------------------------------
\begin{figure*} 
   \centering
    \centerline{
      \includegraphics[width=16cm,angle=0]{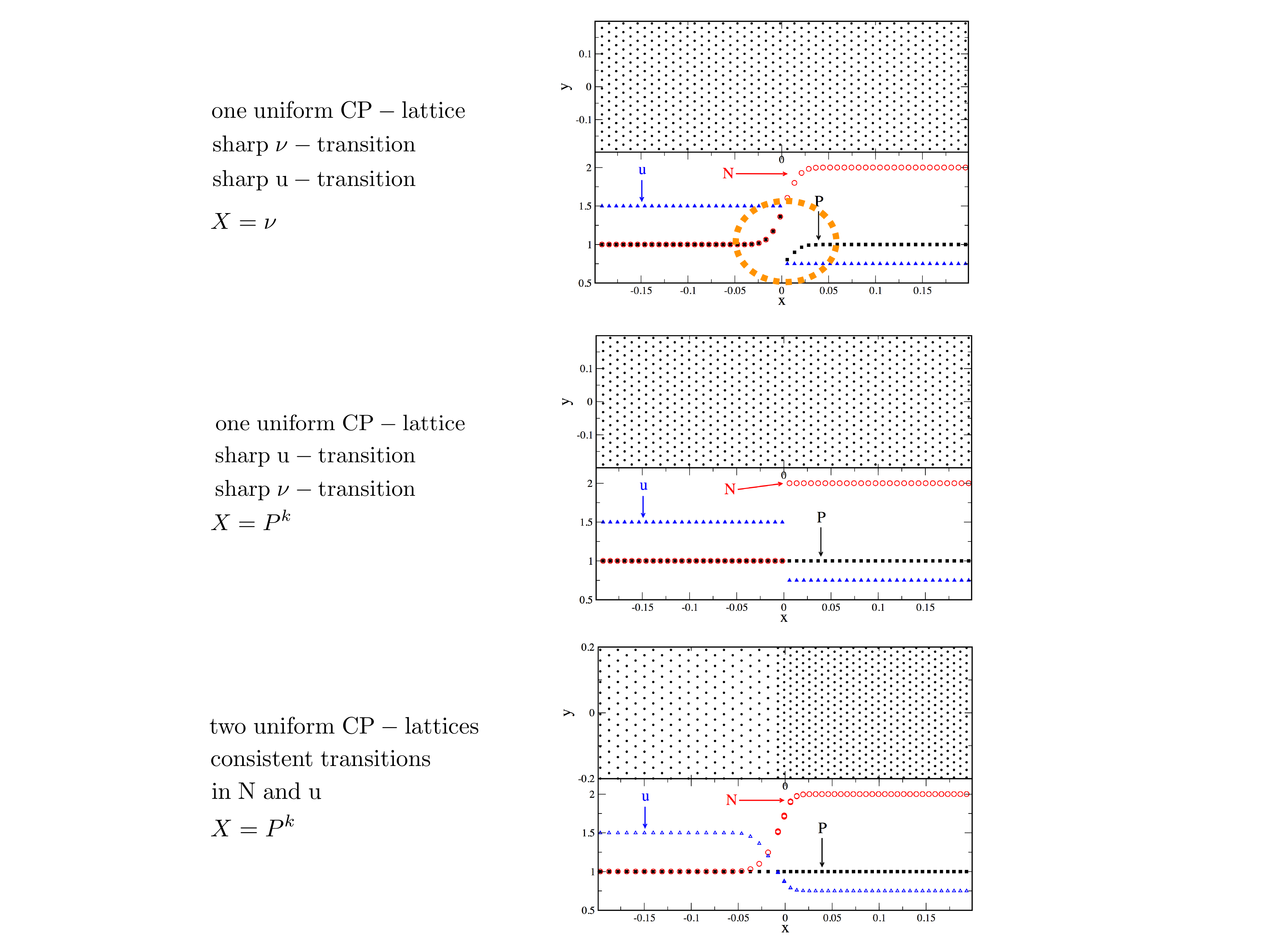}
      }
     \vspace*{0cm}
     \caption{Different ways to set up a contact discontinuity. In the upper panel, particles are
                   placed on a uniform hexagonal lattice (``close packed'', CP) and masses/baryon numbers
                   and internal energies are assigned as to reproduce the constant states, with a sharp
                   transition at $x=0$. Since the subsequent density calculation (with $X= \nu$) produces
                   a smooth density transition, the mismatch between a sharp internal energy $u$ and
                   a smooth density $N$ leads to a ``pressure blip'' (orange oval) that leads to spurious
                   surface tension forces at the interface. If for the same setup $X= P^k$ is used, both
                   $N$ and $u$ have a sharp transition and the pressure blib disappears, see middle panel.
                   The lower panel shows an alternative setup with equal masses/baryon numbers where
                   the density structure is encoded in the particle distribution (here two uniform hexagonal
                   lattices). The corresponding density calculation (with $X=P^k$) also produces a smooth
                   density transition, but the internal energy is consistently (and therefore smoothly) assigned 
                   from the calculated density and the condition of uniform pressure.}
        \label{fig:pressure_at_discont}
\end{figure*}
%-----------------------------------------------------------------------
The standard SPH volume element corresponding to weight factor $X=\nu$ produces a smooth 
density transition,
see upper panel of Fig.~\ref{fig:pressure_at_discont} that together with the sharp change in $u$ 
causes the "pressure blip" (orange oval) that is also frequently seen in SPH  shock tests and that 
is the reason behind the spurious surface tension forces discussed above. The same experiment,
if repeated using $X= P^k$, does not show any noticeable deviation from the desired value of 
unity.  The phenomenon of spurious surface tension is further explored by numerical experiments 
in Sec. \ref{sec:surf_tension}.\\
As an alternative numerical model for a contact discontinuity, we show in the last panel of
Fig.~\ref{fig:pressure_at_discont} a setup for the case of constant particle masses/baryon numbers.
In this case, the information about the density is encoded in the particle distribution, for which we
use here two uniform hexagonal lattices. Also here ($X= P^k$) the density transition is smooth, but
the internal energy is calculated (via an iteration) from the condition of constant pressure, so that the smoothness
of $N$ and $u$ are consistent with each other.
                    
\section{Special-relativistic SPH}
\label{sec:SPH}
We will now apply the techniques discussed in Secs.~\ref{sec:kernel_estimates}-\ref{sec:volume_elements}
to the case of  ideal, special-relativistic fluid dynamics. Excellent descriptions of various aspects of the topic 
(mostly geared towards Eulerian methods) can be found in a number of recent textbooks 
\citep{alcubierre08,baumgarte10,goedbloed10,rezzolla13a}.
In a first step, see Sec.~\ref{sec:gradW_SPH}, we generalize the derivation from the Lagrangian of an ideal 
relativistic fluid to the case of the generalized volume elements as introduced in 
Eq.~(\ref{eq:gen_vol_element}). This leads to a generalization of the kernel gradient based 
equations given in \cite{rosswog10b}. In a second formulation, see Sec.~\ref{sec:IA_SPH}, we use 
integral-based gradient estimates, see Eq.~(\ref{eq:form_replacement}), again for the case of a general volume element. 

\subsection{Special-relativistic SPH with  kernel derivatives}
\label{sec:gradW_SPH} 
We assume a flat space-time metric, $\eta_{\mu \nu}$, with signature (-,+,+,+) and use units in which the 
speed of light is equal to unity, $c=1$. We reserve Greek letters for space-time indices from 0...3
with 0 being the temporal component, $i$ and $j$ refer to spatial components and SPH
particles are labeled by $a,b$ and $k$.\\
Written in  Einstein sum convention the Lagrangian of a special-relativistic perfect fluid can 
be written as \citep{fock64} 
\be
L= - \int T^{\mu \nu} U_\mu U_\nu dV,
\ee
where the energy-momentum tensor reads
\be
T^{\mu \nu}= (e + P) U^{\mu} U^{\nu} + P \eta^{\mu \nu}.
\ee
We can write the energy density as a sum of a rest mass and an internal energy contribution
\be
e= \rho_{\rm rest} c^2 + u \rho_{\rm rest}= n m_0 c^2 (1 + u/c^2),
\ee
where, for now, the speed of light is shown explicitly. The baryon number density $n$ is measured in the 
local fluid rest frame and the average baryon mass is denoted by $m_0$. With the conventions that 
all energies are written in units of $m_0c^2$ and $c=1$,   we can use the normalization
of the four-velocity, $U_\mu U^\mu= - 1$, to simplify the Lagrangian to
\be
L= - \int n(1+u) dV.
\ee
The number density as measured in the "computing frame" (CF), see Eq.~(\ref{eq:CF_density}), is 
--due to length contraction-- increased by a Lorentz factor with respect to the local fluid rest frame
\be 
N= \gamma n \label{eq:N_vs_n}.
\ee
Therefore, the Lagrangian can be written as
\be
L= - \int dV N \left(\frac{1+u}{\gamma}\right)
\ee 
or
\be
L\simeq  - \sum_b V_b N_b \frac{1+u_b}{\gamma_b}= \sum_b \nu_b \frac{1+u_b}{\gamma_b},\label{eq:SR_Lagrangian}
\ee 
where the (fixed) baryon number carried by particle $b$, $\nu_b$, has been introduced.
To obtain the equations of motion from the Euler-Lagrange equations, we need $\nabla_a N_b$ and $d N_b/dt$ 
for which we use (see Eq.~\ref{eq:gen_vol_element})
\bea
\nabla_a \kappa_{X,b}&=& \frac{1}{\tilde{\Omega}_b} \sum_k X_k \nabla_a W_{bk}(h_b) \quad {\rm and}\\
\frac{d \kappa_{X,b}}{dt}&=& \frac{1}{\tilde{\Omega}_b} \sum_k X_k \vec{v}_{bk} \cdot \nabla_b W_{bk}(h_b)
\label{eq:kappa_derivs_SR},
\eea
with the ``grad-h'' terms
\be
\tilde{\Omega}_b= 1 - \frac{\nu_b}{X_b} \frac{\p h_b}{\p N_b} \sum_k X_k \frac{\p W_{kb}(h_b)}{\p h_b}.
\ee
The derivatives of the CF number densities then become
\be
\nabla_a N_b= \frac{\nu_b}{X_b \tilde{\Omega}_b} \sum_k X_k \nabla_a W_{bk}(h_b)\label{eq:nabla_N_SR}
\ee 
and
\be 
\frac{d N_a}{dt}  =  \frac{\nu_a}{X_a \tilde{\Omega}_a} \sum_b X_b \vec{v}_{ab} \cdot \nabla_a 
W_{ab}(h_a).\label{eq:dNdt_SR}
\ee

\subsubsection{The general momentum equation}
From straight forward differentiation of Eq.~(\ref{eq:SR_Lagrangian}) using Eq. (\ref{eq:N_vs_n}), 
the first law of thermodynamics, $\p u_b/\p n_b= P_b/n_b^2$, and $\p (\gamma_b^{-1} )/ \p \vec{v}_a= \gamma_b \vec{v}_b \delta_{ab}$ 
one finds the canonical momentum (for the explicit steps see \cite{rosswog09b})
\be
\vec{p}_a \equiv \frac{\p L}{\p \vec{v}_a}= \nu_a \vec{S}_a= \nu_a \gamma_a \vec{v}_a \left ( 1+u_a + \frac{P_a}{n_a}\right)
\label{eq:S_a}
\ee
and the evolution equation for the canonical momentum per baryon, $\vec{S}_a$, follows directly from the 
Euler-Lagrange equations (Eq. (155) in \cite{rosswog09b})
\be
\frac{d \vec{S}_a}{dt}
= -  \frac{1}{\nu_a}  \sum_{b} \left\{ \frac{P_a V_a^2}{\tilde{\Omega}_a} \frac{X_b}{X_a} \nabla_a W_{ab}(h_a) +
                                                                    \frac{P_b V_b^2}{\tilde{\Omega}_b} \frac{X_a}{X_b} \nabla_a W_{ab}(h_b) \right\}.
\label{eq:gen_mom_SR}
\ee
For the choice $V_k= \nu_k/N_k$ this reduces to the momentum equation given in \cite{rosswog10b}.

\subsubsection{The general energy equation}
The energy derived from the Lagrangian is
\be
E\equiv \sum_a \frac{\p L}{\p \vec{v}_a} \cdot \vec{v}_a - L = \sum_a \nu_a \epsilon_a, 
\ee
where the canonical energy per baryon is
\be
\epsilon_a = \vec{v}_a \cdot \vec{S}_a + \frac{1+u_a}{\gamma_a} = \gamma \left(1 + u_a + \frac{P_a}{n_a} \right) - \frac{P_a}{N_a},
\label{eq:epsilon_a}
\ee
or,
\be
\epsilon_a = \gamma \enth - \frac{P}{N},
\label{eq:en_a}
\ee
where we have used the specific, relativistic enthalpy
\be
\enth= 1 + u + \frac{P}{n}.
\label{eq:enthalpy}
\ee
The subsequent derivation is identical to the one in \cite{rosswog09b} up to their Eq.~(165),
\be
\frac{d \epsilon_a}{dt}= \vec{v}_a \cdot \frac{d \vec{S}_a}{dt} + \frac{P_a}{N_a^2} \frac{d N_a}{dt},
\ee
which, upon using Eqs.~(\ref{eq:dNdt_SR}) and (\ref{eq:gen_mom_SR}), yields the  special-relativistic energy equation
\be
\frac{d \epsilon_a}{dt}= - \frac{1}{\nu_a} \sum_b \left\{ \frac{P_a V_a^2}{\tilde{\Omega}_a} \frac{X_b}{X_a} \vec{v}_b \cdot \nabla_a W_{ab}(h_a)
                                                                                            + \frac{P_b V_b^2}{\tilde{\Omega}_b} \frac{X_a}{X_b} \vec{v}_a \cdot \nabla_a W_{ab}(h_b)\right\}.
\label{eq:gen_ener_SR}
\ee
Again, for $V_k= \nu_k/N_k$ this reduces to the energy equation given in \cite{rosswog10b}.
\\
The set of equations needs to be closed by an equation of state. In all of the tests presented below, we use a
polytropic equation of state, 
\be
P= (\Gamma-1)n u,
\label{eq:poly_EOS}
\ee 
where $\Gamma$ is the polytropic exponent (keep in mind our convention of measuring energies in units of $m_0 c^2$).
The corresponding sound speed is
\be
c_s= \sqrt{\frac{(\Gamma-1) (\enth-1)}{\enth}}.
\label{eq:v_sound}
\ee
The choices of the variables $N$, $\vec{S}$ and $\epsilon$ are suggested by the Lagrangian derivation and they avoid 
problems that have plagued earlier relativistic SPH formulations. For a comparison with Eulerian approaches we refer 
to the literature, e.g. to \cite{marti03} or \cite{keppens12}.

\subsubsection{Consistent values for smoothing lengths, densities and weights}
\label{sec:dens_it}
The smoothing lengths, the CF density and possibly the weight $X$
depend on each other. If the weight depends on the density, say for the
case $X=P^k$, we first perform a few steps to find accurate values of the
new pressure: a) calculate new volumes according to 
Eq.~(\ref{eq:gen_vol_element}) using the smoothing lengths and weights
from the previous time step, b) from the resulting new value of the density
we update the pressure according to Eq.~(\ref{eq:poly_EOS}), c) update the
smoothing length according to Eq.~(\ref{eq:h_vol}) d) once more update the
volume and e) again the pressure. This pressure value is finally used 
to perform an iteration between  volume, Eq.~(\ref{eq:gen_vol_element}), 
and the smoothing length, Eq.~(\ref{eq:h_vol}). Due to the previous steps the
guess values at this stage are already very accurate, so that on average
only one iteration is needed to meet our convergence criterion 
$|N^{(n+1)} - N^{(n)}|/N^{(n+1)} < 10^{-4}$.\\
While this procedure requires a number of iterations, we find that 
it is worth the effort, since a careless update of the smoothing length 
can produce a fair amount of noise that can compromise the quality of 
a simulation. An inaccurate update of the smoothing lengths may be
a largely overlooked source of noise in many SPH codes.

\subsection{Special-relativistic SPH based on integral approximations to derivatives}
\label{sec:IA_SPH}
As an alternative, we suggest a relativistic SPH formulation that is based on the integral 
approximation of gradients given in Eq.~(\ref{eq:integral_gradient}). This generalizes the
Newtonian formulation of \cite{garcia_senz12}. If we use the formal replacement, 
Eq.~(\ref{eq:form_replacement}), we can write alternative, integral-based relativistic SPH 
equations as

\be
\frac{d\vec{S}_a}{dt}= - \frac{1}{\nu_a} \sum_b \left\{ P_a V_a^2 \frac{X_b}{X_a} \vec{G}_{a} +
                                         P_b V_b^2 \frac{X_a}{X_b} \vec{G}_{b} \right\}
\label{eq:momentum_eq_no_diss_integral}
\ee
and
\be
\frac{d \epsilon_a}{dt} = 
- \frac{1}{\nu_a} \sum_b \left\{ P_a V_a^2 \frac{X_b}{X_a} \vec{v}_b\cdot \vec{G}_{a} +
                                         P_b V_b^2 \frac{X_a}{X_b}  \vec{v}_a\cdot \vec{G}_{b} \right\},
\label{eq:ener_eq_no_diss_integral}
\ee 
where
\be
\left(\vec{G}_{a}\right)^k= \sum_{d=1}^D C^{kd}(\vec{r}_a,h_a) (\vec{r}_b - \vec{r}_a)^d W_{ab}(h_a)
\ee
and 
\be
\left(\vec{G}_{b}\right)^k= \sum_{d=1}^D C^{kd}(\vec{r}_b,h_b) (\vec{r}_b - \vec{r}_a)^d W_{ab}(h_b).
\ee
The density calculation remains unchanged from Eq.~(\ref{eq:CF_density}).
Note that contrary to \cite{garcia_senz12} we do not apply "grad-h" terms here since they result from 
derivatives of the kernel function.
Since the functions $\vec{G}_k$ share the same symmetries as the usual kernel derivatives (i.e. they change 
their sign if $a$ and $b$ are interchanged)  this alternative relativistic SPH formulation 
also enforces the {\em numerical conservation of physically conserved quantities by construction}.\\
In this scheme, the smoothing lengths are updated exactly as described above for the kernel-gradient 
based method, see Sec.~\ref{sec:dens_it}.

\subsection{Dissipative terms}
\label{sec:AV}
In order to handle shocks properly, additional measures need to be taken to ensure that appropriate 
amounts of entropy are generated in shocks. This can be done by implementing a 
Riemann solver or by adding explicit, artificial dissipation terms which is the approach that
we follow here.  We use the form of the dissipative terms suggested by \cite{chow97}
\be
\left(\frac{d\vec{S}_a}{dt}\right)_{\rm diss}= - \sum_b \nu_b \Pi_{ab} 
\overline{\nabla_a W_{ab}} 
\ee
\be
\left(\frac{d\epsilon_a}{dt}\right)_{\rm diss}=  - \sum_b \nu_b \vec{\Omega}_{ab} \cdot
\overline{\nabla_a W_{ab}}, 
\ee
where 
\bea
\Pi_{ab}&=& - 
\frac{K_{ab} v_{{\rm sig},ab}}{\bar{N}_{ab}} (\vec{S}_a^\ast-\vec{S}_b^\ast) \cdot\hat{e}_{ab}
\label{eq:diss_mom}\\
\vec{\Omega}_{ab}&=& -
\frac{K_{ab} v_{{\rm sig},ab}}{\bar{N}_{ab}} (\epsilon_a^\ast-\epsilon_b^\ast)\hat{e}_{ab},
\label{eq:diss_en}
\eea
for approaching particles and $\Pi_{ab}=0$ and $\vec{\Omega}_{ab}=0$ otherwise.
The parameter $K_{ab}$ is the arithmetic average of the dissipation parameters of
the particles $a$ and $b$ whose steering is explained in detail in Sec.~\ref{sec:AV_controle}.
We use the symmetrized kernel gradient
\be
\overline{\nabla_a W_{ab}} = \frac{1}{2}\left[\nabla_a W_{ab}(h_a) +  \nabla_a W_{ab}(h_b) \right],
\label{eq:average_W}
\ee
together with
\be
\gamma_k^\ast= \frac{1}{\sqrt{1-(\vec{v}_k\cdot \hat{e}_{ab})^2}},
\ee
\be
\vec{S}_k^\ast= \gamma^\ast_k \left(1+u_k+\frac{P_k}{n_k}\right) \vec{v}_k
\ee
and 
\be
\vec{\epsilon}_k^\ast= \gamma^\ast_k \left(1+u_k+\frac{P_k}{n_k}\right) - \frac{P_k}{N_k}.
\ee
For the formulation based on the integral approximation of gradients 
$\overline{\nabla_a W_{ab}}$ needs to be replaced by
\be
\overline{\vec{G}}_{ab}= \frac{1}{2}\left[\vec{G}_a +  \vec{G}_b \right].
\label{eq:average_G}
\ee
We found that it is actually important to consistently apply either kernel or IA-gradients
in both the non-dissipative and dissipative terms. For example, in geometrically 
complicated tests such as the blast-bubble interactions shown in Sec.~\ref{sec:blast_bubble_I} and
\ref{sec:blast_bubble_II} we found that 
slight mismatches in the gradient estimates can lead to instabilities that are
serious enough to crash a simulation. With consistent gradients in both types of
terms no such instabilities ever were observed.\\
For the signal speed we use \citep{rosswog10b} 
\be
v_{{\rm sig},ab}= {\rm max}(\alpha_a,\alpha_b),\label{eq:vsig}
\ee
where
\be
\alpha_k^{\pm}= {\rm max}(0,\pm \lambda^\pm_k)
\ee
with $\lambda^\pm_k$ being the extreme local eigenvalues of the Euler equations, see e.g. \cite{marti03},

\be
\lambda^\pm_k= \frac{v_\parallel(1-c_{{\rm s},k}^2) \pm c_{{\rm s},k} \sqrt{(1-v^2)(1-v_\parallel^2 - 
v_\perp^2 c_{{\rm s},k}^2)}}{1-v^2 c_{{\rm s},k}^2}
\ee
and $c_{{\rm s},k}$ being the relativistic sound velocity of particle $k$, see Eq.~(\ref{eq:v_sound}). 
In 1 D, this simply reduces
to the usual velocity addition law, $\lambda^\pm_k= (v_k\pm c_{{\rm s},k})/(1\pm v_k c_{{\rm s},k}) $.

\subsubsection{Controlling the amount of dissipation: triggers on shocks and velocity noise}
\label{sec:AV_controle}
% idea of Morris and Monaghan
Artificial dissipation is really only needed under specific circumstances such as to
produce entropy in shocks where it mimics nature's behavior, though on a larger, resolvable
scale. Nevertheless, in older SPH implementations artificial dissipation was (and still often is)
applied everywhere, regardless of whether it is actually needed or not and, as a consequence, 
one is modelling some kind of a viscous fluid rather than the intended inviscid Euler equations.\\
\cite{morris97} suggested as a cure to provide each particle with its own dissipation parameter
$K_a$ and to evolve this dissipation parameter according to an additional ODE\footnote{We express here
everything in terms of our notation where the dissipation parameter is called $K$. In the literature
on non-relativistic SPH the dissipation parameters are usually called $\alpha$ and $\beta$. To translate
our approach to non-relativistic SPH, one may replace $K$ by $\alpha$ and choose $\beta$ as a multiple
(usually = 2) of $\alpha$.}
\be
\frac{dK_a}{dt}= \mathcal{K}^+_{\;a} - \mathcal{K}_{\;a}^-.
\ee
They suggested for the source term
\be
\mathcal{K}_a^+= {\rm max}\left(-(\nabla\cdot\vec{v})_a,0\right)
\label{eq:source_MM_AV}
\ee
and
\be
\mathcal{K}_a^- = \frac{K_a(t) - K_{\rm min}}{\tau_a}
\label{eq:diss_evol}
\ee
for the decay term, where $K_{\rm min}$ represents a minimum, ``floor'' value for the viscosity parameter
and $\tau_a$ is the individual decay time scale. This approach (or slight modifications of it) has been 
shown to substantially reduce unwanted effects in practical simulations \citep{rosswog00,dolag05,wetzstein09}.\\
% Criticism by Cullen and Dehnen
Recently, \cite{cullen10} suggested further improvements to the \cite{morris97} approach. They argued
that a floor value for the viscosity parameter is unnecessary and that the original scheme may, in some
situations, be too slow to reach the required values of the dissipation parameter. They suggest to immediately
raise the viscosity parameter to the desired value rather than obtaining it by integrating the above ODE.
Moreover, and as already noted in the original paper of \cite{morris97}, a scheme with the originally suggested 
source term, Eq.~(\ref{eq:source_MM_AV}), would also spuriously trigger on a constant slow compression with 
$\nabla \cdot \vec{v}$= const. Therefore, \cite{cullen10} suggested to trigger on the time derivative of $\nabla \cdot \vec{v}$.
They further pointed 
out that $\nabla\cdot \vec{v}$ as calculated by standard SPH gradients can have substantial 
errors which trigger unnecessary dissipation in shear flows even if standard shear-limiters 
\citep{balsara95} are used.\\

\noindent{\em Strategy}\\
Before we come to the detailed expressions used here we  want to briefly summarize our strategy 
when to apply dissipation. The challenge is to assign at each time step to each particle an
appropriate dissipation parameter $K_a$. We trigger dissipation by a) shocks and b) (to a lesser extent) 
by velocity noise. The presence of a noise trigger allows us to let the dissipation parameter decay extremely quickly:
if noise should appear, the noise trigger will take care of it.
Like in \cite{cullen10} the current viscosity value is compared to a ``desired'' 
value and, if indicated, it is raised immediately. In our case, the desired parameter value is
the maximum of a shock and noise value
\be
     K_{a,\rm des}= {\rm max}(K_{a,\rm shock},K_{a,\rm noise}), 
\label{eq:K_des}
\ee
and if $K_{a,\rm des} > K_a(t)$, we instantaneously set $K_a= K_{a,\rm des}$, 
otherwise $K_a(t)$ smoothly decays according to
\be
\frac{dK_a}{dt}= - \mathcal{K}_{\;a}^-= - \frac{K_a(t)}{\tau_a}.
\label{eq:K_evol}
\ee
Note that all the velocity gradients that enter our dissipation scheme are calculated via the
accurate gradient prescriptions of the full integral approximation, see Eq.~(\ref{eq:fIA}). This is similar to recent approaches
by \cite{cullen10} and \cite{read12} that also used more accurate, non-standard gradient estimates.\\

\noindent{\em Shock trigger}\\
We use  the temporal change of the compression as a shock indicator \citep{cullen10}
\be
A_{a,\rm shock}= {\rm max} \left( - \frac{d (\nabla \cdot \vec{v})_a}{dt},0 \right).
\label{eq:A_shock}
\ee
We have also performed a number of experiments combining both spatial \citep{read12} and 
temporal changes \citep{cullen10} of the compression, but did not find any obvious advantage
with respect to Eq.(\ref{eq:A_shock}). Since the latter is simple to calculate (as a numerical derivative)
and does not involve second derivatives, we use it in the subsequent tests.
From this shock indicator we calculate the desired shock dissipation parameter
\be
K_{a,\rm shock}=  K_{\rm max} \frac{A_{a,\rm shock}}{A_{a,\rm shock} + (c_{{\rm s},a}/h_a)^2}.
\label{eq:K_shock}
\ee

\noindent{\em Noise trigger}\\
We also wish to have the possibility to apply dissipation in regions of ``velocity noise''.
This is less crucial than the shock trigger and not really required, but it improves
the convergence (as shown at the example of the Gresho-Chan vortex).
Noisy regions are characterized by fluctuations in the sign of $\divv$, i.e. some particles
feel an expansion while their neighbors get compressed. Therefore, the ratio
\be
\frac{S_{1,a}}{S_{2,a}}\equiv \frac{\sum_b (\divv)_{b,\rm fIA}}{\sum_b |\divv|_{b,\rm fIA}} 
\label{eq:noise_trigg}
\ee
can deviate from $\pm 1$  in a noisy region since contributions of different sign are added up 
in $S_{1,a}$ and therefore such deviations can be used as a ``noise indicator 1'':
\be
\mathcal{N}_a^{(1)}= \left| \frac{\tilde{S}_{1,a}}{S_{2,a}} - 1\right|,
\label{eq:N_trigg_1}
\ee
where the quantity
\be
\tilde{S}_{1,a}= \left\{\begin{array}{ll}  -S_{1,a} & {\rm if} \; (\nabla \cdot \vec{v})_{a} < 0 \\
         \quad S_{1,a} & {\rm else}\end{array}\right. .
\ee
If all particles in the neighborhood are either compressed or expanding, $\mathcal{N}_a^{(1)}$ vanishes.\\
The  noise trigger of Eq.~(\ref{eq:N_trigg_1}) actually only triggers on the signs of $\divv$ and does not take 
into account how substantial the compressions/expansions are compared to the ``natural scale'' $c_s/h$.
The above noise trigger therefore switches on even if the fluctuation is not very substantial, but only
causes a very small level of the dissipation parameter ($\sim 0.02$ for our typical parameters, see below).
One might therefore, alternatively, consider to simply apply a constant ``dissipation floor'' $K_{\rm min}$. We chose, however, 
the trigger-version, both for the aesthetic reason that we do not want untriggered dissipation and
since the triggered version shows a slightly higher convergence rate in the Gresho-Chan test.  The 
differences, however, are not very significant.\\
We also want to add a noise trigger that takes the significance of the noise in comparison to the
natural scale $c_s/h$ into account. This turns out to have beneficial effects to get rid of ``post-shock
wiggles'' that can still be present when the shock-triggered dissipation has already decayed. To 
this end we calculate average $\divv$ values separately for each sign:
\bea
\mathcal{S}^+_a &=&   \frac{1}{ N^+} \; \sum_{b, \divv_b>0}^{N^+} \gamma_b \divv_b\\
\mathcal{S}^-_a &=& - \frac{1}{N^-} \; \sum_{b, \divv_b<0}^{N^-} \gamma_b \divv_b,
\eea
where $N^+/N^-$ is number of neighbor particles with positive/negative $\divv$ and $\gamma_b$ is the
Lorentz factor. The quantity
that we trigger on is the product of both quantities. If there are sign fluctuations, but they are small
compared to $c_s/h$, the product is very small, if we have a uniform expansion or compression
one of the factors will be zero. So only for sign changes and significantly large compressions/expansions
will the product have a substantial value. Therefore, our ``noise indicator 2'' reads
\be
\mathcal{N}_a^{(2)}= \sqrt{\mathcal{S}^+_a \mathcal{S}^-_a}  .
\label{eq:N_trigg_2}
\ee
The final noise parameter is then
\be
K_{a,{\rm noise}}= K_{\rm max} \; {\rm max}(\kappa^{(1)}_{a},\kappa^{(2)}_{a}),
\label{eq:K_noise}
\ee
where
\be
\kappa^{(1)}_{a}=  \frac{\mathcal{N}^{(1)}_a}{ \mathcal{N}^{(1)}_a+   \mathcal{N}_{\rm \; \; noise}}
\label{eq:kappa1}
\ee
and
\be
\kappa^{(2)}_{a}=  \frac{\mathcal{N}^{(2)}_a} {\mathcal{N}^{(2)}_a + 0.2 (c_{{\rm s},a}/h_a)}.
\ee
The reference value $\mathcal{N}_{\rm \; \; noise}$ is determined via the Gresho-Chan vortex test, see Sec.~\ref{sec:Gresho}.
The desired dissipation parameter is then chosen as in Eq.~(\ref{eq:K_des}) from 
Eq.~(\ref{eq:K_shock}) and Eq.~(\ref{eq:K_noise}).\\

\noindent{\em Parameters}\\
We determine the dimensionless parameters in the above scheme by a large number of numerical experiments.
If not triggered, the dissipation parameter $K_a$ decays on a time scale
\be
\tau_a= \chi \frac{h_a}{c_{s,a}},
\ee
where we use $\chi=2$ to ensure a very rapid decay. In the Gresho-Chan test we find good results for
$\mathcal{N}_{\rm \; \; noise}= 50$, for the maximally possible dissipation parameter we use 
$K_{\rm max}=1.5$. We found this set of parameters represents a reasonable compromise for all tests.\\
The functioning of our dissipation triggers is illustrated in Fig.~\ref{fig:BiB_diss_switch}.

\subsection{Recovery of the primitive variables}
\label{sec:recovery}
Similar to many grid-based approaches, we have to pay a price for the simplicity of the evolution equations 
(\ref{eq:gen_mom_SR})/(\ref{eq:gen_ener_SR}) or  
(\ref{eq:momentum_eq_no_diss_integral})/(\ref{eq:ener_eq_no_diss_integral}) with the need to recover 
the physical variables from the numerical 
ones \citep{chow97,rosswog10b}.
The strategy is to express all variables in Eq.~(\ref{eq:poly_EOS}) in terms of the updated variables
$\vec{S}, \epsilon, N$ and the pressure $P$, then solve for the new pressure, and finally substitute
backwards until all physical variables are available.\\
First solve the momentum equation, Eq.~(\ref{eq:S_a}), for the velocity, which together with 
Eq.~(\ref{eq:epsilon_a}) provides us with
\be
\vec{v}= \frac{\vec{S}}{\epsilon + P/N} \;\; {\rm and} \;\; 
\gamma(P)= \frac{1}{\sqrt{1 - S^2/(\epsilon + P/N)^2}} \;\; {\rm and} \;\;
n(P)= \frac{N}{\gamma(P)}.
\label{eq:v_gamma_n}
\ee
We now still need to find $u(P)$ so that we can substitute everything in Eq.~(\ref{eq:poly_EOS}).
This can be obtained by solving Eq.~(\ref{eq:enthalpy}) for $u$ and by using Eq.~(\ref{eq:N_vs_n})
\be
u(P)= \frac{\epsilon}{\gamma} + \frac{P}{\gamma N} (1-\gamma^2) -1.
\label{eq:u_P}
\ee
The new value of $P$ is then obtained by a root-finding algorithm for
\be
f(P) \equiv P - (\Gamma-1) \; n(P) \; u(P) = 0.
\ee
Once the new, consistent value of $P$ is found, one successively recovers
a) the Lorentz factor and velocity from Eq.~(\ref{eq:v_gamma_n})
b) the rest frame density from Eq.~(\ref{eq:N_vs_n})
c) the internal energy from Eq.~(\ref{eq:u_P}) and
d) the enthalpy from Eq.~(\ref{eq:enthalpy}).

\subsection{Time integration}
\label{sec:time_stepping}
We use the optimal third-order TVD algorithm \citep{gottlieb98} with global time step to integrate the system of equations.
The time step is simply chosen according to
\be
\Delta t= C \; {\rm min} \left(\Delta t_{\rm C}, \Delta t_{\rm F}   \right),
\ee
where $\Delta t_{\rm C}= h_{\rm min}/v_{\rm sig, max}$,
$\Delta t_{\rm F}= \sqrt{h_{\rm min}/|d\vec{S}/dt|_{\rm max}}$,
$h_{\rm min}$ is the minimum smoothing length and $|d\vec{S}/dt|_{\rm max}$
is the maximum momentum derivative of all particles. For the prefactor we 
conservatively use $C=0.5$.

\section{Multi-dimensional benchmark tests}
\label{sec:tests}
All of the following tests, whether in the Newtonian or special-relativistic regime, are performed
with a new 2D SPH code, called SPHINCS\_SR. At the current stage, it is by no means optimized, its
entire purpose is to explore different (special-relativistic) SPH formulations.\\
In the following, we scrutinize the effects of the different new elements
in number of benchmark tests by using four different SPH formulations:
\bi
 \i {\bf Formulation 1 ($\mathcal{F}_1$):}
    \bi
    \i Density:                      Eqs.~(\ref{eq:CF_density}) and (\ref{eq:gen_vol_element}) with weight $X=P^k$, $k= 0.05$
    \i Momentum equation: Eq.~(\ref{eq:momentum_eq_no_diss_integral})
    \i Energy equation:         Eq.~(\ref{eq:ener_eq_no_diss_integral})
    \i Dissipative terms:       see Sec.~\ref{sec:AV}, use Eq.(\ref{eq:average_G})
    \i Kernel:                         Wendland kernel $W_{3,3}$ with $\eta= 2.2$, see Eq. (\ref{eq:h_vol})
    \ei
    We expect results of similar quality for the $W_{{\rm H},9}$ kernel, but since the
    Wendland kernel $W_{3,3}$ produced less noise in the ``noise box'' and the Gresho-Chan vortex test
    we chose it as our default kernel.\\ 

 \i {\bf Formulation 2 ($\mathcal{F}_2$):}
    \bi
    \i Density:                      Eqs.~(\ref{eq:CF_density}) and (\ref{eq:gen_vol_element}) with weight $X=P^k$, $k= 0.05$
    \i Momentum equation: Eq.~(\ref{eq:gen_mom_SR})
    \i Energy equation:         Eq.~(\ref{eq:gen_ener_SR})
    \i Dissipative terms:       see Sec.~\ref{sec:AV}, use Eq.(\ref{eq:average_W})
    \i Kernel:                         Wendland kernel $W_{3,3}$ with $\eta= 2.2$, see Eq. (\ref{eq:h_vol})
    \ei
    The difference between $\mathcal{F}_1$ and $\mathcal{F}_2$
    measures the impact of the different gradient prescriptions.\\

 \i {\bf Formulation 3 ($\mathcal{F}_3$):}
   \bi
    \i Density:                      Eqs.~(\ref{eq:CF_density}) and (\ref{eq:gen_vol_element}) with weight $X= \nu$
    \i Momentum equation: Eq.~(\ref{eq:momentum_eq_no_diss_integral})
    \i Energy equation:         Eq.~(\ref{eq:ener_eq_no_diss_integral})
    \i Dissipative terms:       see Sec.~\ref{sec:AV}, use Eq.(\ref{eq:average_G})
    \i Kernel:                         Wendland kernel $W_{3,3}$ with $\eta= 2.2$, see Eq. (\ref{eq:h_vol})
    \ei
    Same as $\mathcal{F}_1$, but with the more ``standard'' choice $X= \nu$ to explore the 
    impact of volume element.\\

 \i {\bf Formulation 4 ($\mathcal{F}_4$):}
    \bi
    \i Density:                      Eqs.~(\ref{eq:CF_density}) and (\ref{eq:gen_vol_element}) with weight $X= \nu$
    \i Momentum equation:  Eq.~(\ref{eq:gen_mom_SR})
    \i Energy equation:         Eq.~(\ref{eq:gen_ener_SR})
    \i Dissipative terms:       see Sec.~\ref{sec:AV}, use Eq.(\ref{eq:average_G}),\\
        \hspace*{0.6cm}         constant dissipation parameter $K= 1$ 
    \i Kernel:                         M$_4$ kernel with $\eta= 1.2$, see Eq. (\ref{eq:h_vol})
    \ei
    These are choices close to what is used in many SPH codes. As will become clear below, these
    are rather poor choices, i.e. in many cases the accuracy of a code could be substantially increased by a 
    number of relatively simple measures with respect to ``standard recipes''.
\ei
Usually we will focus on the results of the $\mathcal{F}_1$ formulation, but when larger deviations 
for other formulations occur, we also show a brief comparison. \\
All of the following tests are performed with the special-relativistic code SPHINCS\_SR. But since the
discussed improvements are not specific to special relativity but instead concern SPH techniques 
in general, we discuss both Newtonian and special-relativistic tests. Each of the tests is marked 
accordingly: "N" for Newtonian and "SR" for special relativity.

\subsection{Initial particle distribution (N/SR)}
\label{sec:IC}
%-----------------------------------------------------------------------
\begin{figure*} 
   \centerline{
    \includegraphics[width=14cm,angle=-90]{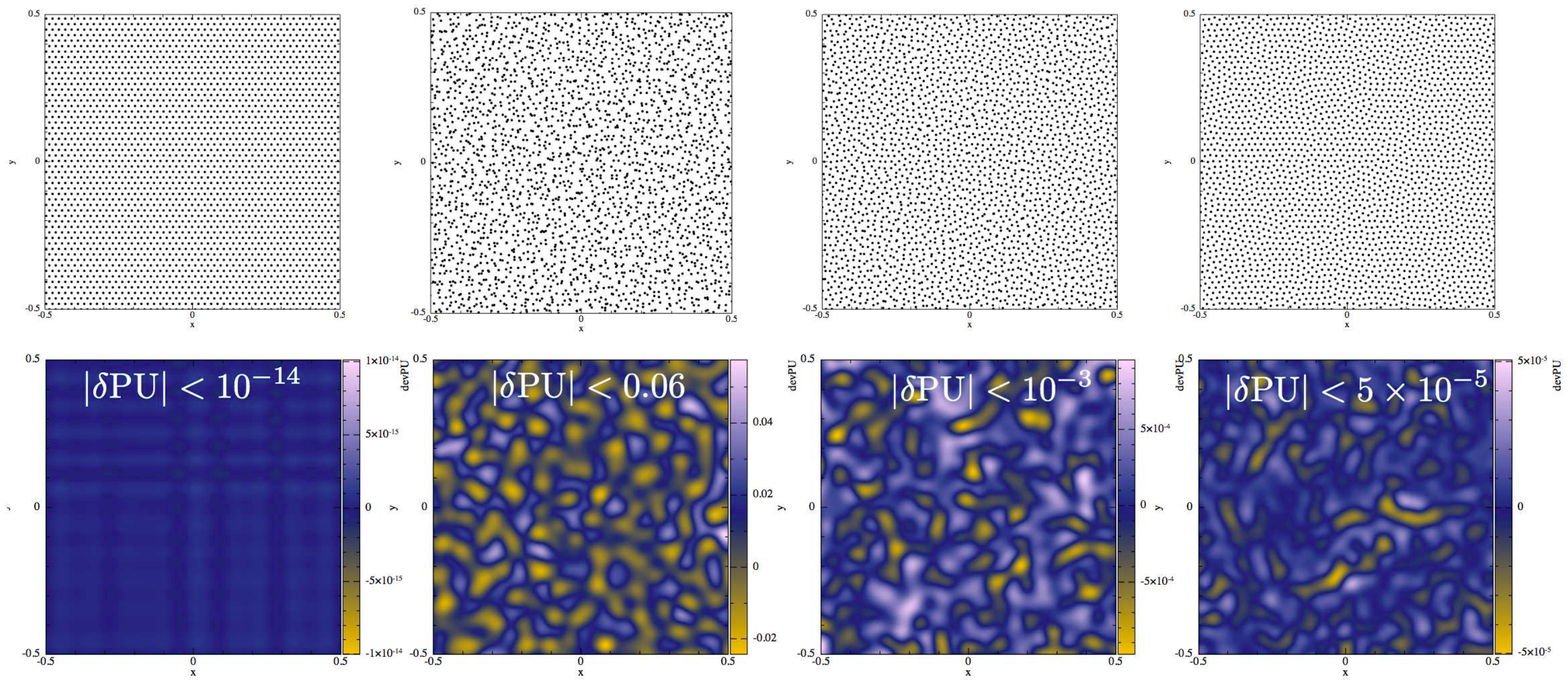}}
   \vspace*{-3cm}
   \caption{Particle distributions (first row): initial hexagonal lattice (left column), 
                  heavily perturbed hexagonal lattice (second column), 
                  after 10 iterations (third column) and final distribution (fourth column). 
                  The second row shows the deviation from a partition of unity, $\delta PU$. Note that 
                  the color bar shows vastly different scales.}
   \label{fig:particle_dist}
\end{figure*}
%-----------------------------------------------------------------------
The initial conditions and in particular the initial particle distribution are crucial for the accuracy
of an SPH simulation. Of course, the quality indicators $\QIone$ - $\QIfour$,
Eqs.~(\ref{eq:quality_SPH_interpolation1}) to (\ref{eq:quality_SPH_gradient2}), should be fulfilled
to good accuracy and this suggests to use some type of lattice. Unfortunately this is not enough
and the particle distribution should have at least two more properties:  a) it should be stable for the 
used SPH formulation-lattice combination, i.e. particles in pressure equilibrium should remain in their configuration 
and b) there should not be preferred directions in the particle distribution. Condition a) means that
the particles should be in a minimum energy configuration and how this relates to the kernel choice
and the form of the SPH equations is from a theoretical point of view poorly understood to date. Simply placing particles
on some type of lattice is usually not good enough: the particles will in most cases begin moving off the lattice
and introduce noise. We will explore this explicitly in Sec.~\ref{sec:noise}. Condition b) is necessary
since preferred directions can lead to artifacts, for example, a shock
travelling along a preferred direction of a lattice will collect preferentially particles in this direction
and this can lead to unwanted ``ringing'' effects. An example of this effect is shown in
Fig.~\ref{fig:2D_Sod_initial_distrib}.\\
In the following tests, we will, as a default, place the SPH particles on a hexagonal lattice, see 
Fig.~\ref{fig:particle_dist}, upper left panel, but if artifacts can occur we will use a ``glass-like''
particle distribution instead (upper rightmost panel). To produce the glass, our strategy is to start 
from a hexagonal lattice, perturb it heavily and subsequently drive that particles into positions 
so that they represent a good partition of unity, see Eq.~(\ref{eq:quality_SPH_interpolation1}). 
We proceed according to  the following steps:
\ben
\i Place the desired number of particles on a hexagonal lattice (upper left panel) corresponding to the closest packing
of spheres with radii $r_s$, where each particle has an effective volume of $2 \sqrt{3} r_s^2$ 
(Fig.~\ref{fig:particle_dist}, first panel upper row).
\i Apart from ``frozen'' boundary particles, perturb the particles heavily by  displacing 
    them by a distance $r_0= 0.9 r_s$ in a random direction (Fig.~\ref{fig:particle_dist}, second panel upper row).
\i The perturbed particle distribution is then driven to a good partition of unity
by applying pseudo-forces proportional to the negative gradient of the quality indicator $\mathcal{Q}_1$, see
Eq.~(\ref{eq:quality_SPH_interpolation1}). 
Since in the end we want a uniform particle distribution so that all volumes $V_b$ should finally be the same, $V_0$,
we take the volumes out of the sum, use a uniform smoothing length $h_0= 2.2 \sqrt{V_0}$ and as pseudo-force simply
\be
\vec{f}_a= - \sum_b \nabla_a W_{ab}(h_a)
\label{eq:pseudo_force}
\ee
together with the Wendland kernel.
The maximum value of all particles at the initial, heavily perturbed distribution is denoted $\vec{f}_{\rm max}^0$.
Obviously, Eq.~(\ref{eq:pseudo_force}) just corresponds to a force opposite to the gradient of the SPH 
particle number density as measured via a kernel summation, so driving the particles to a uniform number density.
\i As a next step an iteration is performed.
Rather than really using an acceleration and a time step, we apply a displacement so that the particle with
the maximum force $\vec{f}_{\rm max}^0$ is allowed to be displaced {\em in the first iteration} by a 
distance $r_0= 0.5 r_s$ and in all subsequent iterations by
\be
\delta \vec{r}_a = \frac{r_0}{f_{\rm max}^0} \vec{f}_a.
\ee
This iteration is performed until max$_a (|\delta \vec{r}_a|)/r_s < 10^{-3}$. Since the algorithm is designed
to take small steps, it takes a large number of steps to meet the convergence criterion (several hundreds for the above
parameters), however, the iterations are computationally inexpensive, since they can all be performed with 
a once (generously large)  created neighbor list, so that neighbors only need to be searched once.
\een
An example for 10 K particles is shown in Fig.~\ref{fig:particle_dist}: after the initial, heavy perturbation the maximum
of the deviation from the partition of unity is
\be
\delta {\rm PU}_a \equiv | 1 - \sum_b V_b W_{ab}(h_a)|
\ee
up to 6 percent, but after only 10 iterations it is below $10^{-3}$ and once the criterion is met it is below  $5 \times 10^{-5}$ 
everywhere. The final particle distribution does not have globally preferred directions, but, of course, this comes at the 
price of a not perfect (but good!) partition of unity.

\subsection{Static I: "Noise box" (N)}
\label{sec:noise}
%-----------------------------------------------------------------------
\begin{figure*} 
\centerline{
   \includegraphics[width=6.5cm,angle=-90]{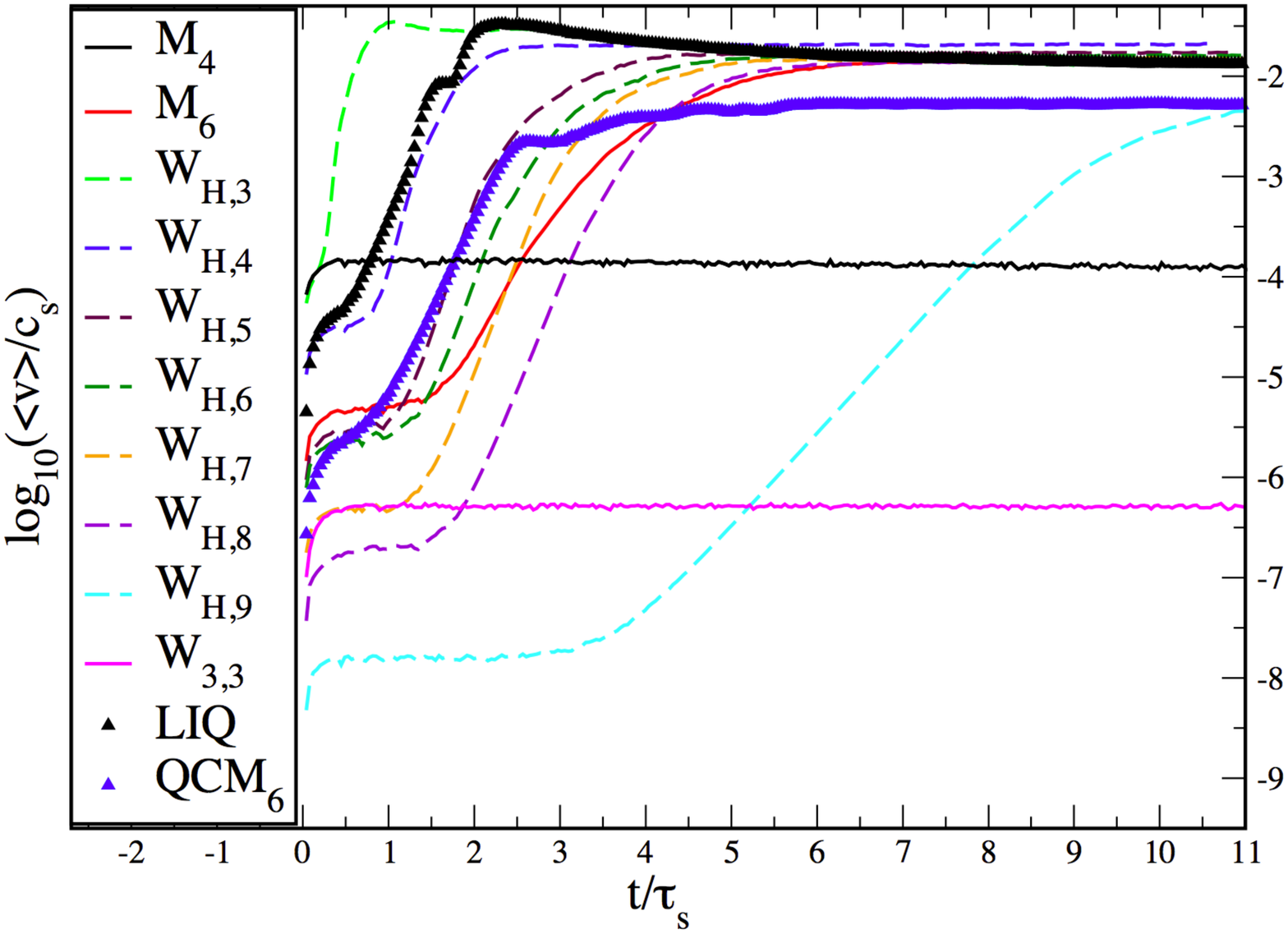}
   \includegraphics[width=6.5cm,angle=-90]{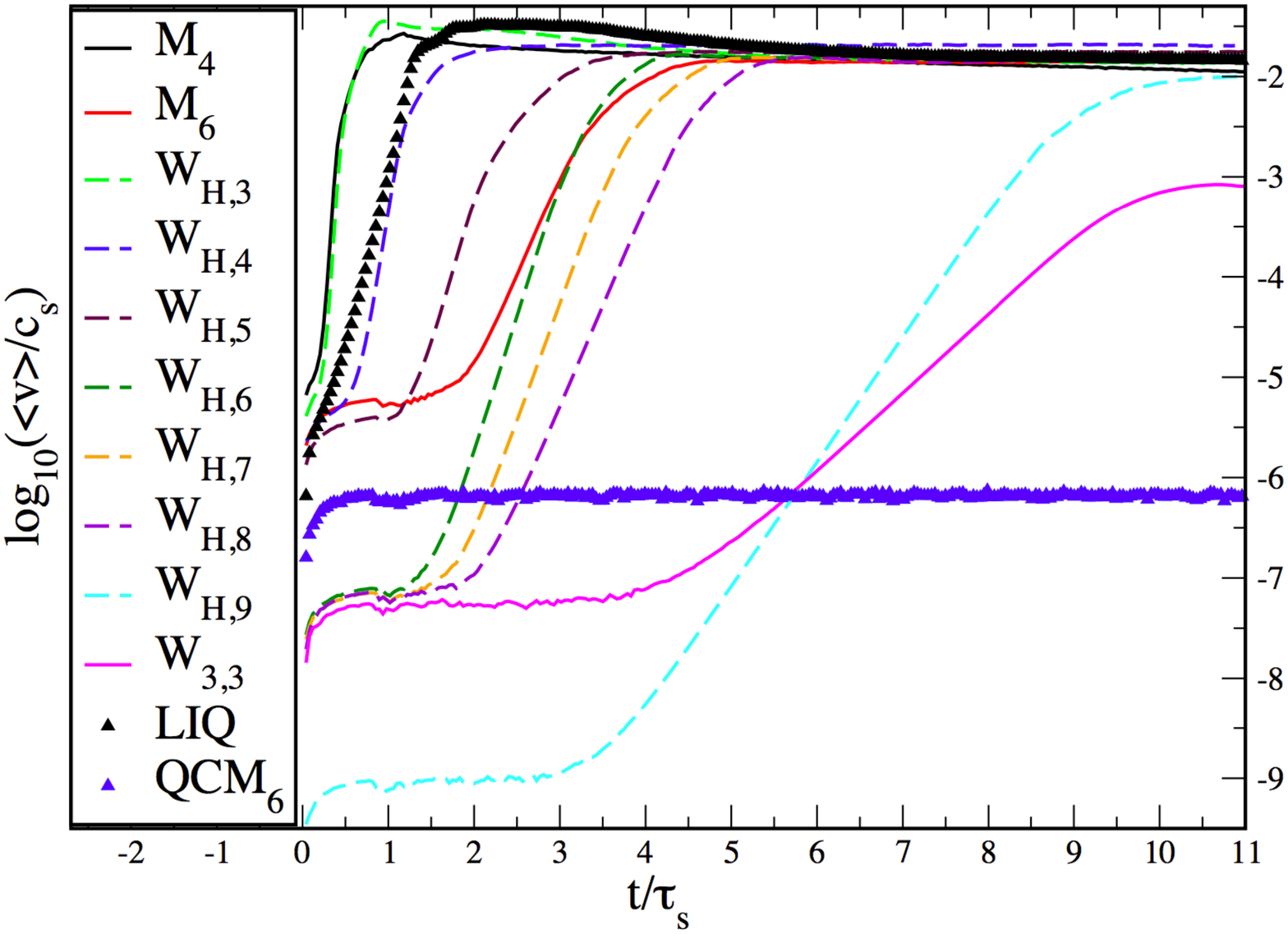} 
    }
   \caption{Results of the "noise box test" for an initial hexagonal (left) and quadratic 
           lattice (right). Shown is the logarithm of the average particle velocity (in 
           units of the sound speed) as a function of time (in units of the sound crossing 
           time through the computational domain). For both initial lattice configurations and 
           nearly all kernel choices the particles start eventually moving off the lattice 
           and move with average velocities of 1-2\% of the speed of sound. 
           For the hexagonal lattice, only the cubic spline kernel  (solid black, right panel; 
           "M$_4$") and the Wendland kernel (W$_{3,3}$) are stable. For the quadratic lattice only 
           the modified quintic spline kernel (blue triangles; "QCM$_6$") retains the particles
           in their original configuration on the time scale of the experiment.}
   \label{fig:noise_box}
\end{figure*}
%-----------------------------------------------------------------------
To date there is still only poor theoretical understanding which particle configurations represent
stable minimum energy configurations and how this depends on the chosen kernel. As a heuristic approach one can apply a 
"relaxation method" where artificial damping is used to drive the particles into a near-optimal 
configuration. However, this can be very time consuming, it is not necessarily clear when the 
equilibrium has been reached and the approach can become very challenging in practice for 
complicated initial conditions.\\
Often the particles are simply placed on a lattice which ensures accurate interpolation 
properties, see Sec.\ref{sec:SPH_interpolation} and \ref{sec:SPH_gradient}, but this does 
not guarantee that the particles are in a stable equilibrium 
position. In practice, for most kernel-lattice combinations the particles will start 
"moving off the lattice" and keep moving unless they are explicitly damped. It has been
observed \citep{springel10a} that this moving-off-the-lattice can hamper proper convergence 
in KH instabilities and, as we will see below, it also plays a major role in the convergence of 
the Gresho-Chan vortex problem.\\
We perform here a simple experiment where we place 10K particles in the domain [0,1] $\times$ [0,1]
so that their density is $N=1$ and their pressure is $P=100$ everywhere ($\Gamma= 5/3$). We
add margins of three smoothing lengths with of "frozen boundary particles" at each side. 
We perform the experiment twice, once with a hexagonal or ``close-packed'' (``CP-lattice''; left panel
in Fig.~\ref{fig:noise_box}) and once with a quadratic lattice (``Q-lattice''; right panel in 
Fig.~\ref{fig:noise_box}). Ideally, these configurations should be perfectly preserved. Subsequently 
we let the inner particles evolve freely (in practice we set our noise parameter to $\mathcal{N}_{\rm \; \; noise}= 10^6$,
see Eq.~(\ref{eq:kappa1}), so that the noise trigger does not switch on and no dissipation is applied) 
and thereby monitor the average particle velocity (in units of  the sound speed, $\langle v \rangle/c_{\rm s}$) 
as a function of time (in units of the sound crossing time, $\tau_{\rm s}$). \\
There are a number of interesting conclusions from this experiment. First, for almost
all kernel choices, the particles eventually move off the initial lattice and move with 
average velocities of 1-2\% of the sound speed. Interesting exceptions are the cubic spline 
kernel for which the hexagonal lattice seems to be  a stable minimum energy 
configuration ($\langle v \rangle/c_{\rm s}\approx 10^{-4}$). The Wendland kernel
$W_{3,3}$ remains to an even higher  accuracy ($\langle v \rangle/c_{\rm s}\approx 5 \times 10^{-7}$)
on the CP-lattice. The W$_{\rm H,9}$-kernel remains for around three sound crossing times perfectly
on the CP lattice, but then the particles begin to move and settle to a noise level comparable to
the lower-order kernels ($\langle v \rangle/c_{\rm s}\approx 10^{-2}$). It is also interesting
that kernels that are naively expected to be close approximations to each other show a very different
behaviour in this test. The W$_{\rm H,3}$-kernel, which closely approximates M$_4$, see Fig.~\ref{fig:smooth_kernels},
shows a very different noise behavior: at the chosen $\eta$ (=1.2, like M$_4$) it already starts forming
pairs while M$_4$ stays nearly perfectly on the CP-lattice. Also the W$_{\rm H,5}$-kernel is substantially 
more noisy than the original M$_6$-kernel. The modified quintic spline kernel, 
QCM$_6$, although not very accurate in previous tests, see Figs.~\ref{fig:dens_accuracy} and \ref{fig:grad_accuracy},
shows actually only little noise. It does not stay on the CP-lattice, but nevertheless only produces
little noise ($\langle v \rangle/c_{\rm s}\approx 5 \times 10^{-3}$), and it is the only kernel 
that remains exactly on the Q-lattice ($\langle v \rangle/c_{\rm s} <  10^{-6}$). Both the 
W$_{\rm H,9}$- and the W$_{3,3}$-kernels move off the Q-lattice after about three sound crossing
times, but W$_{3,3}$ shows a much lower noise level ($\langle v \rangle/c_{\rm s} <  10^{-3}$ vs. 
$\approx 10^{-2}$).\\
The behavior in this ``noisebox test'' is consistent with the results in the Gresho-Chan vortex,
see Sec.~\ref{sec:Gresho}, where noise is one of the accuracy-limiting factors. Also in this latter test
the W$_{3,3}$ shows the least noise, followed by the QCM$_6$-kernel, which performs even better than
$W_{\rm H,9}$. Interestingly, there seems to be no clear relation  between the degree of noise and  the 
kernel order, at least not for the explored initial configurations. The stability properties of particle distributions
deserve more theoretical work in the future.\\

\subsection{Static II: surface tension test (N)}
\label{sec:surf_tension}
\begin{figure*} 
   \centerline{
    \includegraphics[width=20cm,angle=0]{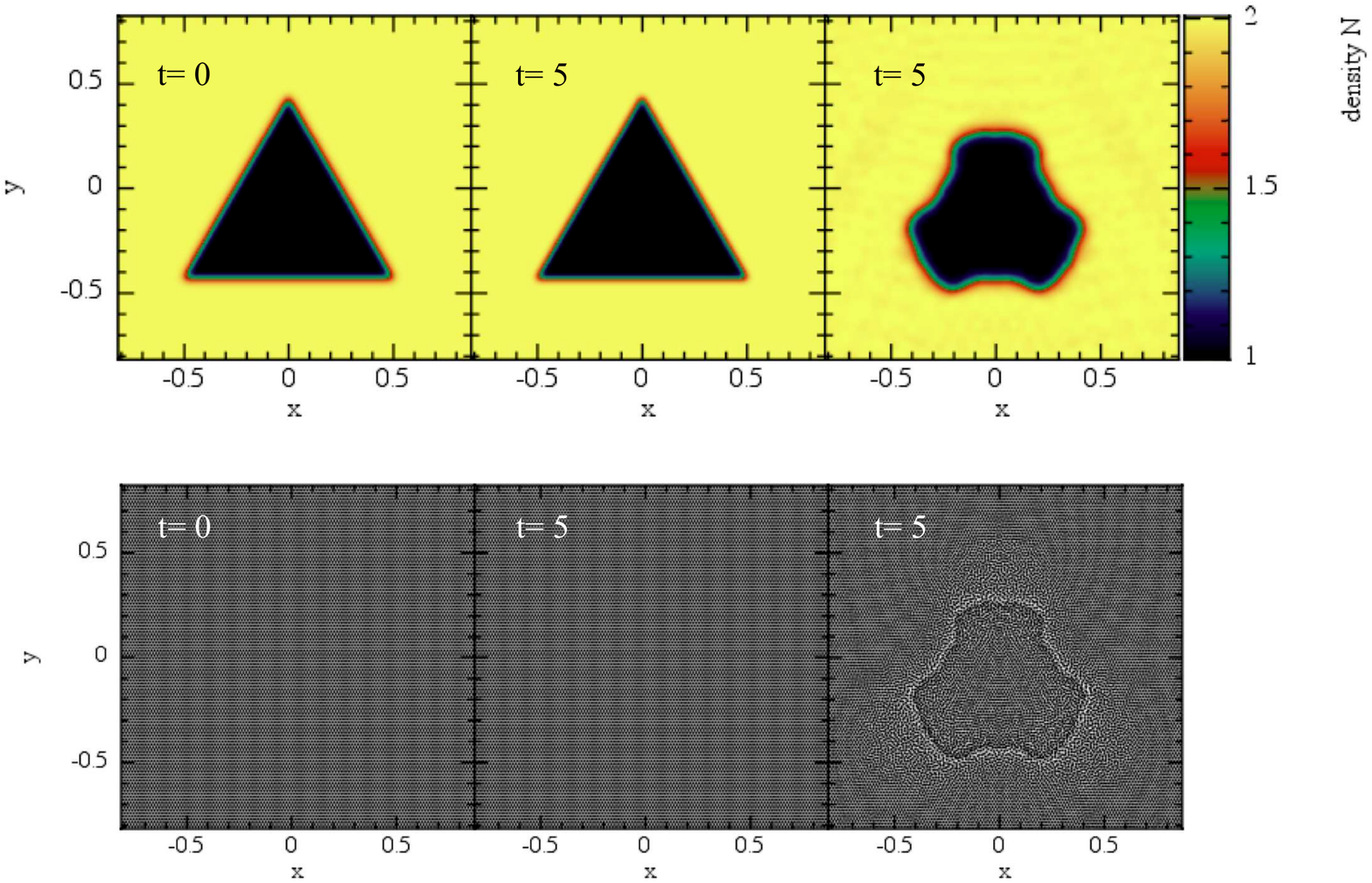}  
    }
   \vspace*{-1cm}
   \caption{Surface tension test: evolution of a triangular region in pressure equilibrium with its surroundings. 
            The first row shows the density $N$, the second the corresponding particle distribution. 
            The first column shows the initial condition, the second column the result for weighting factor
            $X=P^k$ (visually identical to the results obtained with $X=1$) and the last column shows the 
            result for the standard SPH volume choice ($X=\nu$ corresponding to $V_b= \nu_b/N_b$, or, 
            $V_b= m_b/\rho_b$ in Newtonian language). The standard SPH volume choice shows at $t=5$ 
            already strong deformations that are the result of spurious surface tension forces. The alternative
            choices $X=P^k$ and $X=1$ (not shown) perfectly preserve the original shape.}
   \label{fig:surface_tension}
\end{figure*}

As discussed in Sec.~\ref{sec:volume_elements}, depending on the initial setup, 
the standard choice for the SPH volume 
element $V_b= \nu_b/N_b$ (or $V_b= m_b/\rho_b$ in the Newtonian case) can lead to spurious 
surface tension forces across contact discontinuities which can prevent subtle 
instabilities from growing.\\
To test for the presence of such a spurious surface tension for the different choices 
of the volume element we set up the following experiment. We distribute 20K particles 
homogeneously on a hexagonal lattice within an outer box of $[-1,1] \times [-1,1]$. For 
the triangular central region with edge points ($-0.5,-\sqrt{3}/4$), ($0.5,-\sqrt{3}/4$) 
and ($0,\sqrt{3}/4$) we assign the density $N_i= 1.0$, the outer region has density 
$N_o= 2.0$, the pressure is $P= P_0= 2.5$ everywhere and the polytropic exponent is chosen 
as $\Gamma= 5/3$. Ideally the system should stay in exactly this state if it is allowed to 
evolve. Possibly present spurious surface tension forces would have the tendency to 
deform the inner triangle into a circle. We perform this test with formulation $\mathcal{F}_1$
and $\mathcal{F}_3$, but the only difference that matters here is the volume element.
In fact, after only $t= 5$ or about 2.6 sound crossing times the standard $X=\nu$-discretization 
has already suffered a substantial deformation, see Fig.~\ref{fig:surface_tension}, while 
$\mathcal{F}_1$ does not show any sign of surface tension and is indistinguishable 
from the original configuration (left panel). The choice $X=1$, by the way, yields identical
results to $X= P^k$.
%-----------------------------------------------------------------------
%-----------------------------------------------------------------------

\subsection{Gresho-Chan-like vortex (N)}
\label{sec:Gresho}
\begin{figure*} 
   \hspace*{-1cm}\includegraphics[width=14cm,angle=-90]{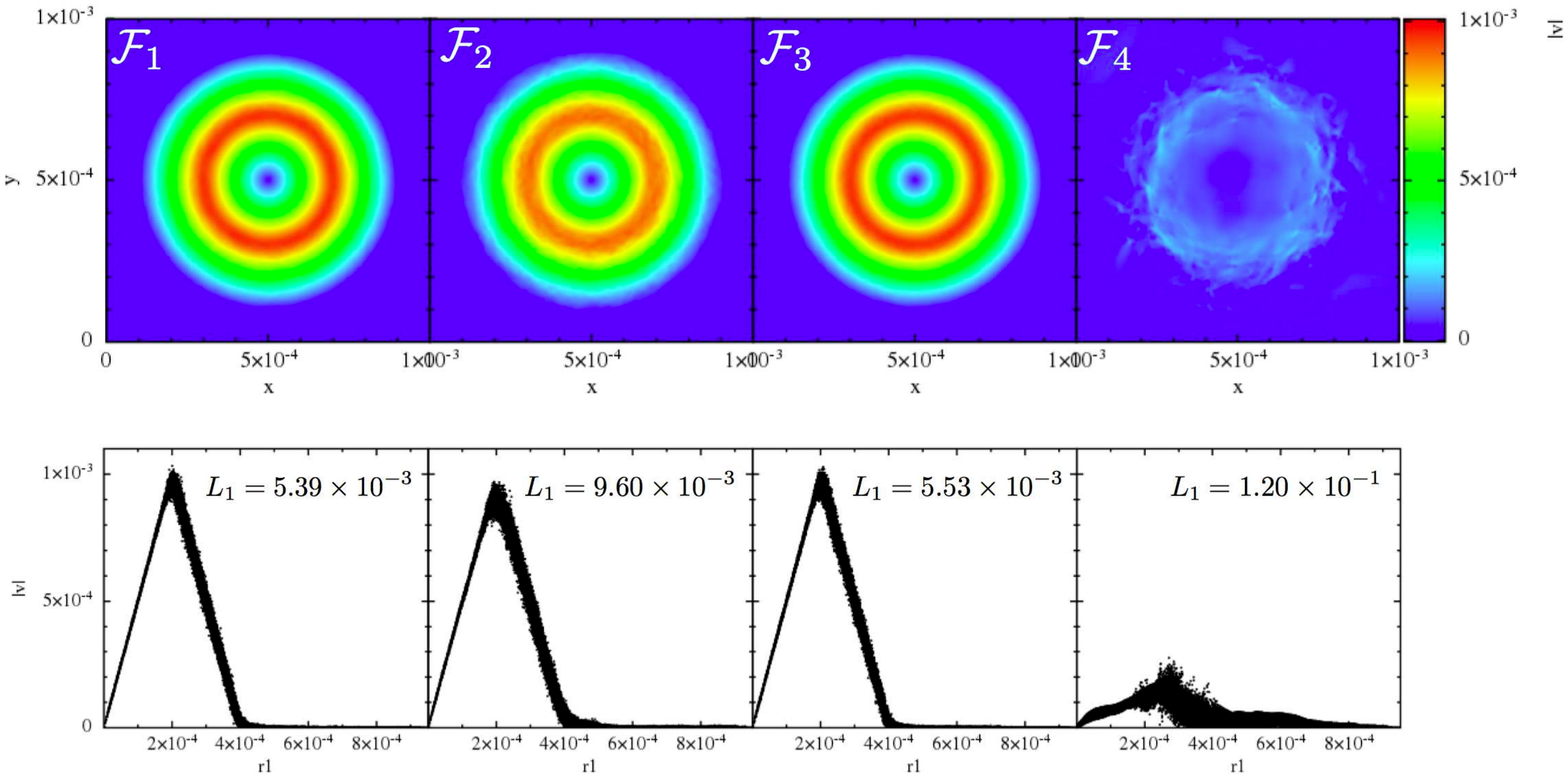}
   \vspace*{-2cm}
   \caption{Performance of the different SPH formulations in the Gresho-Chan vortex test (from left to right column: 
            $\mathcal{F}_1$ to $\mathcal{F}_4$; snapshots are shown at $t=1$, all 50K
            particles are displayed). The upper row shows $|\vec{v}|$ color-coded in the x-y-plane,
            the lower row shows the velocity as a function of the distance to the vortex center.
            In the lower row also the $L_1$-error is indicated. The constant high dissipation
            in $\mathcal{F}_4$ seriously deteriorates the result compared to $\mathcal{F}_1$ to 
            $\mathcal{F}_4$. The impact of the integral-based vs. kernel-derivative gradients
            can be judged by comparing $\mathcal{F}_1$ (integral-based; first column) to $\mathcal{F}_2$
            (kernel gradient based; otherwise identical; second column). The choice of the volume element
            hardly makes a difference in this test (see $\mathcal{F}_1$ vs $\mathcal{F}_3$).}
   \label{fig:Gresho_formulation}
\end{figure*}

The Gresho-Chan vortex \citep{gresho90} is considered a particularly difficult test
in general and in particular for SPH. As shown in \cite{springel10a}, standard SPH 
shows very poor convergence in this test. 
The test deals with a stationary vortex that should be in stable equilibrium. Since 
centrifugal forces and pressure gradients balance exactly, any 
deviation from the initial configuration that develops over time is spurious and 
of purely numerical origin. The azimuthal component of the velocity in this test rises 
linearly up to a maximum value of $v_0$ which is reached at $r= R_1$ and subsequently 
decreases linearly back to zero at 2$R_1$
\be
v_\varphi (r)=  v_0 \left\{
  \begin{array}{ l l l}
     u \hspace*{1cm} {\rm for \quad } u \le 1\\
     2 - u  \hspace*{0.5cm} {\rm for \quad } 1 < u \le 2\\
     0   \hspace*{1cm} {\rm for \quad } u > 2,\\
   \end{array} \right.
\ee
where $u= r/R_1$. If we require that centrifugal and pressure accelerations balance, 
the pressure becomes
\be
P(r)= P_0 + \left\{
  \begin{array}{ l l l}
     \frac{1}{2} v_0^2 u^2  \hspace*{2.8cm} {\rm for \quad } u \le 1\\
     4 v_0^2 \left(\frac{u^2}{8} - u + \ln{u} + 1 \right)  \hspace*{0.65cm} {\rm for \quad } 1 < u \le 2\\
     4 v_0^2 \left(\ln 2 - \frac{1}{2}\right) \hspace*{1.8cm} {\rm for \quad } u > 2.\\
   \end{array} \right.
\ee
In the literature on non-relativistic hydrodynamics \citep{liska03,springel10a,read12,dehnen12} usually $v_0= 1$ 
is chosen together with $R_1= 0.2$, a uniform density $\rho= 1$ and a polytropic exponent of 5/3. 
Since we want to run this Newtonian test with our special-relativistic code, we choose $R_1= 2 \times 10^{-4}$, 
$P_0= 5 \times 10^{-7}$ and  $v_0= 10^{-3}$ to be safely in the non-relativistic limit. For this test, the particles
are placed on a hexagonal lattice in the domain $[0,10^{-3}] \times [0,10^{-3}]$.\\
The differential rotation displaces the particles from their original hexagonal lattice positions 
and therefore introduces some amount of noise. The noise trigger $\mathcal{N}_{\rm \; \;1}$, see 
Eq.~(\ref{eq:N_trigg_1}), triggers local values of $K_a$ of up to 0.04, $\mathcal{N}_{\rm \; \;2}$ does 
not switch on, just as it should.
Also the shock trigger works very well and does practically not switch on at all: in an initial transient 
phase it suggests (still negligible) values for the dissipation parameter $K_a$ of $\sim 10^{-3}$ and 
then decays quickly to $< 10^{-6}$. 
We have used this test to gauge our noise triggers. We find good results for a noise reference value 
of $\mathcal{N}_{\rm \; \; noise} = 50$, this will be explored further below.\\ 
As a first test we compare the performance of the different formulations $\mathcal{F}_1$ to
$\mathcal{F}_4$ for 50K particles, see Fig.~\ref{fig:Gresho_formulation}.
%-----------------------------------------------------------------------
%-----------------------------------------------------------------------
Both $\mathcal{F}_1$ and $\mathcal{F}_3$ perform nearly identically well in this test, indicating
that (as expected) the choice of the volume element has essentially no influence on the result.
The integral-based gradients ($\mathcal{F}_1$ and $\mathcal{F}_3$) are substantially less noisy
than the kernel-gradient based ones ($\mathcal{F}_2$). Consistent with the findings of \cite{springel10a},
the $\mathcal{F}_4$ formulation performs very poorly, mainly due to the unnecessarily high dissipation, 
but also due to the cubic spline kernel, see below. In the lower row of the figure we display 
$|\vec{v}|$ as a function of the distance from the vortex center $r1$ and also measure the $L_1$-error 
norms for the velocity,
\be
L_1= \frac{1}{N v_0} \sum_{a=1}^{N} |\vec{v}_a - \vec{v}(\vec{r}_a)|,
\ee
with $\vec{v}(\vec{r}_a)$ being the exact stationary solution at position of particle $a$. The error
values for the different formulations are indicated in the lower row of the figure.\\
%-----------------------------------------------------------------------
\begin{figure*} 
   \hspace*{-1cm}\includegraphics[width=14cm,angle=-90]{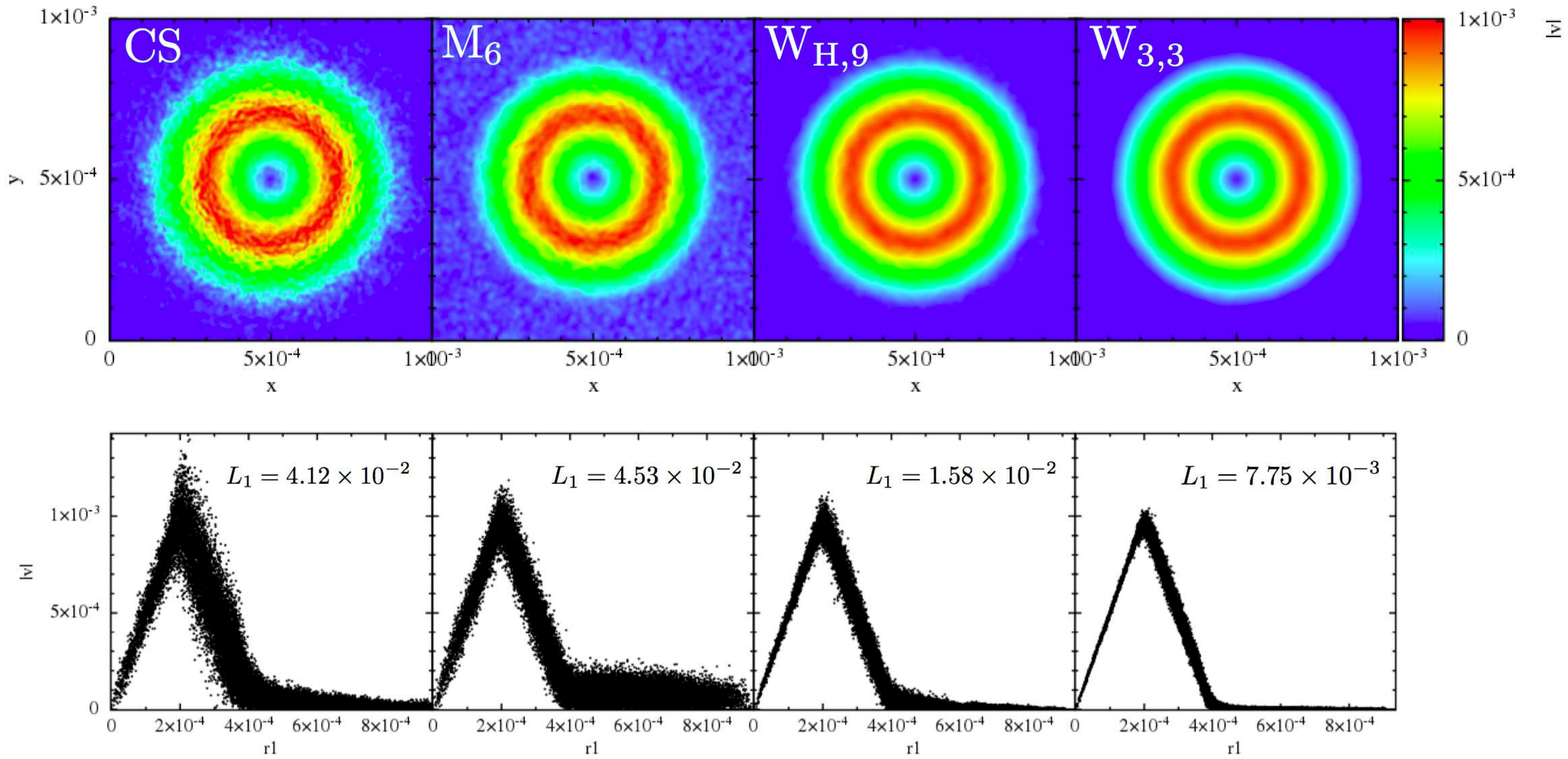}
   \vspace*{-2cm}
   \caption{Performance of different kernels in the Gresho-Chan vortex test. All tests used 25K particles
           and the SPH formulation $\mathcal{F}_1$ (apart from the kernel choice, of course), results are shown
           at $t=1$.
           Upper row: $|\vec{v}|$ in x-y-plane, lower row: $|\vec{v}|$ as a function of the distance from the
           vortex center.
           From left to right: cubic spline, quintic spline, W$_{\rm H,9}$ and Wendland kernel.
           In the lower row also the $L_1$ error norm of the velocity is given. Clearly, the
           Wendland kernel produces the most accurate result.}
   \label{fig:Gresho_kernel}
\end{figure*}
%-----------------------------------------------------------------------
%-----------------------------------------------------------------------
\begin{figure} 
   \hspace*{-1cm}\includegraphics[width=10cm,angle=0]{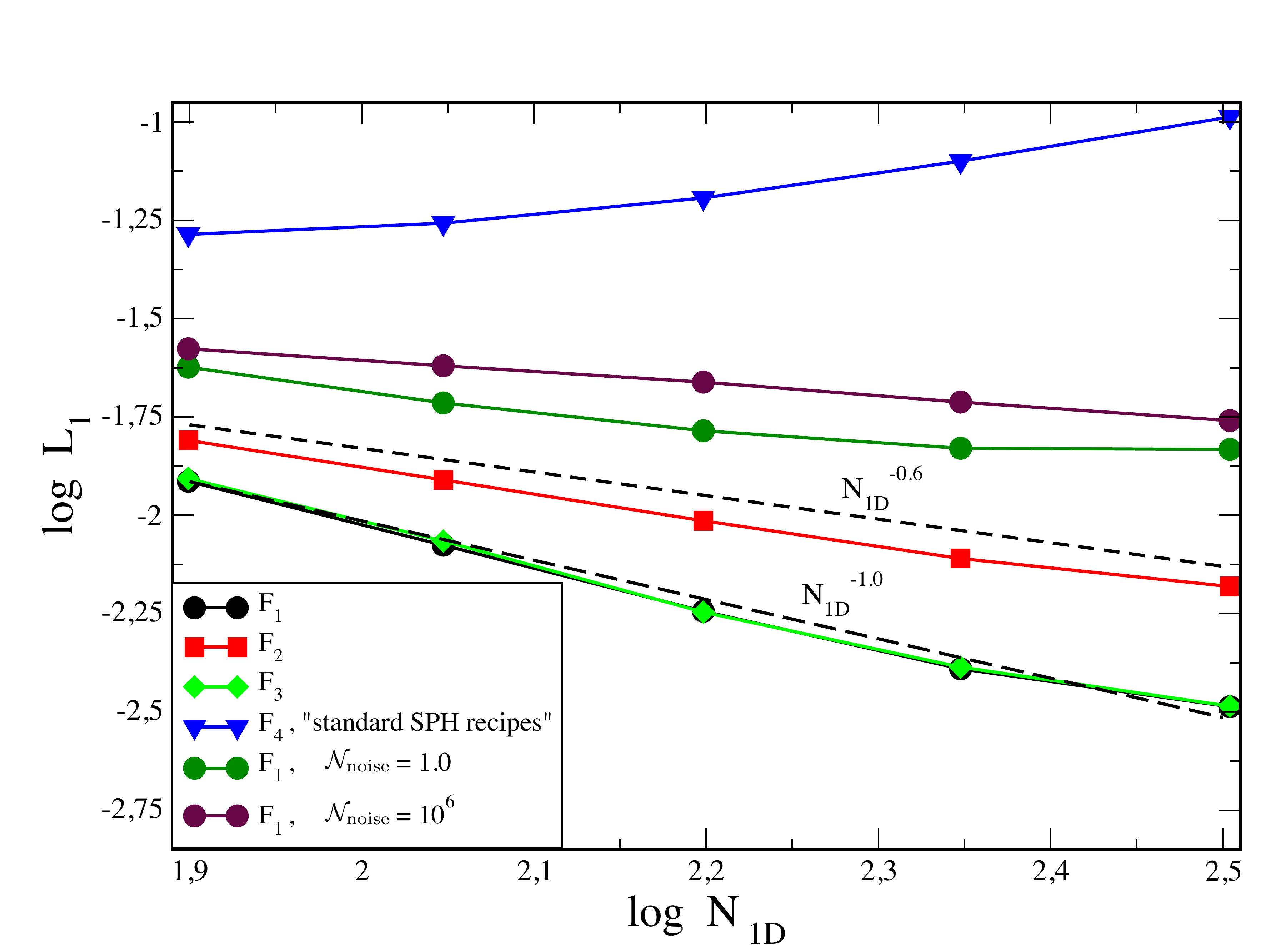}
   \vspace*{0cm}
   \caption{Convergence in Gresho-Chan vortex test. Our best SPH formulation $\mathcal{F}_1$
            converges close to linearly with the effective, one-dimensional particle number $N_{\rm 1D}$ 
            (long-dashed line), $\mathcal{F}_3$ performs very similar.
            Our version with kernel derivatives, $\mathcal{F}_2$, converges somewhat slower, roughly
            $\propto N_{\rm 1D}^{0.6}$ (short-dashed line), consistent with the findings of Dehnen 
            and Aly (2012). The noise trigger clearly improves the convergence, both too little 
            ($\mathcal{N}_{\rm \; \; noise}= 10^6$) and too high sensitivity ($\mathcal{N}_{\rm \; \; noise}= 1.0$) to noise 
            deteriorate the convergence rate. The ``standard SPH recipe'' formulation,  $\mathcal{F}_4$, 
            fails this test rather catastrophically.}
   \label{fig:Gresho_convergence}
\end{figure}
%-----------------------------------------------------------------------
To better understand the role of the chosen kernel function, we perform an additional experiment,
where we use the best SPH formulation, $\mathcal{F}_1$, but vary the kernel (with the smoothing lengths
as indicated in Tab.~\ref{tab:eta_crit}). As already noticed in the "noise-box" test, different kernels 
exhibit different levels of noise which is crucial for the accuracy in this test. In Fig.~\ref{fig:Gresho_kernel}
we show the results for 25K particles for  the CS, M$_6$, $W_{\rm H,9}$ and the $W_{3,3}$ kernel. 
The upper row shows the velocity, color-coded as function of the position, the lower row displays the
velocity as function of the distance to the vortex center. For our chosen noise reference
value $\mathcal{N}_{\rm \; \; noise}$ only very little dissipation is released in response to particle noise. Therefore, at the shown
time $t= 1$ the triangular velocity shape is still reasonably well captured in all cases, though with
substantial noise overlaid in all cases but $W_{\rm H,9}$ and $W_{3,3}$. A large fraction of the error 
is due to the particles moving off their original hexagonal lattice. These results are consistent
with the findings from the ``noise box'' test, see above, and show once more that the $W_{3,3}$ kernel 
only produces a very small amount of noise in comparison to the other explored kernels. We have
run this test for the peaked kernels, each time with the same support as the $W_{3,3}$ kernel 
($\eta=2.2$), not shown. While the LIQ kernel performs  poorly and produces very noisy results
with a large error ($L_1= 6.06 \times 10^{-2}$), the QCM$_6$ kernel performs again astonishingly well, 
though not as good as $W_{3,3}$. It produces symmetric, relatively noise-free results with an error of
only $L_1= 1.06 \times 10^{-2}$, actually even slightly better than the $W_{\rm H,9}$ kernel 
($L_1= 1.58 \times 10^{-2}$).\\
We also explore the convergence with the effective, one-dimensional particle number $N_{\rm 1D}$
of the different formulations in Fig.~\ref{fig:Gresho_convergence}.
Clearly, the ``standard SPH recipe'' formulation $\mathcal{F}_4$ (blue triangles) performs very poorly,
consistent with the findings of \cite{springel10a} who concludes that (standard) SPH does not converge
to the correct solution. Both our formulations $\mathcal{F}_1$ (black circles) and $\mathcal{F}_3$ 
(green diamonds) converge roughly linearly (long-dashed line). The kernel-gradient version $\mathcal{F}_2$ 
converges roughly $\propto N_{\rm 1D}^{-0.6}$, consistent with the findings of \cite{dehnen12}. Recently,
also \cite{hu14a} have implemented a number of worthwhile modifications into the GADGET code 
\citep{springel05a} that substantially improve the convergence in the Gresho-Chan vortex. They 
measured a convergence rate close to 0.8. \cite{read12} report a convergence rate close to 0.9.\\
We also perform experiments to explore the impact of the noise trigger on the convergence: once we  
run the test without dissipation (in practice $\mathcal{N}_{\rm \; \; noise}= 10^6$) and once we deliberately apply 
too much dissipation ($\mathcal{N}_{\rm \; \; noise}= 1.0$). In the first case there is steady (though slower, 
approximately $\propto N_{\rm 1D}^{-0.6}$) convergence, whereas the excessive dissipation in the 
second case seems to hamper convergence. The moderately {\em triggered dissipation by noise therefore
increases the convergence rate}. This is our main motivation for using the noise trigger $\mathcal{N}^{(1)}$,
see Eq.~(\ref{eq:N_trigg_1}). For many tests, however, we would expect to obtain very similar results 
even if this (small) source of dissipation was ignored.

\subsection{Advection tests (SR)}
For our Lagrangian schemes advection tests do not represent any particular challenge since the code 
actually has not much to do apart from accurately recovering the primitive variables and advancing the 
particle positions. Therefore, advection is  essentially perfect. We briefly demonstrate the excellent 
advection properties in the two following tests.

\subsubsection{Advection I}
\begin{figure*} 
   \centerline{
    % triangle
    \includegraphics[width=10cm,angle=-90]{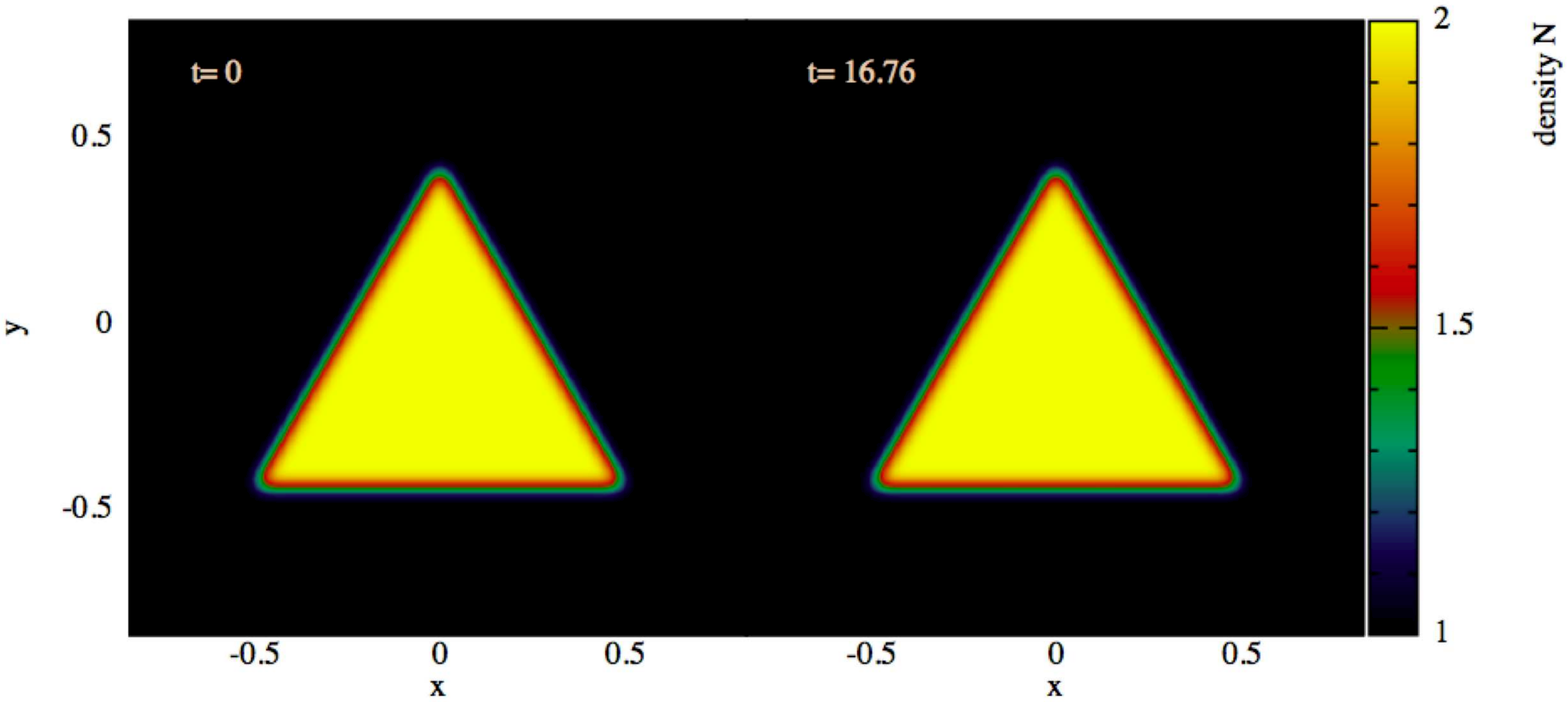} 
    }
    \vspace*{-2cm}
   \caption{Advection test 1: the initial density pattern shown on the left, constant 
            pressure everywhere, is advected with $v_x= 0.9999$ ($c=1$) corresponding to a Lorentz 
            factor $\gamma= 70.7$ through the computational domain with periodic 
            boundary conditions. Right panel: shape after ten times crossing
            the computational domain.}
   \label{fig:advection1}
\end{figure*}

In this first test we start from the configuration of the surface tension test, see 
Sec.~\ref{sec:surf_tension}, with 7K particles placed inside the domain
$[-1,1] \times [-1,1]$. We now give each fluid element a velocity in x-direction of 
$v_x= 0.9999$ corresponding to a Lorentz factor of $\gamma\approx 70.7$ and apply 
periodic boundary conditions. In Fig.~\ref{fig:advection1} we show the results for 
the $\mathcal{F}_1$ equation set. After ten times crossing the computational 
domain\footnote{The numerical value, $t= 16.76$, comes from our periodic boundaries being 
4 smoothing length inside the nominal box boundaries.} the shape of the triangle has not 
changed in any noticeable way. The $\mathcal{F}_2$ results are visually identical, 
$\mathcal{F}_3$ and $\mathcal{F}_4$ (not shown) deform the triangle like in the 
previous surface tension test.
%-----------------------------------------------------------------------
%-----------------------------------------------------------------------
\subsubsection{Advection II}
\begin{figure*} 
   \centerline{ \hspace*{-0.9cm}
   \includegraphics[width=6cm,angle=0]{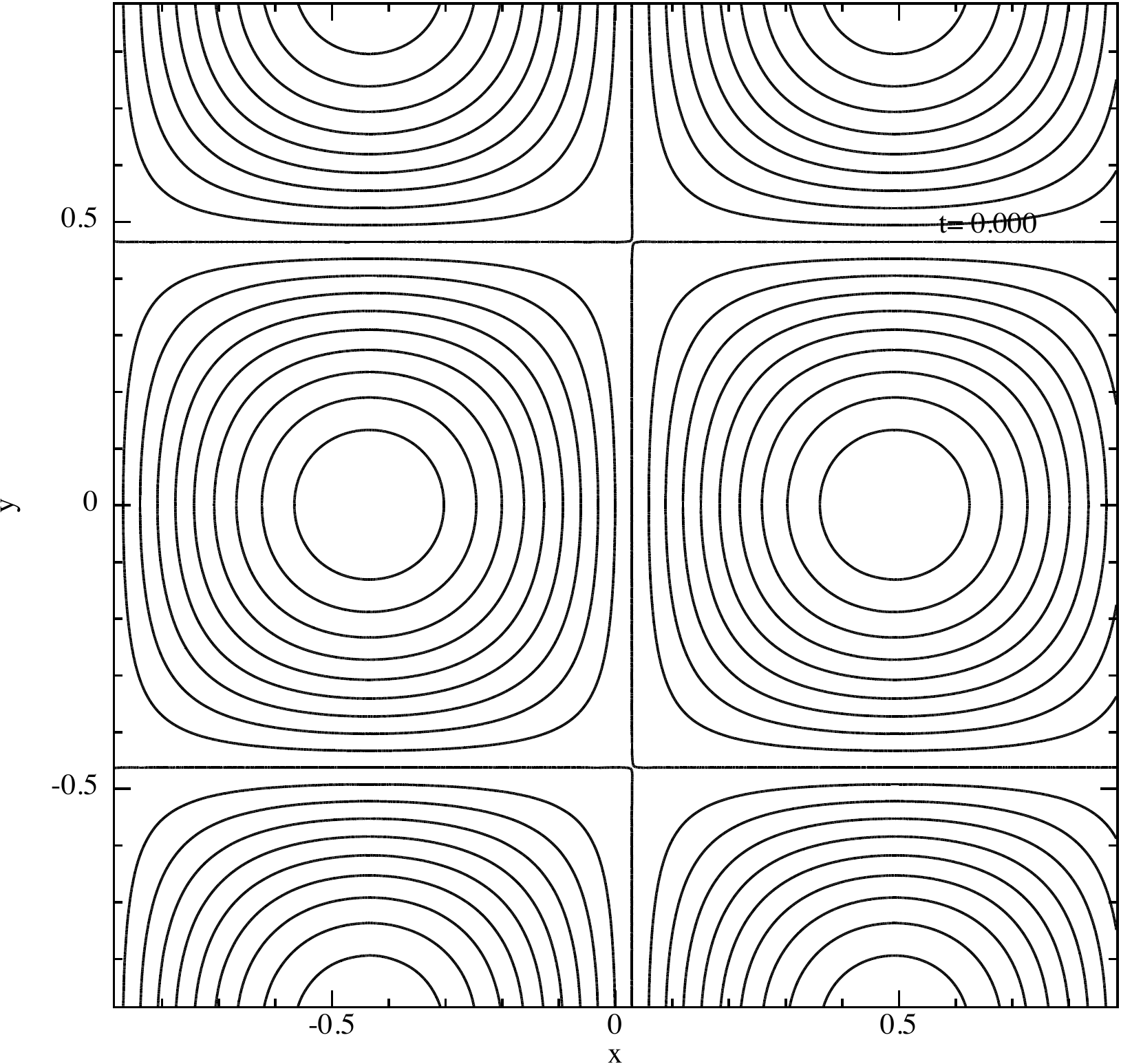} 
   \includegraphics[width=6cm,angle=0]{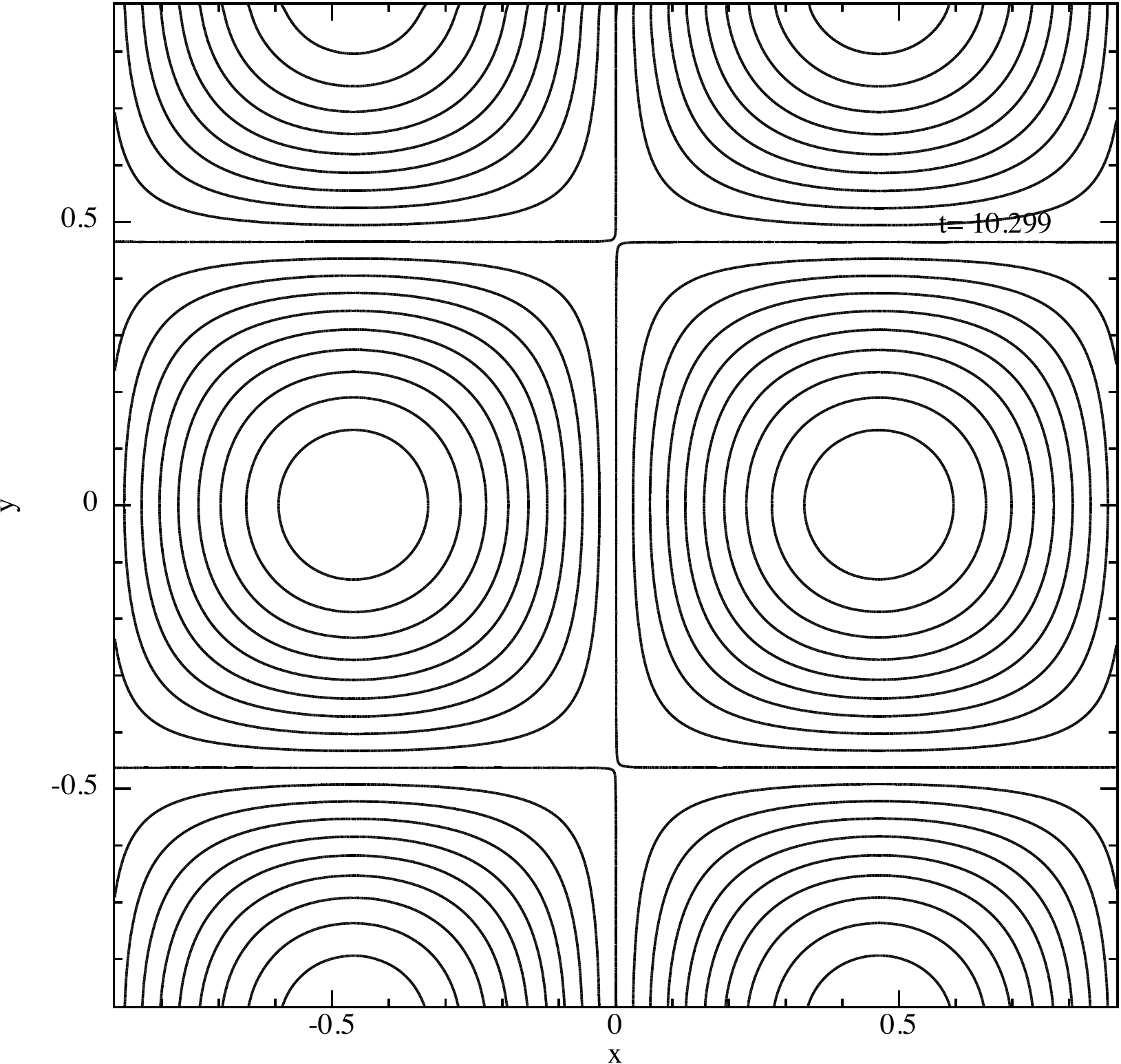}}  
    \caption{Advection test 2: a density pattern is advected with a velocity of 0.9 c ($\gamma= 2.29$)
    through a box with periodic boundaries. The initial condition ($t= 0.000$) is shown in the left panel, 
    the result after crossing the box 5 times ($t= 10.299$) is shown on the right.}
   \label{fig:2D_advection}
\end{figure*}

In this second test we distribute 40K particles in $[-1,1] \times [-1,1]$. We place our periodic
boundary conditions four smoothing length inside of this domain and denote their coordinates
by $x_1/x_2$ and $y_1/y_2$. The particles inside are the ``core'' particles, the particles outside
are appropriate copies that enforce the periodic boundary conditions. 
We set up a density pattern
\be
N(x,y)= 1 + 0.5 \sin[x \; 2\pi/(x_2-x_1)] \cos[y \; 2\pi/(y_2-y_1)],
\ee
use $\Gamma=5/3$ and adjust the internal energy so that the pressure is equal to unity 
everywhere. All particles receive a constant boost velocity of $0.9 c$ in positive x-direction,
corresponding to a Lorentz factor of $\gamma= 2.29$.
The initial density is shown in the left panel of Fig.~\ref{fig:2D_advection}, the density after 
crossing the  box five times is shown in the right panel, both patterns are virtually identical.
%-----------------------------------------------------------------------
%-----------------------------------------------------------------------

\subsection{Riemann problems (SR)}
\label{sec:riemann_probs}
Since we have tested a large variety shock benchmark tests in previous work \citep{rosswog10b,rosswog11a} 
and the results of the new formulations do not differ substantially in shocks, we restrict ourselves here to a 
few standard shock tests and focus on multi-dimensional tests where either the differences due to the new numerical
elements are more pronounced or where SPH has been criticized to not perform well.

\subsubsection{Riemann problem I: Sod-type shock}
\label{sec:riemann_I}
%-----------------------------------------------------------------------
\begin{figure*} 
   \centering
    \hspace*{-1.5cm}\includegraphics[width=12cm,angle=-90]{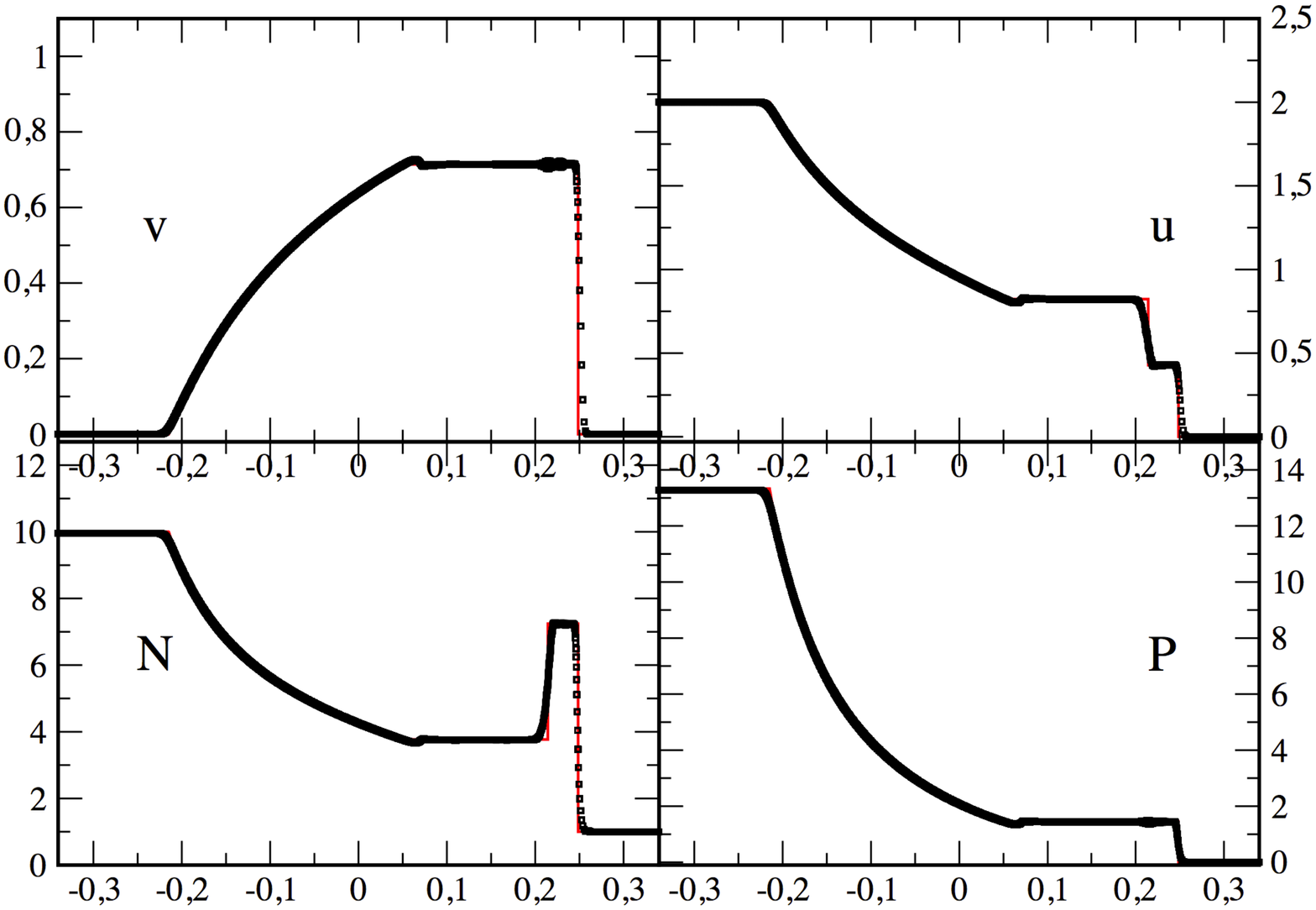}
     \caption{2D relativistic Sod-type shock tube at $t=$ 0.3. The SPH result ($\mathcal{F}_1$ formulation,
                   50K particles; note that every particle is shown, i.e. no averaging has been applied) is shown as black squares, the 
                   red line is the exact solution.}
\label{fig:rel_Sod_2D}
\end{figure*}
\begin{figure*}
   \centering
       \hspace*{-1cm}\includegraphics[width=16cm,angle=0]{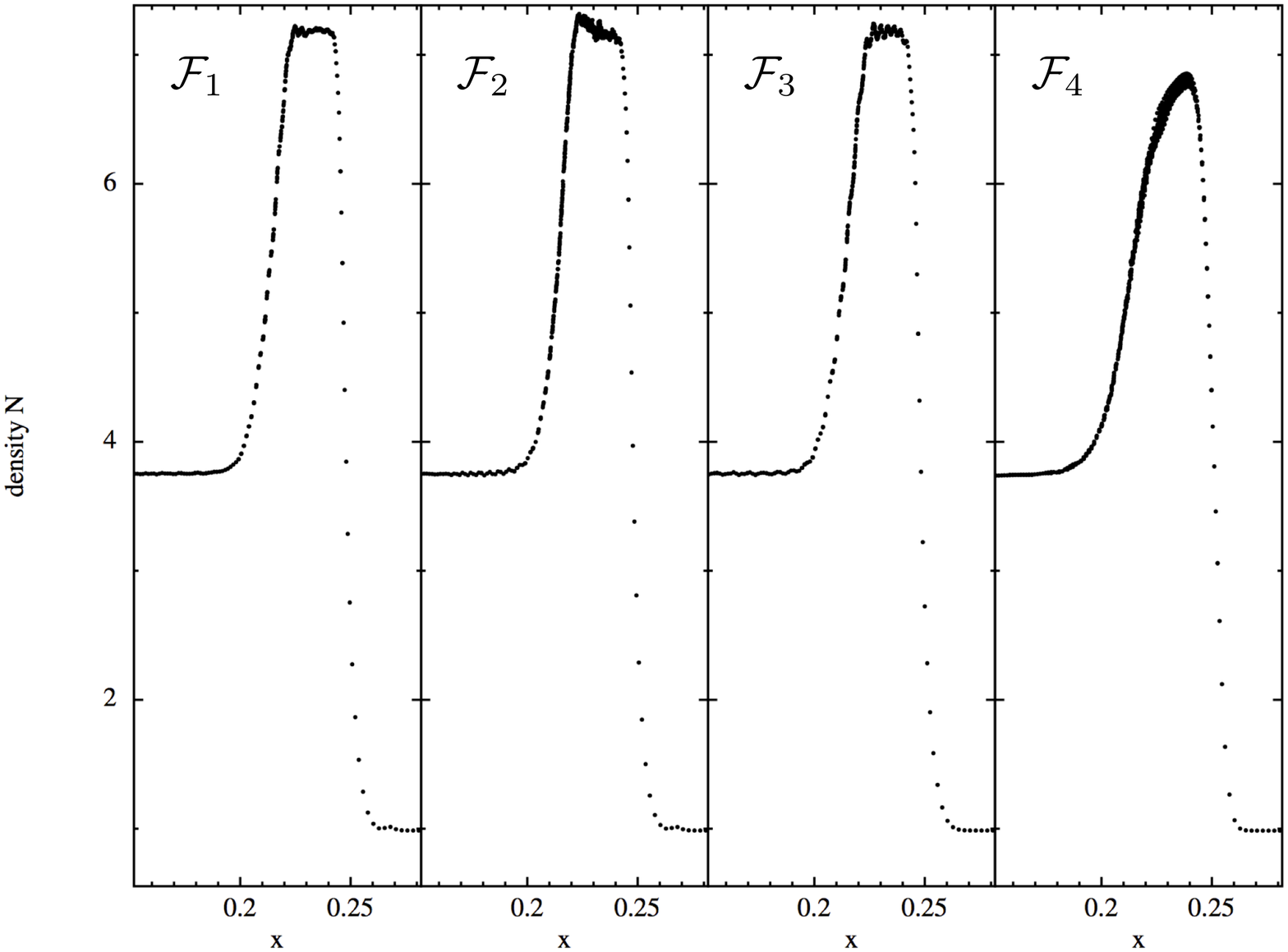} 
     \caption{Low-resolution (20K particles) 2D Sod shock tube test. Shown is the post-shock density plateau
     at $t=0.3$ for the SPH formulations $\mathcal{F}_1$ to $\mathcal{F}_4$ (left to right). The plateau is 
     best captured by the formulations that use the integral-based kernel ($\mathcal{F}_1$ and $\mathcal{F}_3$),
     among them the version with volume weight $X=P^k$ ($\mathcal{F}_1$) shows fewer oscillations. The
     version with weight $X=P^k$ but direct kernel gradients ($\mathcal{F}_2$) still works reasonably well,
    but  the  "standard SPH recipe" formulation ($\mathcal{F}_4$) is excessively dissipative. Note that the
    small oscillations in the plateau could be further reduced, though at the expense of more dissipation.}
\label{fig:Sod_F1_to_F4}
\end{figure*}
\begin{figure*}
   \centering
   \centerline{\hspace*{0.cm}\includegraphics[width=18cm,angle=0]{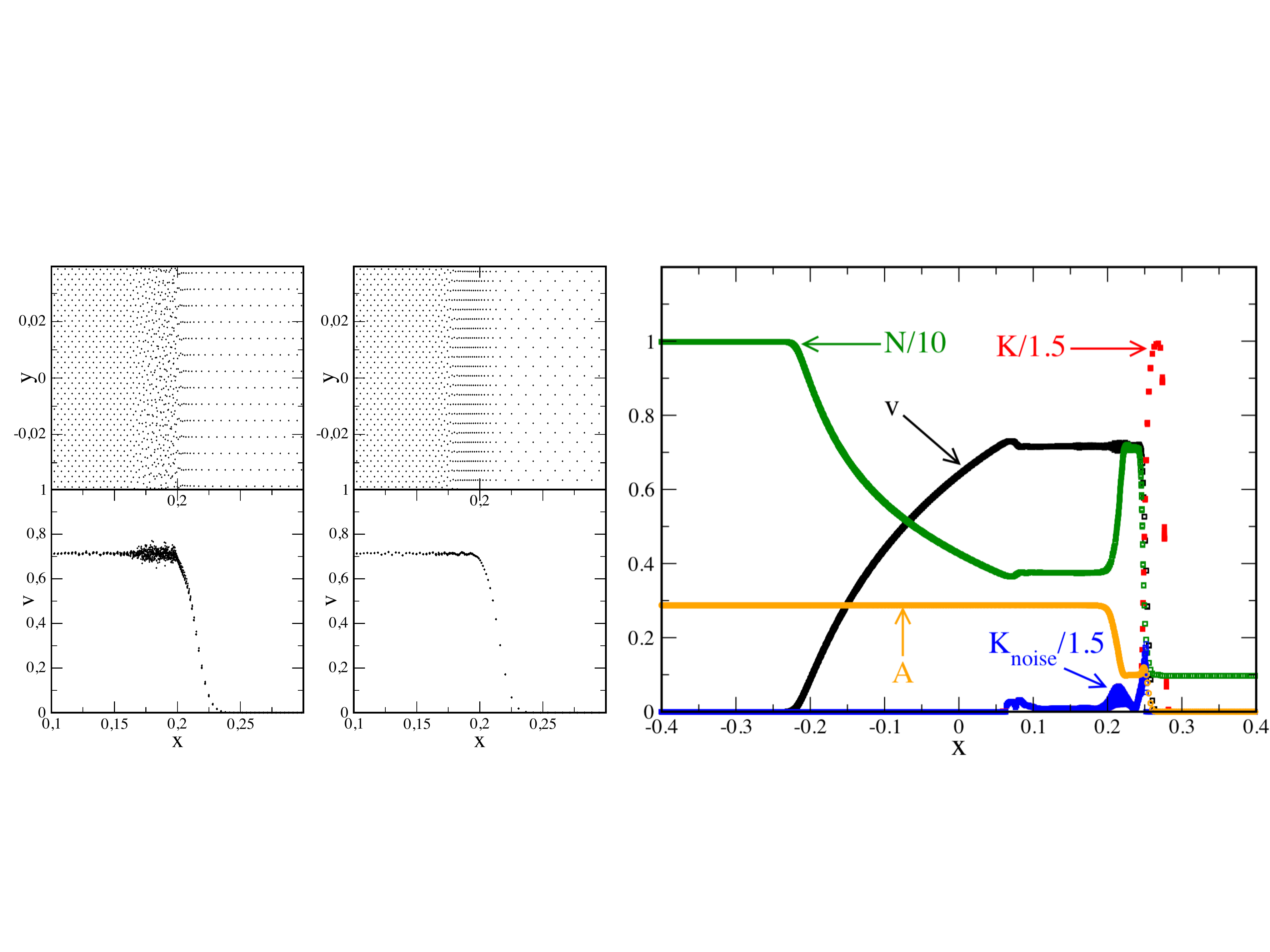} }
\vspace*{-2.5cm}
   \caption{Left: the importance of the initial particle lattice is illustrated by zooming into
   the shock front of a low-resolution, 2D, relativistic Sod simulation (5K particles). 
   The left column shows the result (particle distribution, above, and velocity, below) for the 
   case where 
   on both sides of the shock a hexagonal lattice with a symmetry axis along the shock 
   motion direction was used. The only difference in the right column is that the initial 
   lattice on the right-hand side was rotated by 30$^\circ$. This substantially reduces
   the "remeshing noise" in the velocity (near $x\approx 0.2$) behind the shock.
   Right: 2D shock tube test (30K particles) where also the dissipation parameter $K$ is shown (red).
              The blue circles show the parameter values suggested by the noise trigger, the orange
              symbols show the values of the entropy function $A=P/n^{\Gamma}$.}
   \label{fig:2D_Sod_initial_distrib}
\end{figure*}

%-----------------------------------------------------------------------
This mildly relativistic shock tube ($\gamma_{\rm max} \approx 1.4$) has become a widespread benchmark
for relativistic hydrodynamics codes \citep{marti96,chow97,siegler00b,delZanna02,marti03}. It uses a polytropic
exponent of $\Gamma= 5/3$, vanishing initial velocities everywhere, the left state has a pressure
$P_L=40/3$ and a density $N_L=10$, while the right state is prepared with $P_R=10^{-6}$ and $N_R=1$.\\
We place 50K particles initially on a hexagonal lattice in  $[-0.4,0.4] \times [-0.02,0.02]$, with 
particles in the low density region being rotated by 30 degrees, see below, and use
our standard parameter set. The SPH result of the $\mathcal{F}_1$ formulation (squares, at t= 0.3) agrees 
very well with the exact solution (solid line), see Fig.~\ref{fig:rel_Sod_2D}. Only directly after 
the shock front some "remeshing noise" is visible. This is unavoidable since the particles have to move from their initial configuration into a new one
which also involves small velocity components in y-direction. This remeshing noise could be further 
reduced by applying more dissipation, though at the price of reducing the height of the density plateau near 
$x= 0.25$. 
Note, that with the presented dissipation scheme  the theoretical density peak is reached
with only 50K particles while in earlier work \citep{rosswog10b} it was hardly reached even with 
140K particles. This is mainly due to the new shock trigger where the dissipation parameter peaks $\sim 2$
smoothing lengths ahead of the shock and starts decaying immediately after, similar to the 
case of the \cite{cullen10} shock trigger.\\
A comparison between the different formulations for only 20K particles (at $t=0.3$) is 
shown Fig.~\ref{fig:Sod_F1_to_F4}. 
We focus on the post-shock plateau of $N$ since here dissipation effects are particularly visible. 
The $\mathcal{F}_1$ formulation delivers the cleanest and "edgiest" result.
Note that the choice of $X= \nu$ ($\mathcal{F}_3$) introduces additional density oscillations.
This ``lattice-ringing'' phenomenon at low dissipation using the standard volume element (i.e. for 
$\mathcal{F}_3$) has been observed in a number of the following tests (KH-instabilities, Sec.~\ref{sec:KH}),
the blast bubble interactions and the ``blast-in-a-box'' problem, see Sec.~\ref{sec:combined}).
The clearly worst, excessively dissipative result is obtained with formulation $\mathcal{F}_4$. 
%-----------------------------------------------------------------------
%-----------------------------------------------------------------------
\\
In the previous cases we had rotated the particle distribution on the RHS
by 30$^\circ$ to avoid continuously collecting particles along the direction of motion. We illustrate the
impact of this measure in the experiment shown in Fig.~\ref{fig:2D_Sod_initial_distrib}, left panel: we set up a low 
resolution (5K particles) version of the above test, once we use the same lattice orientation on both sides
(left column, left panel) and once we rotate the distribution by 30$^\circ$ (right column, left panel). While both cases capture 
the overall solution well, the case without rotation shows a substantial ``remeshing noise'' behind the 
shock while in the rotated case the particles remain well-ordered without much  velocity noise. 
Such noise could be removed by adding more dissipation, but rotating the particle distribution 
is certainly the better option. In the right panel of Fig.~\ref{fig:2D_Sod_initial_distrib} we show the results
of another experiment (30K particles) where we demonstrate the working of our dissipation triggers.
The dissipation parameter $K$ is shown together with the desired values from the noise trigger. We also
show the values for the entropy function $A$, as reconstructed from pressure and density via $A= P/n^{\Gamma}$.
%-----------------------------------------------------------------------
%-----------------------------------------------------------------------

\subsubsection{Riemann problem II: Relativistic Planar Shock Reflection}
\begin{figure*} 
   \centering
   \hspace*{-1.cm}\includegraphics[width=12cm,angle=-90]{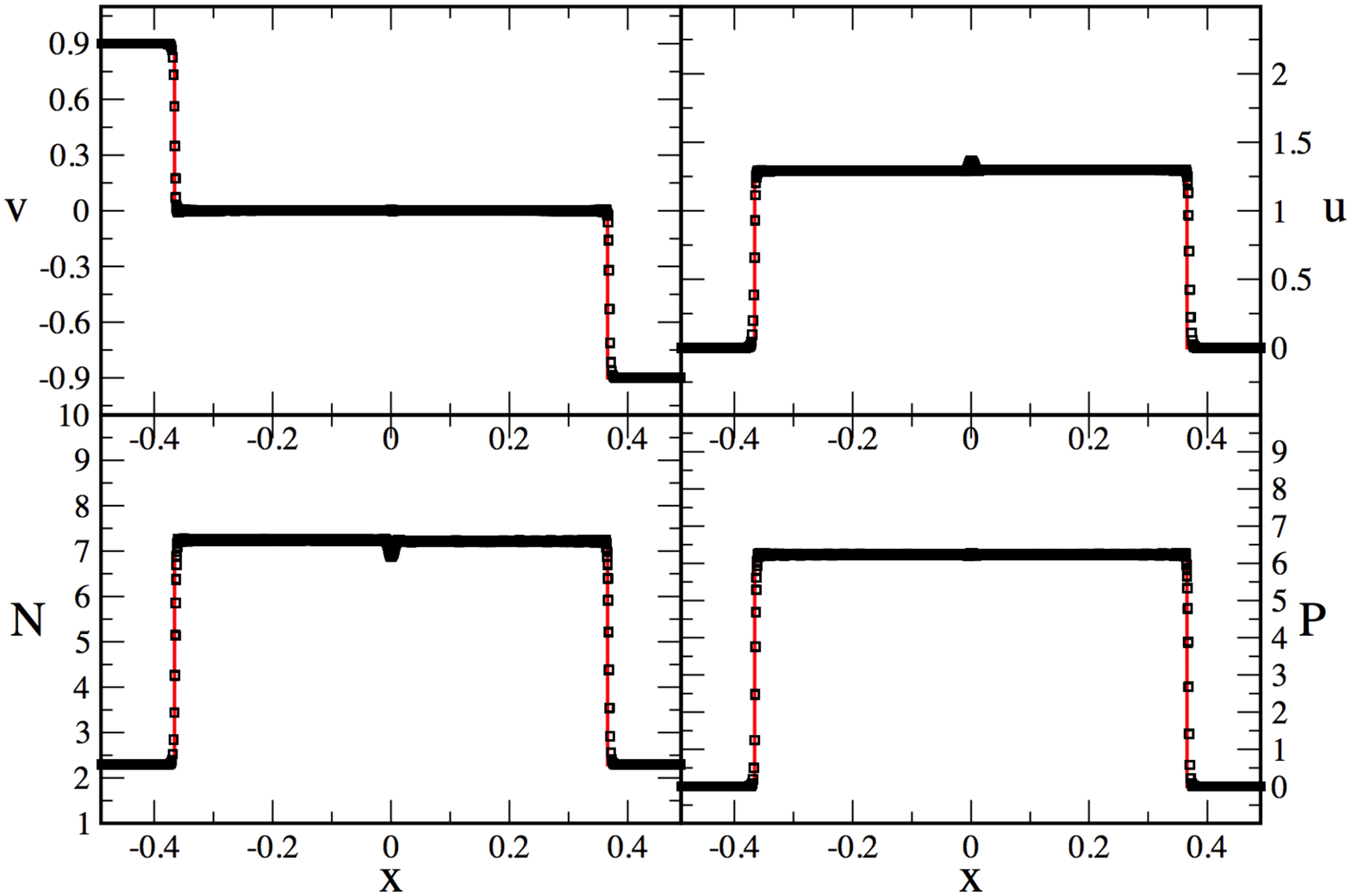} 
   \caption{2D, relativistic planar shock reflection test where two gas streams collide with $v_x= \pm 0.9$ c. 
                 The exact solution is shown by the solid, red line the SPH solution ($\mathcal{F}_1$ 
                 formulation, $t=0.875$) is shown as black squares.}
   \label{fig:RPSR_v0.9}
\end{figure*}

\begin{figure*} 
   \centering
   \hspace*{-1.cm}\includegraphics[width=12cm,angle=-90]{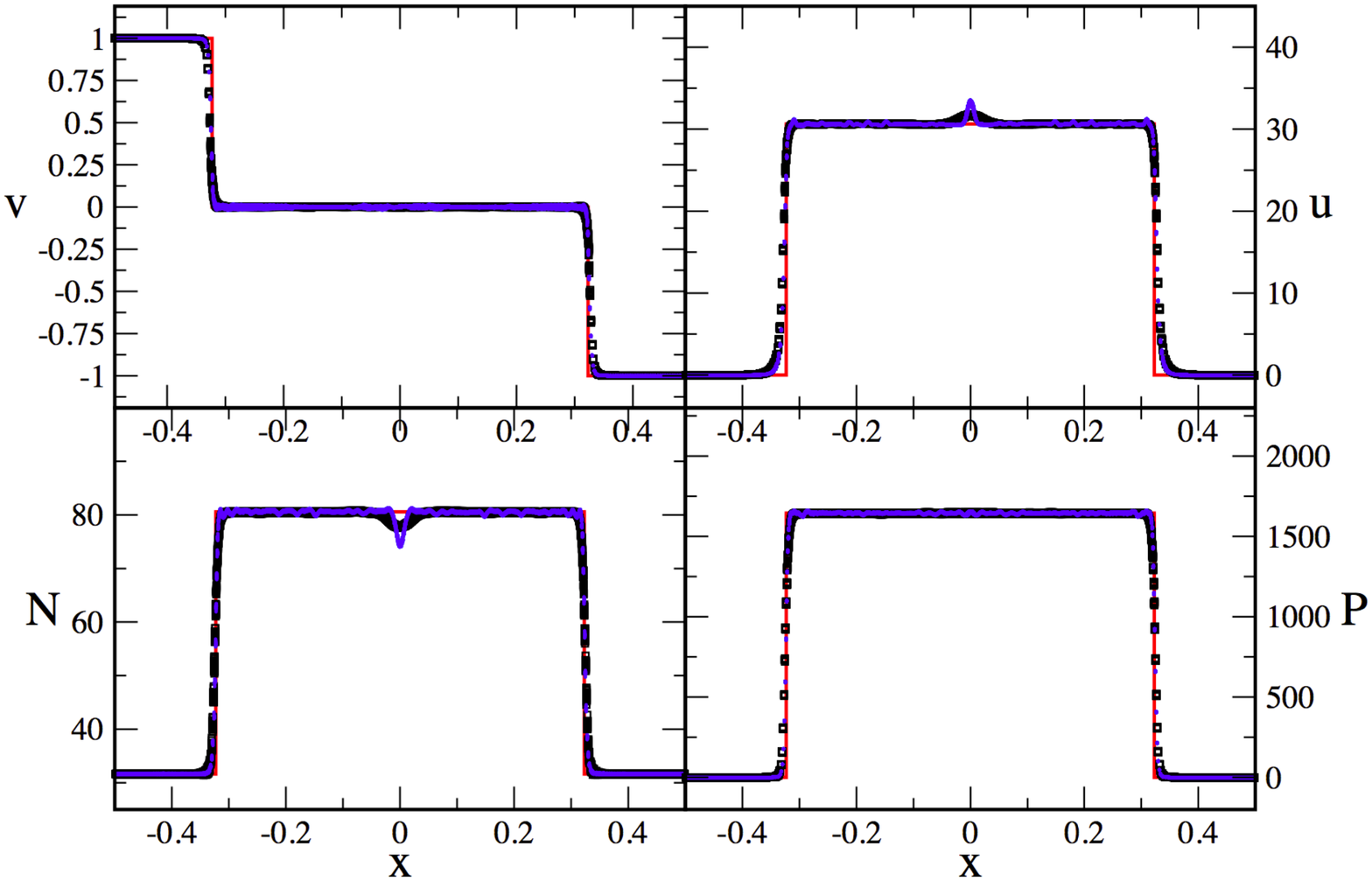} 
   \caption{Two-dimensional, relativistic planar shock reflection test where two gas streams collide with $v_x= \pm 0.9995$ c. 
                 The exact solution is shown by the solid, red line the SPH solution ($\mathcal{F}_1$ 
                 formulation, $t=0.5$) is shown as blue circles for our standard parameter set and as 
                 black squares for the same parameters but a dissipation floor $K_{\rm min}= 1$, see 
                 Eq.~(\ref{eq:diss_evol}).}
   \label{fig:RPSR_v0.9995}
\end{figure*}

As a second problem we show the result of another standard relativistic shock benchmark test
where two gas streams collide,
see e.g. \citep{hawley84a,eulderink95,falle96,aloy99b}. We use  a polytropic exponent of $\Gamma=5/3$
and the left and right state are given by $[n,v_x,v_y,u]_{\rm L}= [1.0,0.9,0.0,2.29 \times 10^{-5}]$ 
and $[n,v_x,v_y,u]_{\rm R}= [1.0,-0.9,0.0,2.29 \times 10^{-5}]$, where the incoming velocities
correspond to Lorentz factors of $\gamma= 2.29$. The result from a simulation with 160K 
particles initially placed on a hexagonal lattice between $[-2.0,2.0] \times [-0.05,0.05]$ at $t=0.875$
is shown in Fig.~\ref{fig:RPSR_v0.9}. Overall, there is excellent agreement with the exact solution (red line),
only at the center an ``overheating'' or ``wall heating'' phenomenon occurs in the density and internal
energy. This is a well-known phenomenon \citep{norman86,noh87,eulderink95,mignone05} that plagues a 
large number (if not all) shock-capturing schemes (including, e.g., the Roe, HLLE and HLLC Riemann solvers).\\
%-----------------------------------------------------------------------
%-----------------------------------------------------------------------
As a more extreme version of this test, we use initial velocities of 0.9995c, i.e.
$[n,v_x,v_y,u]_{\rm L}= [1.0,0.9995,0.0,2.29 \times 10^{-5}]$ and 
$[n,v_x,v_y,u]_{\rm R}= [1.0,-0.9995,0.0,2.29 \times 10^{-5}]$. Here the incoming
velocities correspond to Lorentz factors of $\gamma= 31.6$. We use the same
setup as above, but only 80K particles in the same computational domain.
Also in this extreme test the solution is robustly and accurately captured, see 
Fig.~\ref{fig:RPSR_v0.9995}. The solution with the $\mathcal{F}_1$ formulation
 and our standard parameter set is shown (at $t=0.5$) as blue circles and agrees 
very well with the exact solution (red line). Only at the origin a wall heating phenomenon
occurs in the internal energy and the density. As an experiment, we repeat the same
test with exactly the same setup and parameters, but now we use a dissipation floor 
$K_{\rm min}= 1$, see Eq.~(\ref{eq:diss_evol}). The continued dissipation reduces 
the amount of wall heating. This
is consistent with earlier studies, e.g. \cite{noh87,rosswog07c}, that find that wall 
heating is reduced by applying artificial conductivity. The other formulations perform
similar well, with differences consistent with those seen in Riemann problem I (see 
Fig.~\ref{fig:Sod_F1_to_F4}): $\mathcal{F}_2$ and $\mathcal{F}_3$ show slightly larger
oscillations in the shocked region, $\mathcal{F}_4$ leads to rounder edges, but also,
like the black square solution in Fig.~\ref{fig:RPSR_v0.9995}, to a slightly quieter shocked
region and reduced wall heating due to the larger dissipation.
%-----------------------------------------------------------------------
%-----------------------------------------------------------------------

\subsubsection{Riemann problem III: Einfeldt-type rarefaction test}
\begin{figure*} 
   \centering
   \hspace*{-1.cm}\includegraphics[width=14cm,angle=0]{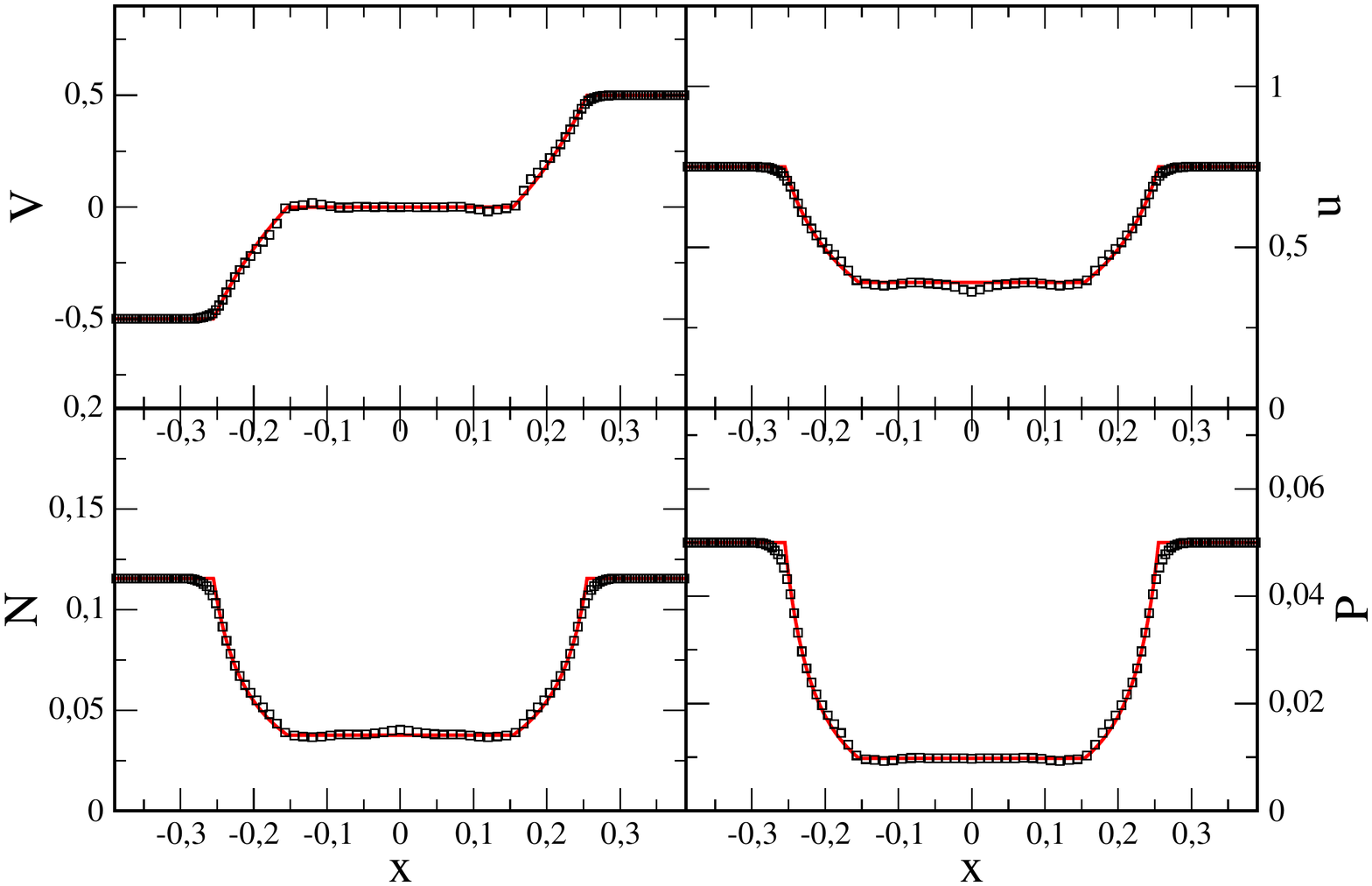} 
   \caption{Result of a 2D, Einfeldt-type rarefaction test. The SPH solution (at $t= 0.3$) is shown as black squares,
                 the exact result as solid, red line. Note that only 5 000 particles have been used for this 
                 two-dimensional test.}
   \label{fig:Einfeldt}
\end{figure*}

Here we explore the ability to properly capture rarefaction waves by means of an
Einfeldt-type \citep{einfeldt91} test. The initial conditions\footnote{Note that we
are specifying here local rest frame densities rather than computing frame densities.
This is simply for a straight-forward comparison with the analytical result obtained 
by the code ``riemann-vt.f'' from \cite{marti03}.} are given by 
$[n,v_x,v_y,P]_{\rm L}= [0.1,-0.5,0.0,0.05]$ and $[n,v_x,v_y,P]_{\rm R}= [0.1,0.5,0.0,0.05]$,
so that two rarefaction waves ($\gamma \approx 1.15$) are launched in opposite directions. This test had
originally been designed to point out a failure mode of Riemann solvers that can return
negative densities or pressures in strong rarefaction waves. We setup this test with
only 5K particles, placed on a hexagonal lattice between $[-0.45,0.45] \times [-0.05,0.05]$.
Note that despite the very low particle number, the results of the numerical simulation 
($\mathcal{F}_1$-formulation; squares; $t= 0.3$) agree very well with the exact solution (red, solid line),
see Fig.~\ref{fig:Einfeldt}.
%-----------------------------------------------------------------------
%-----------------------------------------------------------------------

\subsection{Fluid instabilities  (N)}
\label{sec:instabilities}

\subsubsection{Kelvin-Helmholtz instabilities}
\label{sec:KH}
\begin{figure*} 
   \centerline{\includegraphics[width=11cm,angle=0]{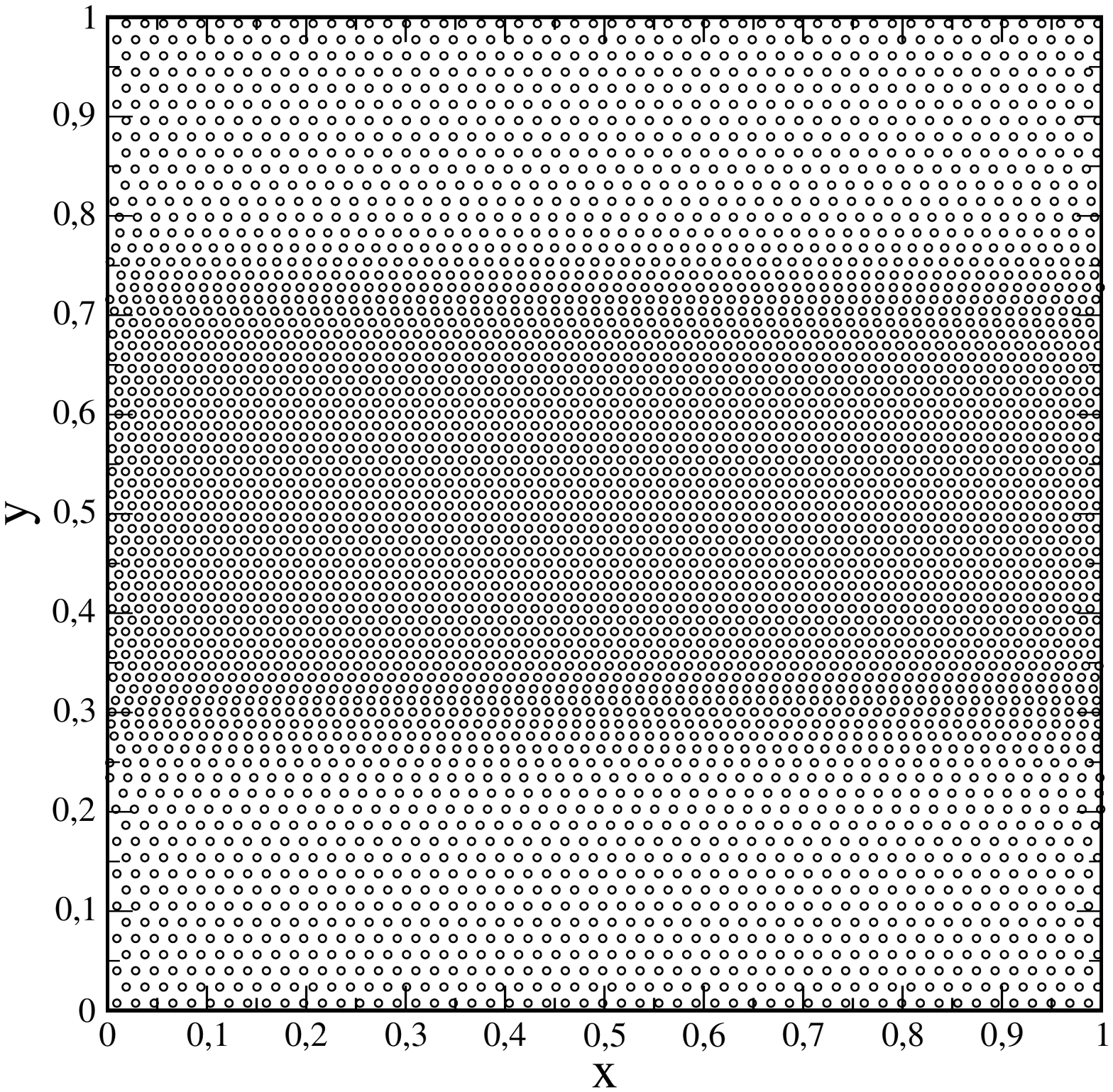} \hspace*{-1.5cm}
                      \includegraphics[width=11cm,angle=0]{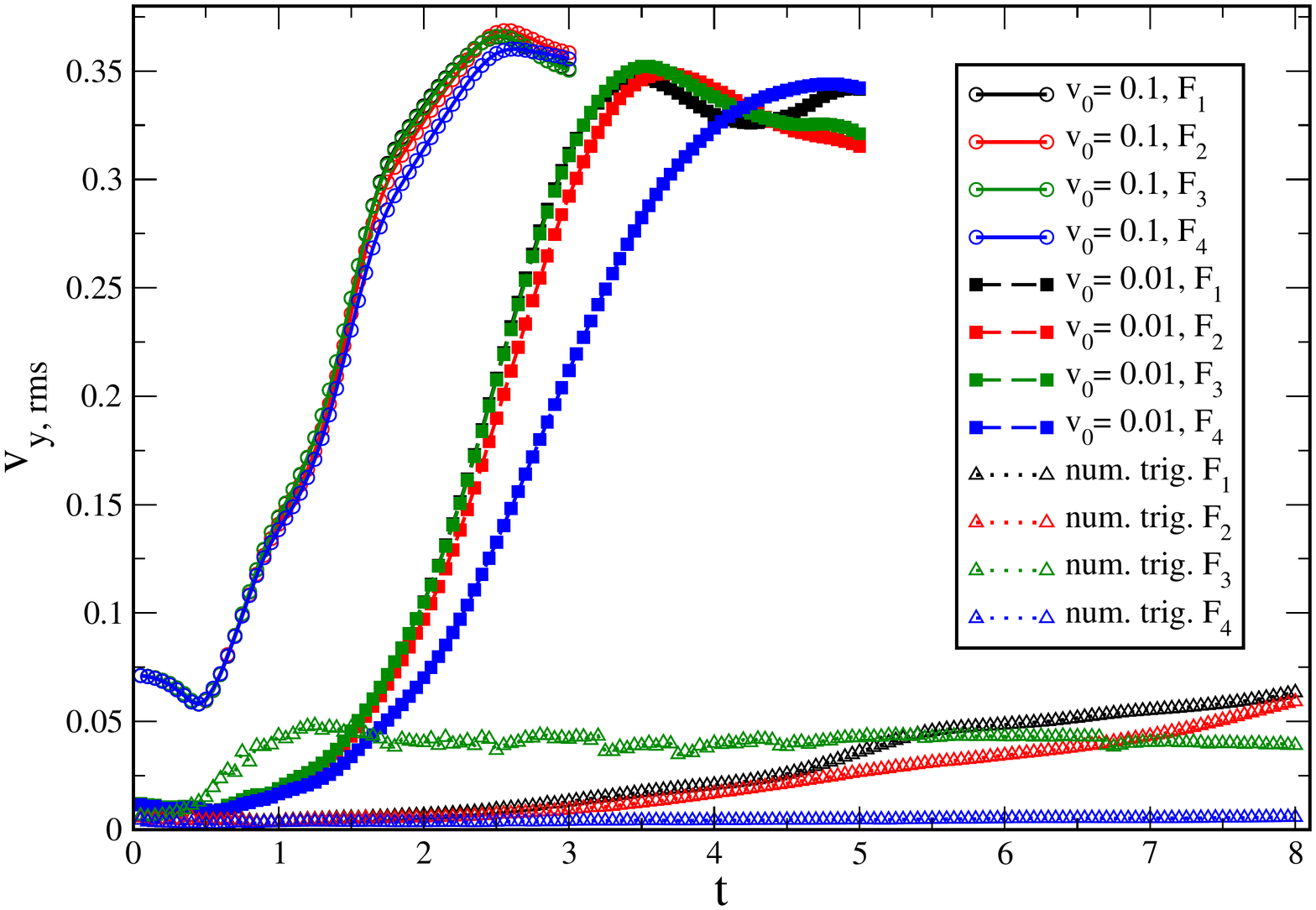}}
   \caption{Left panel: example of a ``quasi-close packed'' particle distribution (8K particles) for a Kelvin-Helmholtz
                 experiment that produces for equal mass particles a double-step density distribution.
                 Right panel: growth of the rms y-velocity component as a function of time in the different experiments. 
                  The solid lines with circles refer to the strongly triggered cases ($v_{y,0}=0.1$; 
                  up to $t=3$), dashed lines with squares to the weakly triggered experiments 
                  ($v_{y,0}=0.01$; up to $t=5$) and dotted lines with triangles to the numerically triggered
                  experiments (up to $t=8$).}
   \label{fig:initial_stripes_KH}
\end{figure*}

\begin{figure*} 
   \centerline{
  \includegraphics[width=12cm,angle=-90]{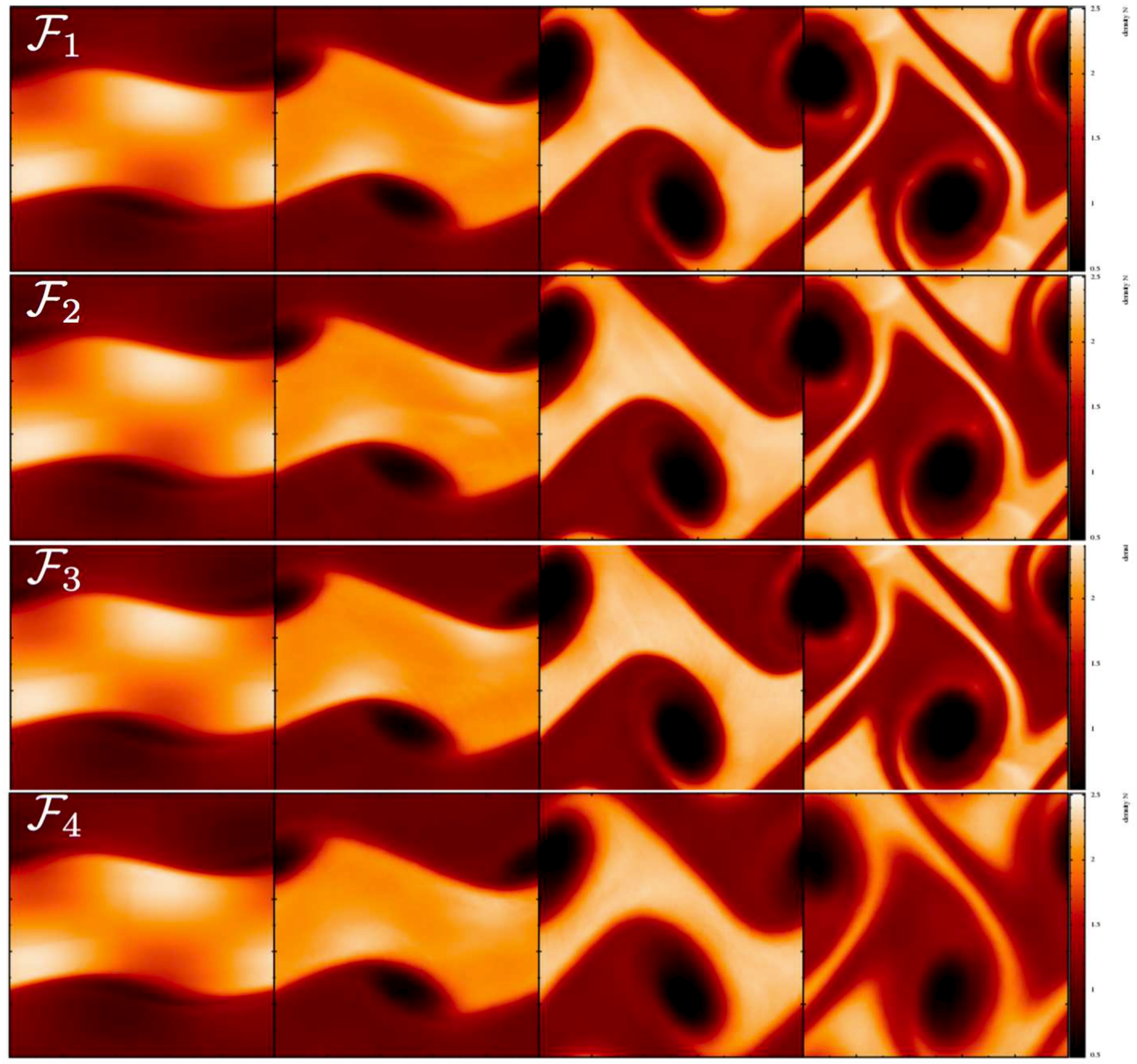}}
  \vspace*{-0cm}
   \caption{Comparison of the performance of the different formulation (with 50K particles; color-coded is the density $N$) 
                  in a Kelvin-Helmholtz test triggered by a moderate initial velocity perturbation of $v_{y,0}= 0.1$.
                  Each row corresponds to the result of one formulation (top to down: $\mathcal{F}_1$ 
                  to $\mathcal{F}_4$), each of the panels in one row refers to t= 0.5, 1.0, 2.0 and 3.0. 
                  The $\mathcal{F}_1$ formulation performs best, but the differences between the different
                  	formulations are overall only moderate.}
   \label{fig:KH_comparison_v0.1}
\end{figure*}

\begin{figure*}
   \centerline{
  \includegraphics[width=12cm,angle=-90]{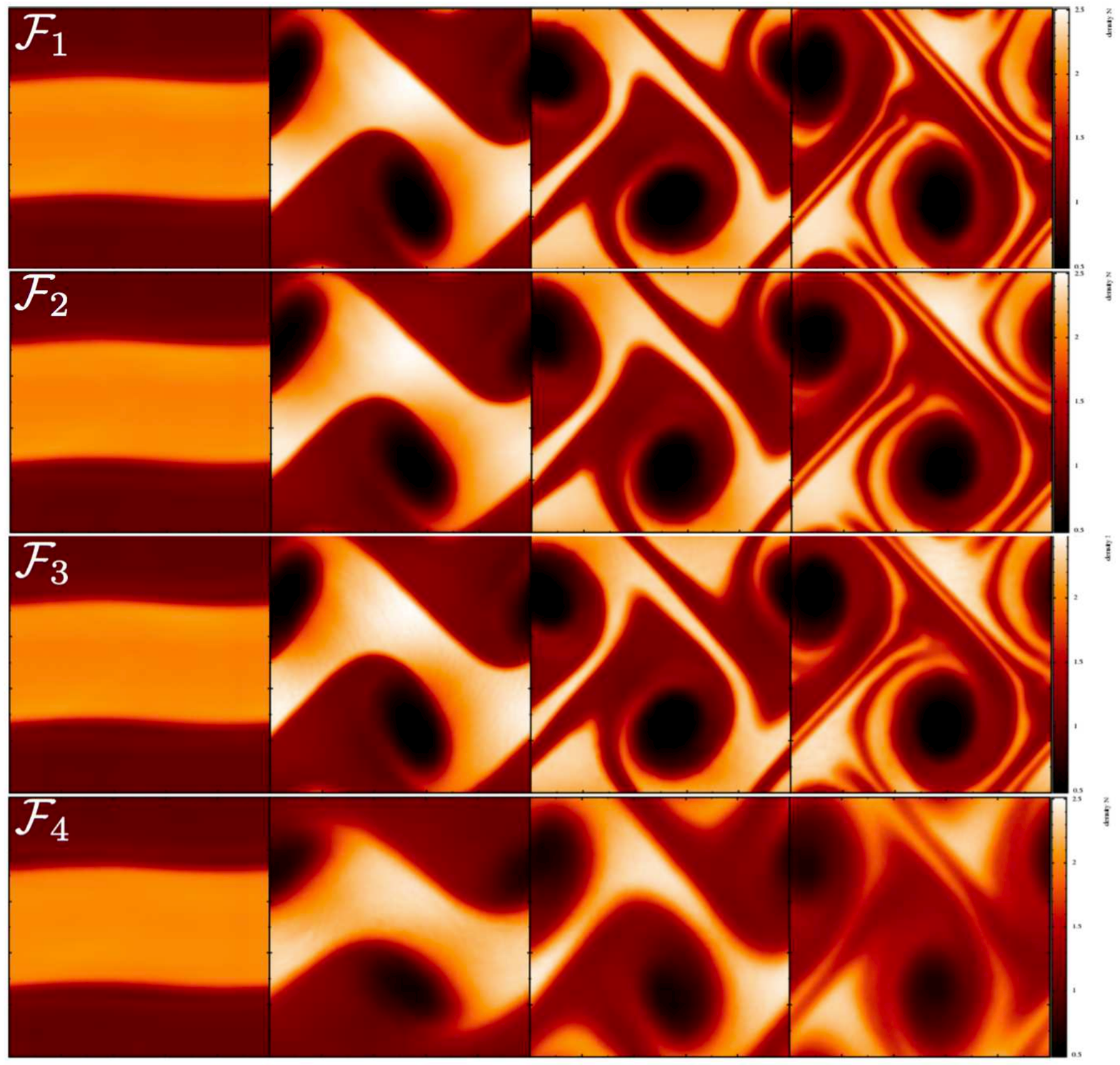}}
  \vspace*{-0cm}
   \caption{Comparison of the performance of the different formulation (with 50K particles; color-coded is the density $N$) 
                  in a Kelvin-Helmholtz test triggered
                  by a small initial velocity perturbation amplitude of only $v_{y,0}= 0.01$.
                  Each row corresponds to the result of one formulation (top to down: $\mathcal{F}_1$ 
                  to $\mathcal{F}_4$), each of the panels in one row refers to t= 1.0, 3.0, 4.0 and 5.0.  All cases show a
                  healthy growth of the instability despite the only very small perturbation.}
   \label{fig:KH_comparison_v0.01}
\end{figure*}

\begin{figure*} 
   \centerline{
   \includegraphics[width=29cm,angle=0]{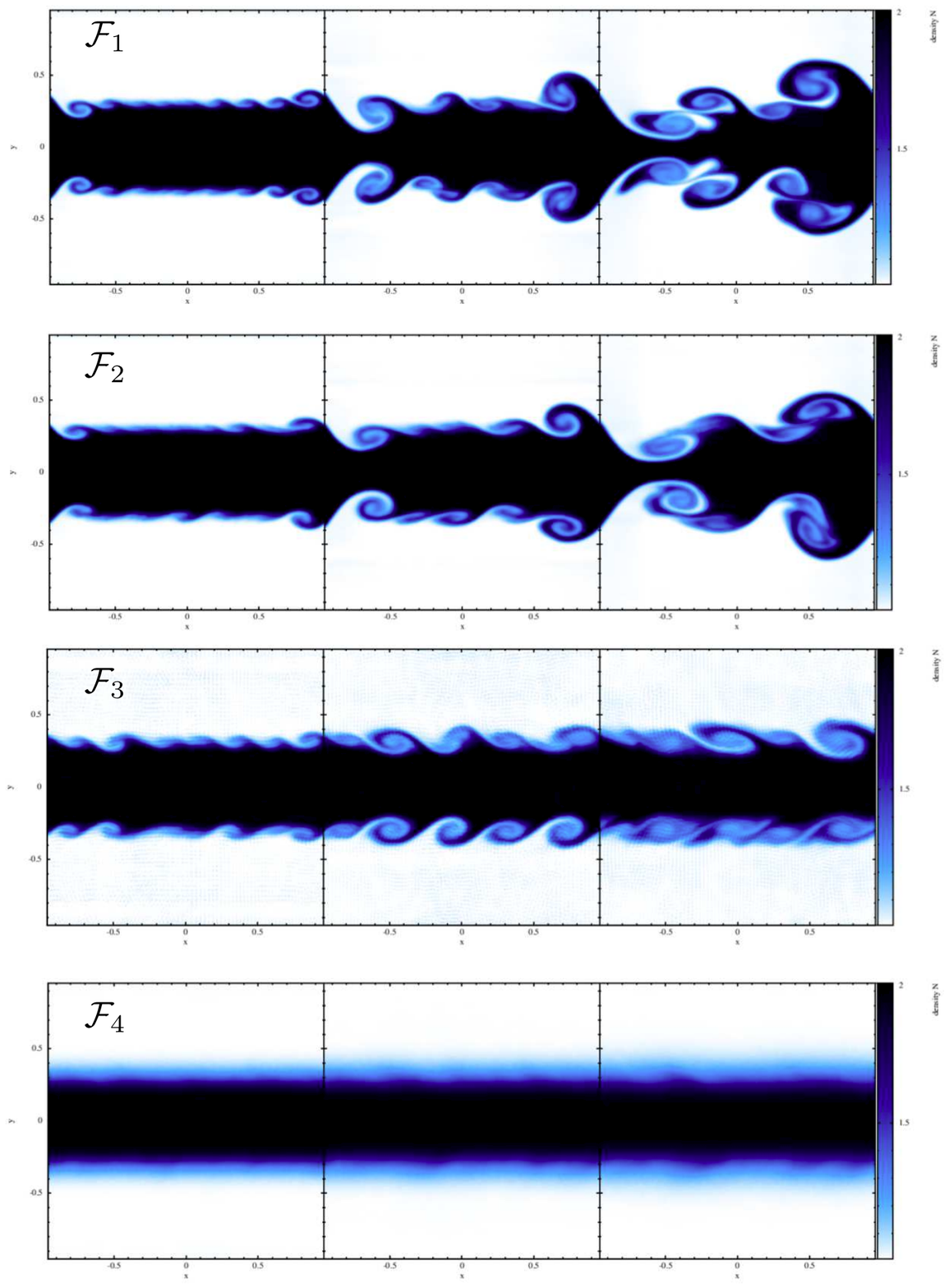}}
   \caption{Numerically triggered Kelvin-Helmholtz instability (200K particles; color-coded is the density $N$). 
                 The high-density band ($N=2$)
                 initially moves at -0.2c to the left, while the low-density bands ($N=1$) move at 0.2c 
                 to the right. No particular mode is excited, the instability grows from small numerical perturbations
                 as the particles at the interface pass along each other. Snapshots are shown at $t=$ 4.0, 6.0 and
                 8.0 (left to right) for the SPH formulations $\mathcal{F}_1$ (top) to $\mathcal{F}_4$(bottom). 
                 Note that the grid-like structure for the $\mathcal{F}_3$ result is not an artifact of the visualization,
                 here the particles vibrate around their lattice positions.}
   \label{fig:KH_num_trigger}
\end{figure*}

\begin{figure*} 
   \centerline{
   \includegraphics[width=14cm,angle=-90]{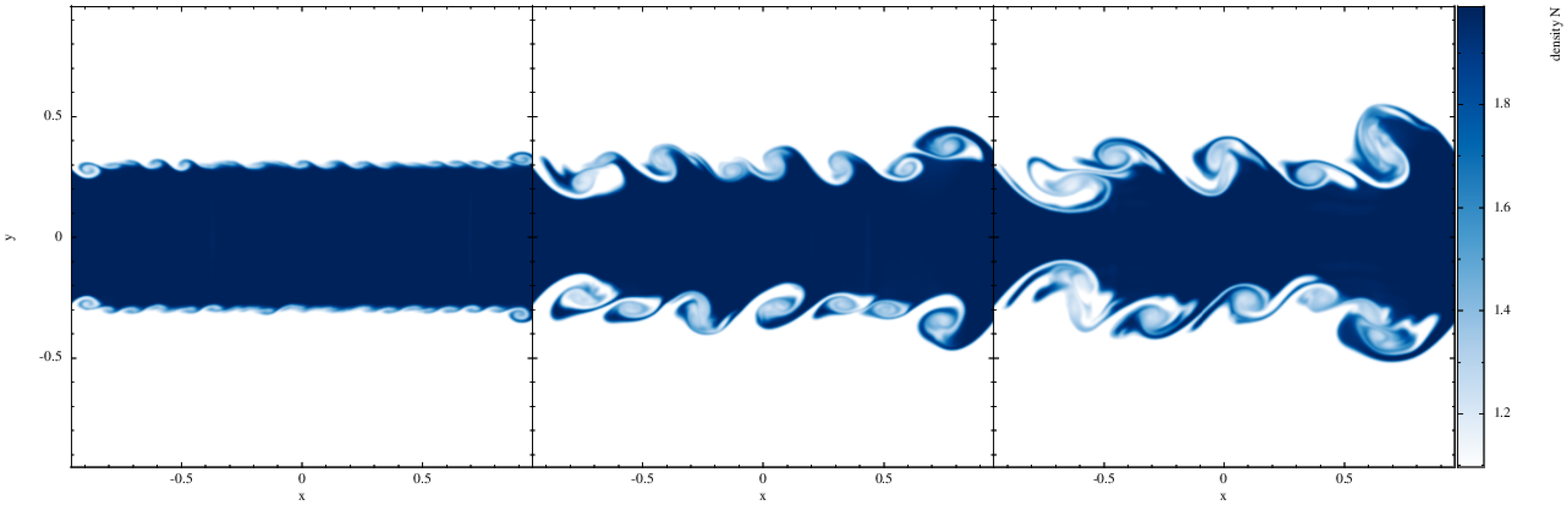}}
   \vspace*{-4cm}
   \caption{Growth of a Kelvin-Helmholtz instability ($\mathcal{F}_1$-formulation) that has not been triggered explicitly (1000K particles). 
                 The density $N$ is shown at $t= 2.0, 4.0$ and 5.25.}
   \label{fig:KH_num_trigg_1000K}
\end{figure*}

To scrutinize the ability of the different formulations to capture Kelvin-Helmholtz instabilities we set
up an experiment similar to,  for example, \cite{robertson10} and \cite{garcia_senz12}. We place 50K equal mass
particles in the domain $[0,1] \times [0,1]$ so that the horizontal stripe between $y^{\rm h}_1=  0.25$
and $y^{\rm h}_2=  0.75$ has density $N=2$, while the upper and lower stripes have a density $N=1$. 
Periodic boundary conditions are applied everywhere, the polytropic exponent is $\Gamma= 5/3$ and 
the pressure is $P= P_0= 2.5$. The middle, high-density stripe moves with $v_x=0.5$ to the right, while 
the other stripes with $v_x=-0.5$ to the left. Since we favor equal-mass particles, we setup a 
``quasi-close packed'' particle distribution where the effective sphere radius varies with the y-coordinate. 
Similar to earlier work \citep{rosswog10b}, we make use of Fermi-functions to create a resolvable 
transitions. For the ``double-step'' of this test where a function $A$ changes at $y^{\rm h}_1$ smoothly
from a value $A_1$ to a value $A_2$ and at $y^{\rm h}_2$ from a value $A_2$ to a value $A_3$ we use
\be
A(y)= F(A_1,A_2,y^{\rm h}_1,\Delta y,y) + F(A_2,A_3,y^{\rm h}_2,\Delta y,y) - A_2,
\label{eq:double_step}
\ee
where
\be
 F(A,B,y_{t},\Delta y,y)= \frac{A - B}{\exp((y-y_t)/\Delta y) + 1} + B.
 \ee
For the characteristic transition width $\Delta y$ we choose the sum of the sphere radii $r_s^h$/$r_s^l$
in the high/low density region,  which we consider a natural choice. In this setup, equal mass particles reproduce 
the desired density pattern via their spatial distribution. An example of such a particle distribution (with 8K 
particles) is shown in Fig.~\ref{fig:initial_stripes_KH}, left panel.
%-----------------------------------------------------------------------
%-----------------------------------------------------------------------
Like in \cite{garcia_senz12} we perturb  the interface with a velocity component in y-direction
\be
 v_y(x)= v_{y,0} \sin(2 \pi x),
\ee
for $v_x(y)$ a double-step transition according to Eq.~(\ref{eq:double_step}) is used.
Note that  the particles in the transition region are not necessarily in an equilibrium configuration 
and subsequent particle re-configurations may trigger additional Kelvin-Helmholtz modes apart from
the desired one (at least for the high-accuracy formulation $\mathcal{F}_1$).  Therefore, we perform a 
``relaxation on the fly'', i.e. for $t < 0.5$ we keep the dissipation parameter $K_a$ on a value of unity 
and only subsequently we let it evolve freely. This procedure works very well and only triggers the desired 
mode. Following \cite{garcia_senz12}, we perform this test in two flavors: once with a substantial initial 
perturbation of $v_{y,0}=0.1$ and the second time we only use $v_{y,0}=0.01$.\\
The results for $v_{y,0}=0.1$ for the different formulations $\mathcal{F}_1$ (top row) to $\mathcal{F}_4$ (bottom row)
are shown in Fig.~\ref{fig:KH_comparison_v0.1}, each time at $t=$ 0.5, 1.0, 2.0 and 3.0. 
Overall, there is good agreement between the different SPH formulations, all show a healthy 
Kelvin-Helmholtz growth. Only $\mathcal{F}_4$ is, as expected, excessively diffusive.
In the $v_{y,0}=0.01$ case the instability grows slower, therefore we show in
Fig.~\ref{fig:KH_comparison_v0.01} snapshots at $t=$ 1.0, 3.0, 4.0 and 
5.0. Once more, all formulations, even the worst and most diffusive $\mathcal{F}_4$, 
are able to capture the instability. This is different from the findings of \cite{garcia_senz12}, their 
standard SPH formulation does not show a healthy growth, despite their slightly larger particle 
number (62.5K compared to our 50K). The difference may come from subtleties, for example, they 
use particles of different masses while we use equal-mass particles. Their particles are placed on a quadratic
lattice (which is not an equilibrium configuration, see Sec.~\ref{sec:noise}) while ours are on a hexagonal
lattice. Moreover, since we apply the dissipation to the numerical variables, see Sec.~\ref{sec:AV}, which 
contain also the specific energy, this introduces a small amount of conductivity, which might help
the instability to grow \citep{price08a}. Last but not least, we have found that the exact algorithm 
for the smoothing length update can introduce a fair amount of noise. In our algorithm, we have taken 
particular care to avoid noise and to have consistent values to a very high accuracy, see Sec.\ref{sec:dens_it}.\\
%-----------------------------------------------------------------------
%-----------------------------------------------------------------------
%-----------------------------------------------------------------------
%-----------------------------------------------------------------------
In a further Kelvin-Helmholtz test no particular mode is excited and the instability is seeded numerically.
To this end we place 200K particles in three stripes of cubic lattices in the computational 
domain [-1,1] $\times$ [-1,1], with high density strip of $N=2$ moving with $v= 0.2$ to the right, while
the low density strips with $N=1$ move with $v= -0.2$ to the left, the pressure is $P=10$ and the polytropic 
exponent $\Gamma= 5/3$. The results are shown in Fig.~\ref{fig:KH_num_trigger} at times $t=$ 4.0, 6.0 and 8.0.
In all cases apart from $\mathcal{F}_4$ a Kelvin-Helmholtz instability with the characteristic billows develops. 
In the very diffusive $\mathcal{F}_4$ case perturbations still grow at the end of the simulation, but at an
extremely slow pace. Note that in the case of the ``standard'' SPH volume element, $\mathcal{F}_3$, 
small vibrations in the particle distributions emerge which appear in the plot as an overlaid ``grid-like'' 
pattern (this is not an artifact of the visualization). Such vibrations have also been observed in other tests 
with $\mathcal{F}_3$, see for example Fig.~\ref{fig:Sod_F1_to_F4}. Comparing $\mathcal{F}_1$ and $\mathcal{F}_2$ one sees that the more 
accurate integral-based gradients resolve sharper features in comparison to the kernel based variant. 
%-----------------------------------------------------------------------
%-----------------------------------------------------------------------
We measure the instability growth via the rms y-velocity component
\be
v_{\rm y, rms}= \sqrt{\frac{1}{N} \sum_b^N v_{y,b}^2}
\ee
as a function of time, see Fig.~\ref{fig:initial_stripes_KH}, right panel, for all three sets of experiments
($v_{y,0}=0.1$ with solid lines with circles; $v_{y,0}=0.01$ with dashed lines with squares;
numerically triggered with dots and triangle symbols). Generally, the $\mathcal{F}_1$-formulation 
grows fastest, although in the triggered experiments the differences to the $\mathcal{F}_3$ is 
small. The $\mathcal{F}_2$ formulation always grows slower than $\mathcal{F}_1$. 
Although the choice of the volume element makes a difference (comparison $\mathcal{F}_1$ and 
$\mathcal{F}_3$) the major effect seems to come from the more accurate, integral-based 
gradients, at least for our setup of the test problem. This might be different for alternative setups. 
In the numerically triggered experiments, $\mathcal{F}_3$ very early develops a moderate value 
for the y-velocity component, which is related to the short-wavelength noise mentioned above,
visible in the third row of Fig.~\ref{fig:KH_num_trigger} as a grid-like pattern underlying the 
overall density distribution. Such particle noise could possibly hamper the growth of weakly triggered
instabilities \cite{springel10a}.
Clearly, the formulation with standard methods, $\mathcal{F}_4$, generally shows
the slowest growth or --if not triggered explicitly-- hardly grows at all.\\
As a last example, we follow the evolution of  1000K particles with the $\mathcal{F}_1$ formulation, again 
without explicitly triggering a particular mode, we simply wait until small numerical perturbations grow 
into healthy Kelvin-Helmholtz billows, see Fig.~\ref{fig:KH_num_trigg_1000K}.\\
%-----------------------------------------------------------------------
%-----------------------------------------------------------------------

\subsubsection{Rayleigh-Taylor instabilities}
\label{sec:RT}
\begin{figure*} 
   \centerline{
   \includegraphics[width=22cm,angle=0]{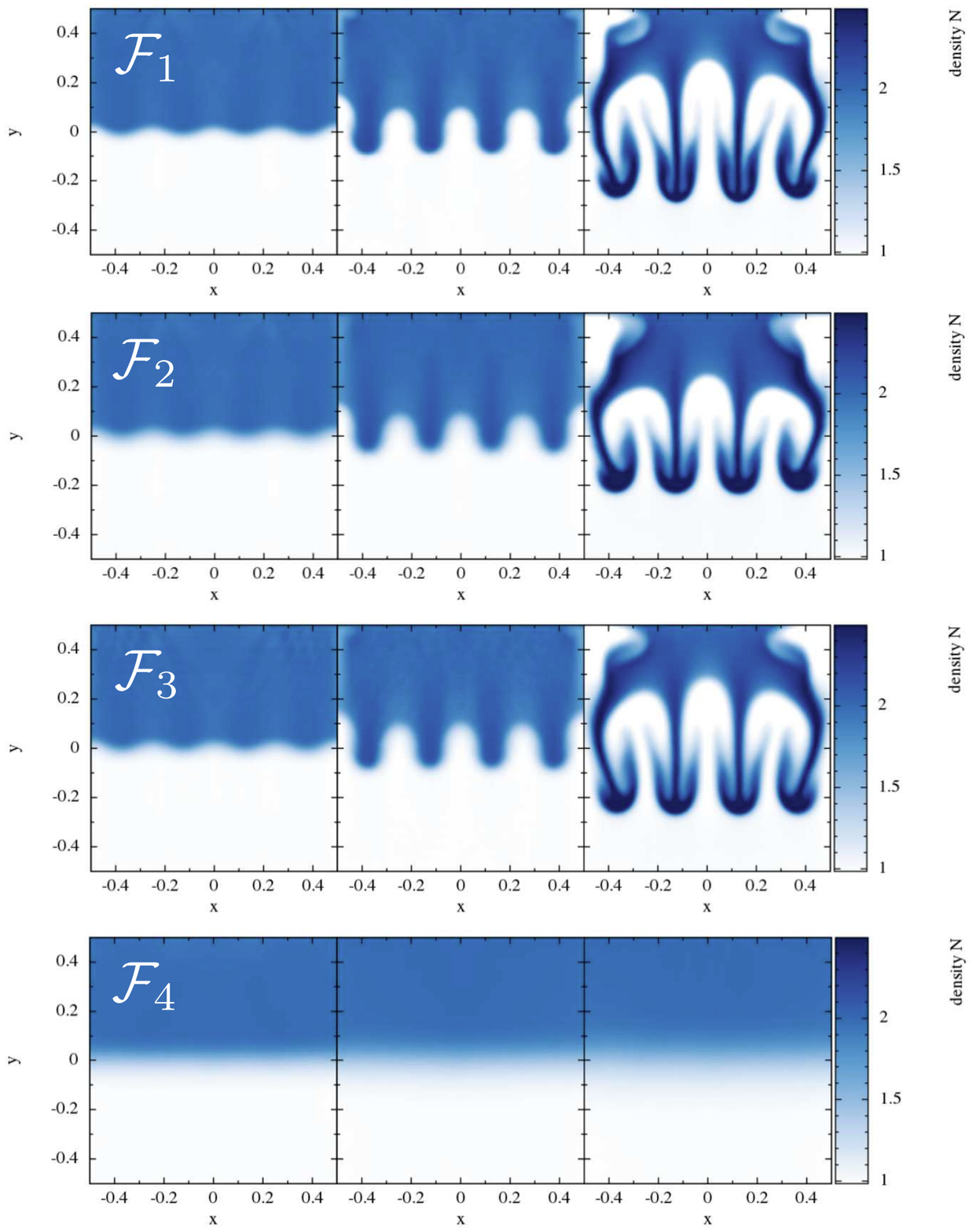}}
   \vspace*{-0.5cm}
   \caption{Growth of a weakly triggered Rayleigh-Taylor instability (50K particles) for the SPH 
                 formulations $\mathcal{F}_1$ to $\mathcal{F}_4$.
                 For each formulation the snapshots are taken at $t= 2.5, 5.0$ and 8.25.}
   \label{fig:RT_evol_50K}
\end{figure*}

The Rayleigh-Taylor (RT) instability is another classical test case. In its simplest form, it 
occurs when a denser fluid rests on top of a lighter one in a gravitational field. When perturbed, 
the denser fluid begins to sink down and the release of gravitational energy triggers a characteristic, 
``mushroom-like'' flow pattern, both for the lighter
fluid rising up and the heavier one sinking down. Generally,  SPH is thought to be seriously
challenged in dealing with RT instabilities, but as we will show below, the suggested measures 
provide a major improvement also in this case.\\
We consider two fluid layers with density $N_1$ and $N_2$ ($N_2 > N_1$), separated by a 
transition located at $y_t$ in an external gravitational field $\vec{g}= g \hat{e}_y$. Since the equilibrium
configuration is at rest and the initial velocity perturbation is tiny, we do not distinguish in this 
setup between local rest frame and computing frame densities. We model the
external gravitational field as an additional, constant acceleration term in the momentum equation
(either Eq.~(\ref{eq:gen_mom_SR}) or (\ref{eq:momentum_eq_no_diss_integral})). To have a well-defined
problem, we model the interface by a narrow, but resolvable transition with a width that tends to
zero as the resolution increases. Explicitly, we use a Fermi function
\be
N(y)= N_2 - \frac{\Delta N}{1 + \exp{\left(\frac{y - y_t}{\Delta y}\right)}},
\label{eq:dens_transition}
\ee  
where $\Delta N= N_2 -N_1$ and $\Delta y$ characterizes the transition width of the density step, 
similar to the Kelvin-Helmholtz case, see Sec.~\ref{sec:KH}. The hydrostatic equilibrium condition 
then yields the pressure distribution as a function of height as
\be
P(y)= g \; \Delta y \; \Delta N \; \ln \left[\frac{1 + \exp \left(-\frac{y-y_t}{\Delta y}\right)} {1 + \exp \left(\frac{y_t}{\Delta y}\right)}\right] + g N_2 y  + P(0)
\ee
To set up a particle distribution with equal baryon numbers (``masses'') for each SPH particle, we set up
a close-packed lattice in which the sphere radius $r_s$  varies with $y$, similar to the Kelvin-Helmholtz test above. 
Again, the transition between the two regions is mediated by varying  $r_s$ 
according to a Fermi function with transition width $\Delta y= r_s^{(1)} + r_s^{(2)}$.
We choose the following numbers: the particles are placed in the domain [-0.5,0.5] $\times$ [-1,1],
densities are $N_1= 1$, $N_2=2$, the polytropic exponent $\Gamma= 5/3$, the pressure $P_0$= 1 and 
the external acceleration $g= -0.5$. These numbers are oriented at the RT test in \cite{garcia_senz12}, but our setup
differs from theirs with respect to a) our transition region is smooth, b) becomes infinitely sharp with 
particle number going to infinity and c) in our case the density information is encoded in the particle 
distribution (due to our equal baryon number particles) rather than in the particle baryon numbers as
in the work of \cite{garcia_senz12}. Finally, to
trigger the instability, we perturb the interface region slightly (like in \cite{abel11} and \cite{garcia_senz12}) 
by
\be
\delta v_y= \left\{\begin{array}{ll} \frac{v_0}{4} \left[1 + \cos\{8 \pi (x + \frac{1}{4}) \} \right] \left[    1 + \cos \{5 \pi (y - \frac{1}{2}\} \right]  & |y| < 0.25 \\
         0 & {\rm else}\end{array}\right.,
\ee
with a very low perturbation amplitude of $v_0= 0.01$. Periodic boundary conditions are applied
at $x= \pm 0.5$, in the y-direction all derivatives are enforced to vanish for $|y| > 0.8$.\\
The hydrodynamic evolution for $\mathcal{F}_1$ to $\mathcal{F}_4$ is shown (each time 
at $t= 2.5, 5.0$ and 8.25) in Fig.~\ref{fig:RT_evol_50K}. Consistent with the findings of  other tests,
$\mathcal{F}_1$ develops the finest density structures and the instability grows fastest. The results
for $\mathcal{F}_2$ and $\mathcal{F}_3$ are not very different, though, they also show a healthy
growth of RT mushrooms. Like in the untriggered KH case, the ``standard method approach'' 
$\mathcal{F}_4$ fails completely and does not allow the weakly triggered instability to grow.\\
In modern adaptive mesh refinement simulations, e.g. \cite{keppens12}, secondary Kelvin-Helmholtz
events occur on the falling spikes/pillars. The resolution in our  tests is not large enough to resolve 
them. This is mainly due to the large smoothing length required for the Wendland kernel and due to
the use of equal mass/baryon number particles which lead to particularly low resolution in the low 
density  regions.
%-----------------------------------------------------------------------
%-----------------------------------------------------------------------

\subsection{Combined tests}
\label{sec:combined}
We also perform a number of more complex tests for which no exact solutions are available
and the results need to be compared to other methods documented in the literature. 
These tests illustrate the robustness and flexibility of the new approaches and demonstrate
that a complex interplay between shocks, rarefaction and fluid instabilities can be handled 
with ease.

\subsubsection{Relativistic blast-bubble interaction I (SR)}
\label{sec:blast_bubble_I}
\begin{figure*} 
   \includegraphics[width=18cm,angle=0]{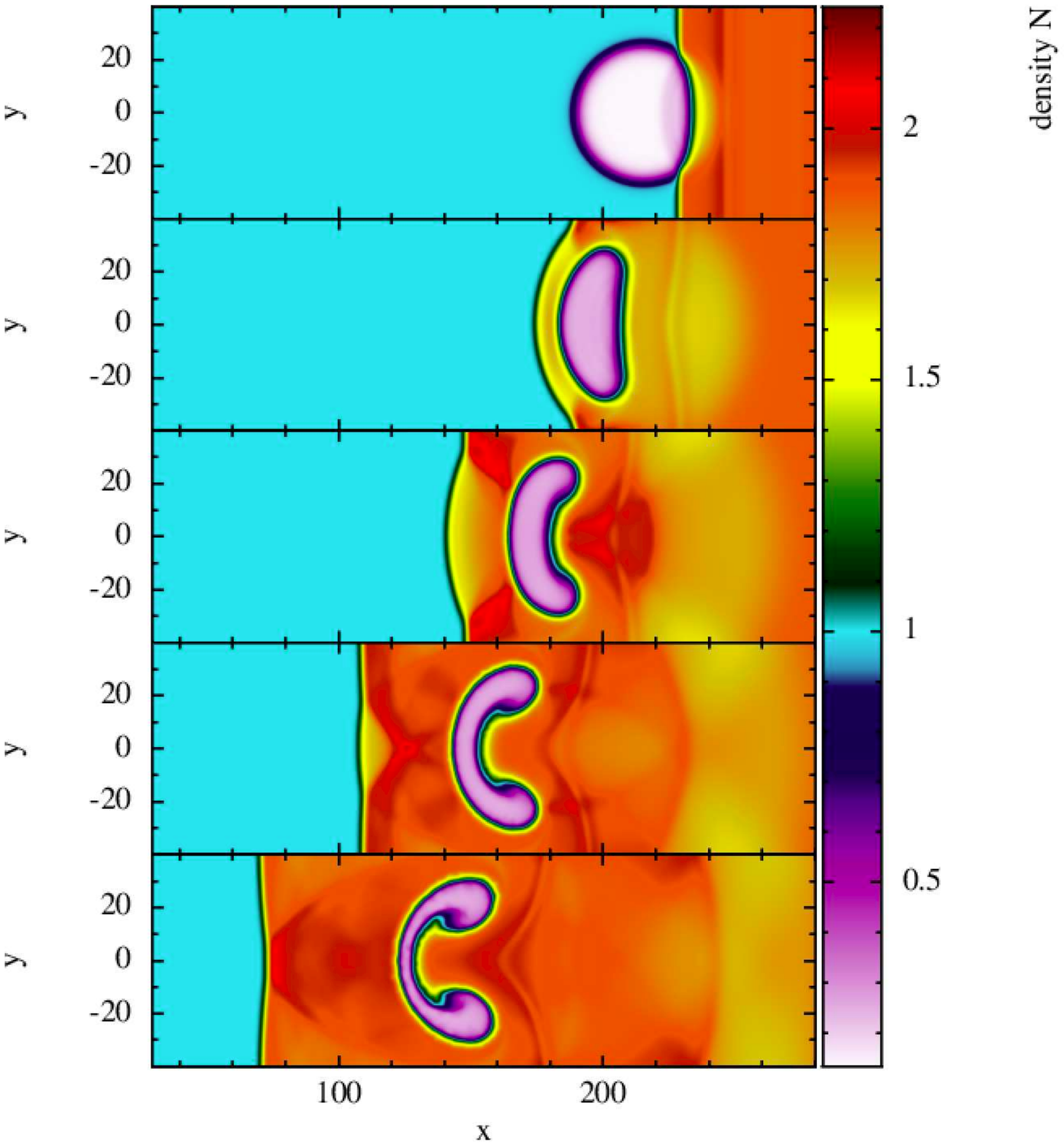}
   \vspace*{-3cm}
   \caption{Blast-bubble interaction I: a blast wave is impacting from the right onto a low-density bubble
                 in pressure equilibrium. This test has been performed with the $\mathcal{F}_1$ formulation 
                 and 100K particles. Snapshots are shown at $t= 90, 180, 270, 360$ and 450.}
   \label{fig:blast_bubble_I}
\end{figure*}

\begin{figure*} 
   \includegraphics[width=16cm,angle=0]{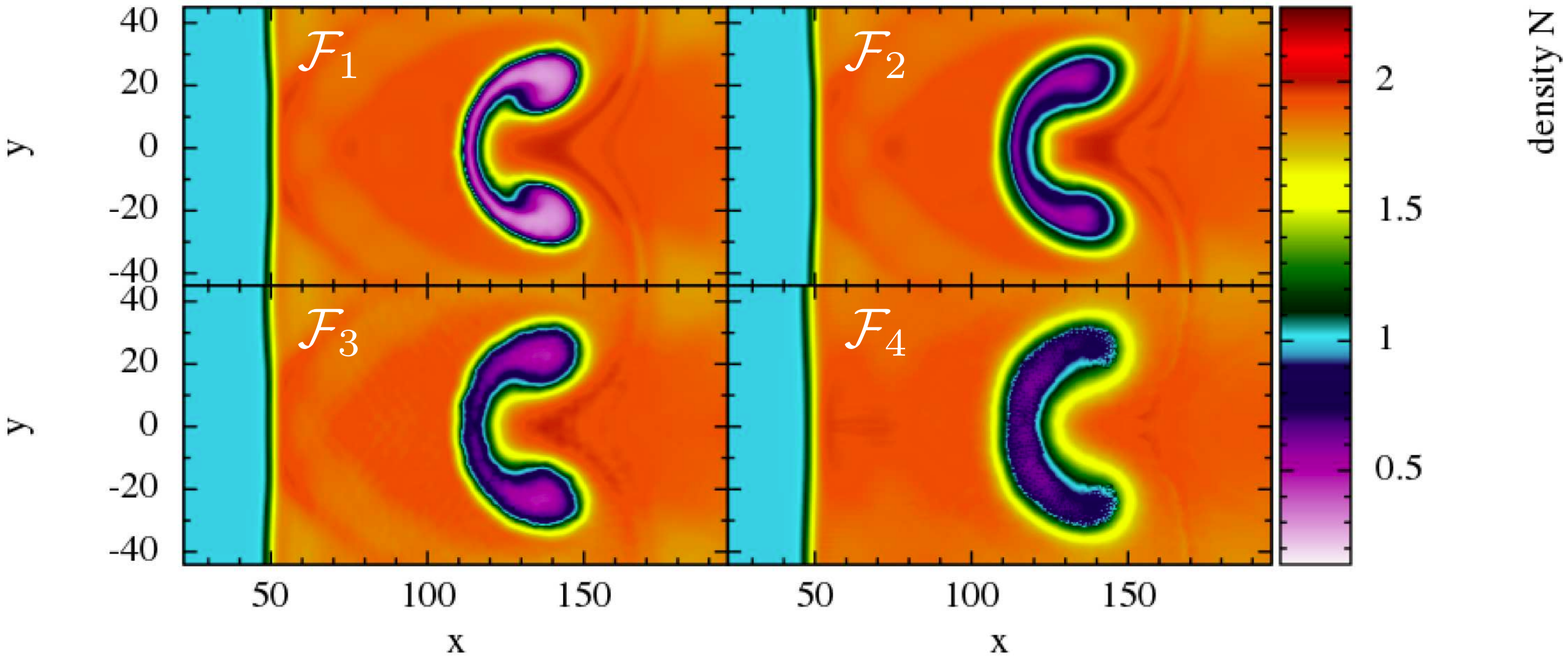}
   \vspace*{-4cm}
   \caption{Comparison of different approaches for blast-bubble interaction I for low resolution (50K particles; at $t=500$). 
                  Once more, the $\mathcal{F}_1$ formulation clearly performs best, both gradient accuracy (compare $\mathcal{F}_1$
                  and $\mathcal{F}_2$) and the volume element (upper row vs lower row) impact on the result.}
   \label{fig:compar_blast_bubble_I_v2}
\end{figure*}

The interaction of a relativistic blast wave with a spherical bubble that is in pressure equilibrium
with its surroundings is a challenging test that probes the ability of capturing shocks, allowing for instabilities and to 
adapt geometrically. Here we set up the initial conditions exactly as in \cite{he12}, their example 6.8,
who performed this test with their 2D, special-relativistic adaptive mesh refinement code. This setup
results in mildly relativistic velocities ($v\approx 0.35$, $\gamma \approx 1.07$).
To this end we place $10^5$ particles on a hexagonal lattice in the domain [0, 325] $\times$ [-45, 45] with
reflective boundaries at $y= \pm 45$ and the states at the left and right end of the domain are frozen.
The left-moving shock is initially located at $x= 265$ and the left/right state are given by
\be
(N,P,v_x,v_y)=  \left\{
  \begin{array}{ l l}
     (1, 0.05, 0, 0) \hspace*{2.6cm} {\rm for \quad } x \le 265\\
     (1.865225, 0.15, -0.196781, 0)  \hspace*{0.3cm} {\rm for \quad } x > 265.\\
   \end{array} \right.
\ee
A polytropic equation of state is used with an adiabatic exponent $\Gamma=5/3$. 
In front of the shock wave  we place a spherically symmetric bubble
with radius $R_{\rm B}= 25$, centered around $(215,0)$. The low-density bubble is in pressure 
equilibrium with its surroundings and in the state given by $[N,P,v_x,v_y]_{\rm B}= [ 0.1358, 0.05, 0, 0]$.\\
In Fig.~\ref{fig:blast_bubble_I} we show the density at $t=$ 90, 180, 270, 360, 450, as in 
the original paper, for the $\mathcal{F}_1$ formulation. The 
results agree excellently with those found in the relativistic adaptive mesh refinement approach of
He \& Tang (2012), see their Figs. 13 and 14, including the shape and location of the various travelling waves.\\
%-----------------------------------------------------------------------
%-----------------------------------------------------------------------
To explore the impact of the various new ingredients we repeated this test for 
formulations $\mathcal{F}_1$ to $\mathcal{F}_4$ with only 50K particles.
A comparison of the results at $t= 500$ is displayed in Fig.~\ref{fig:compar_blast_bubble_I_v2}. 
$\mathcal{F}_1$ clearly performs best, $\mathcal{F}_4$  does not capture any of the ``curling in'' 
of the upper and lower lobe. Here the gradient accuracy seems crucial (compare $\mathcal{F}_1$ and
$\mathcal{F}_3$ with $\mathcal{F}_2$ and $\mathcal{F}_4$).
%-----------------------------------------------------------------------
%-----------------------------------------------------------------------

\subsubsection{Relativistic blast bubble interaction II (SR)}
\label{sec:blast_bubble_II}
\begin{figure*} 
   \vspace*{-1.5cm}
   \includegraphics[width=17cm,angle=0]{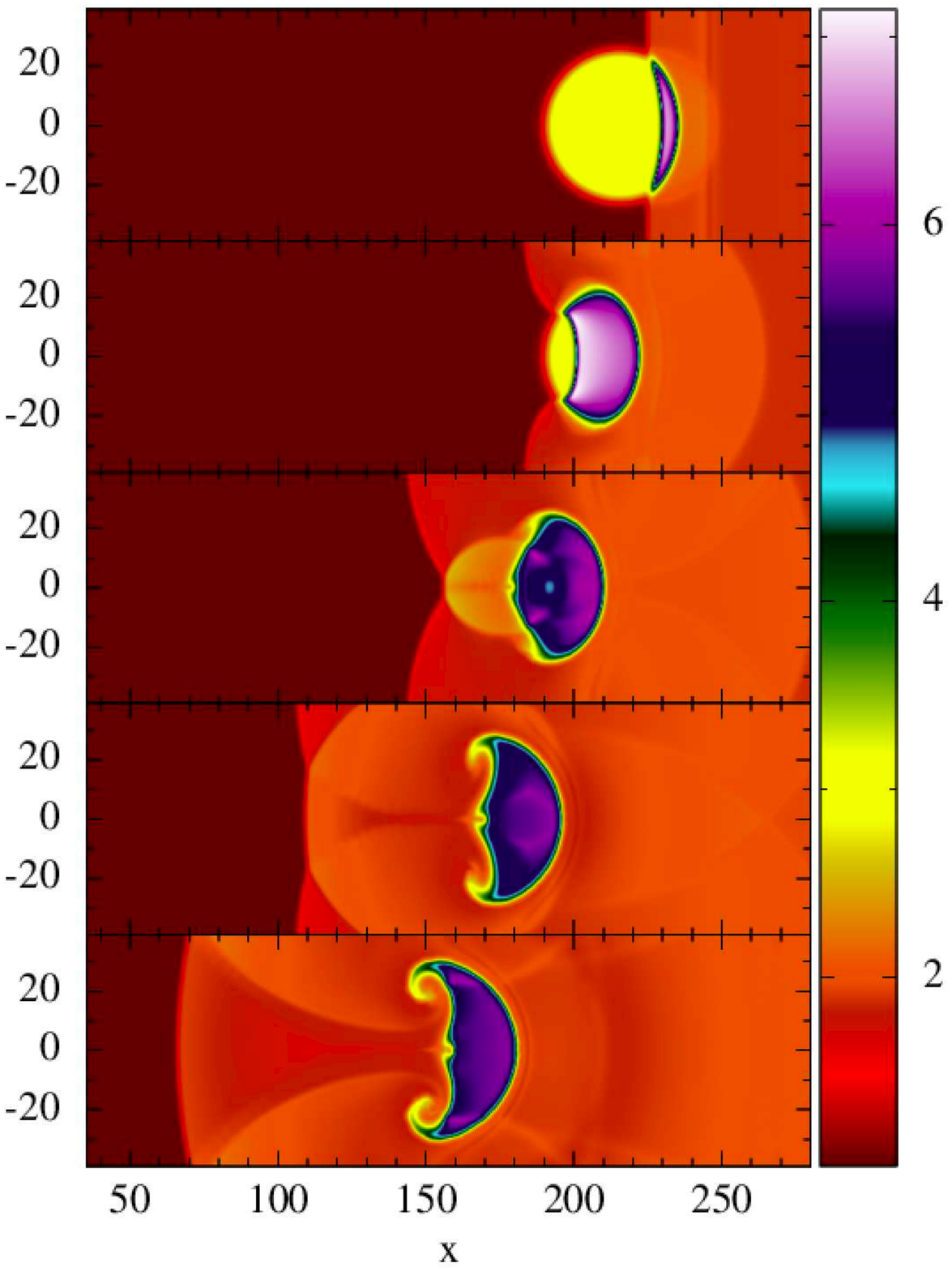}
   \vspace*{-2cm}
   \caption{Blast-bubble interaction II: a blast wave is impacting from the right onto a high-density bubble
                 in pressure equilibrium. This test has been performed with the $\mathcal{F}_1$ formulation 
                 and 200K particles. Snapshots are shown at $t= 100, 200, 300, 400$ and 500. Color-coded
                 is the computing frame baryon density $N$.}
   \label{fig:blast_bubble_II}
\end{figure*}

\begin{figure*} 
   \includegraphics[width=14cm,angle=0]{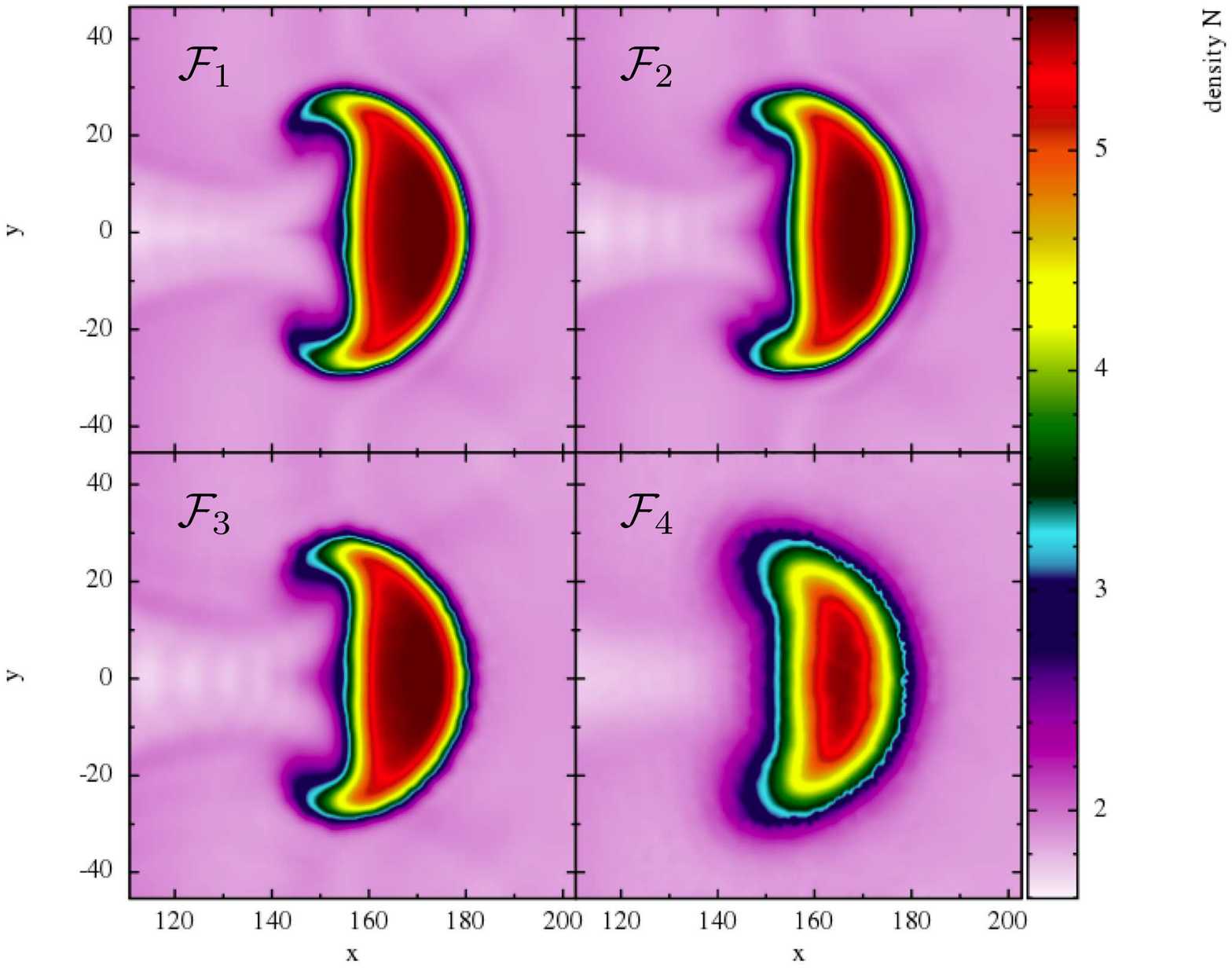}
   \vspace*{0cm}
   \caption{Comparison of different approaches $\mathcal{F}_1$ to $\mathcal{F}_4$ for 
                 a low resolution  comparison of blast-bubble interaction II test (50K particles; $t= 500$).}
   \label{fig:compar_blast_bubble_II_v2}
\end{figure*}

This test is similar to the previous one, but now the bubble has a larger density of $N= 3.1538$,
as in He \& Tang (2012), their example 6.9. The higher density results in a different flow pattern, just 
as in non-relativistic hydrodynamics. This setup
results in mildly relativistic velocities ($v\approx 0.35$, $\gamma \approx 1.07$).
Fig.~\ref{fig:blast_bubble_II} shows snapshots (200K particles)
at $t= 100, 200, 300, 400,$ and 500, just as in the original paper. Again, we find excellent 
agreement with their results.
%-----------------------------------------------------------------------
%-----------------------------------------------------------------------
We again explore the sensitivity to our new ingredients by re-running this test with
formulations $\mathcal{F}_1$ to $\mathcal{F}_4$ with only 50K particles. The results at
$t= 500$ are displayed in Fig.~\ref{fig:compar_blast_bubble_II_v2}.
The tendencies are similar to the previous test. While $\mathcal{F}_1$ to $\mathcal{F}_3$
show reasonable agreement, $\mathcal{F}_1$ shows the cleanest, oscillation-free bubble surface.
$\mathcal{F}_1$ and $\mathcal{F}_3$ at least show a vague reminiscence of  the ``hole'' near $x= 160$, $y= 0$
that is found in high-resolution case. For such small scale features the gradient accuracy seems
crucial. Again,  some oscillations appear for $\mathcal{F}_3$ near $x=180$). Not much 
structure is visible in the case of  $\mathcal{F}_4$ due to the very large dissipation and surface tension effects.
%-----------------------------------------------------------------------
%-----------------------------------------------------------------------

\subsubsection{"Blast in a box" (SR)}
\label{sec:BiB}
This test simulates an over-pressured central region that expands in a perfectly, spherically symmetric manner.
Once the blast is reflected by the boundaries a complicated interaction between shock-shock and shock-contact 
discontinuities sets in. In the Newtonian setup of, e.g. \cite{stone08}, these interactions create the Richtmyer-Meshkov 
instabilities in the central low-density region. 
The Richtmyer-Meshkov instabilities in this test are a serious challenge for SPH simulations
since they occur in the lowest density regions which are very poorly resolved. Therefore we run this
test at a relatively large resolution (600K particles), but also compare low-resolution simulations (200K)
of the different formulations.\\
In this test we have experimented somewhat with the initial particle setup, since lattices may lead to 
a ``pile up''  of particles in certain directions. We have experimented with a particle setup as described in
Sec.~\ref{sec:IC}, but since the combination of Wendland kernel and IA-gradient produces also very 
symmetrical results for a hexagonal lattice we use for simplicity such a setup in the following tests.\\
As the actual test problem we set up 600K particles in the domain $[-0.5, 0.5] \times [-0.75, 0.75]$ 
with periodic boundary conditions everywhere. We use a polytropic EOS with $\Gamma= 5/3$. The initial 
state is characterized by
\be
(N, P, v_x, v_y)=  \left\{
  \begin{array}{ l l}
     (1,  10, 0, 0) \hspace*{0.8cm} {\rm for \; } \sqrt{x^2+y^2} \le 0.1\\
     (1, 0.1, 0, 0) \hspace*{0.7cm} {\rm for \; } \sqrt{x^2+y^2} > 0.1.\\
   \end{array} \right. 
\ee
This setup results in maximum Lorentz factors of $\gamma\approx 3.6$.
%-----------------------------------------------------------------------
\begin{figure*} 
\vspace*{-7cm}
\includegraphics[width=15cm,angle=0]{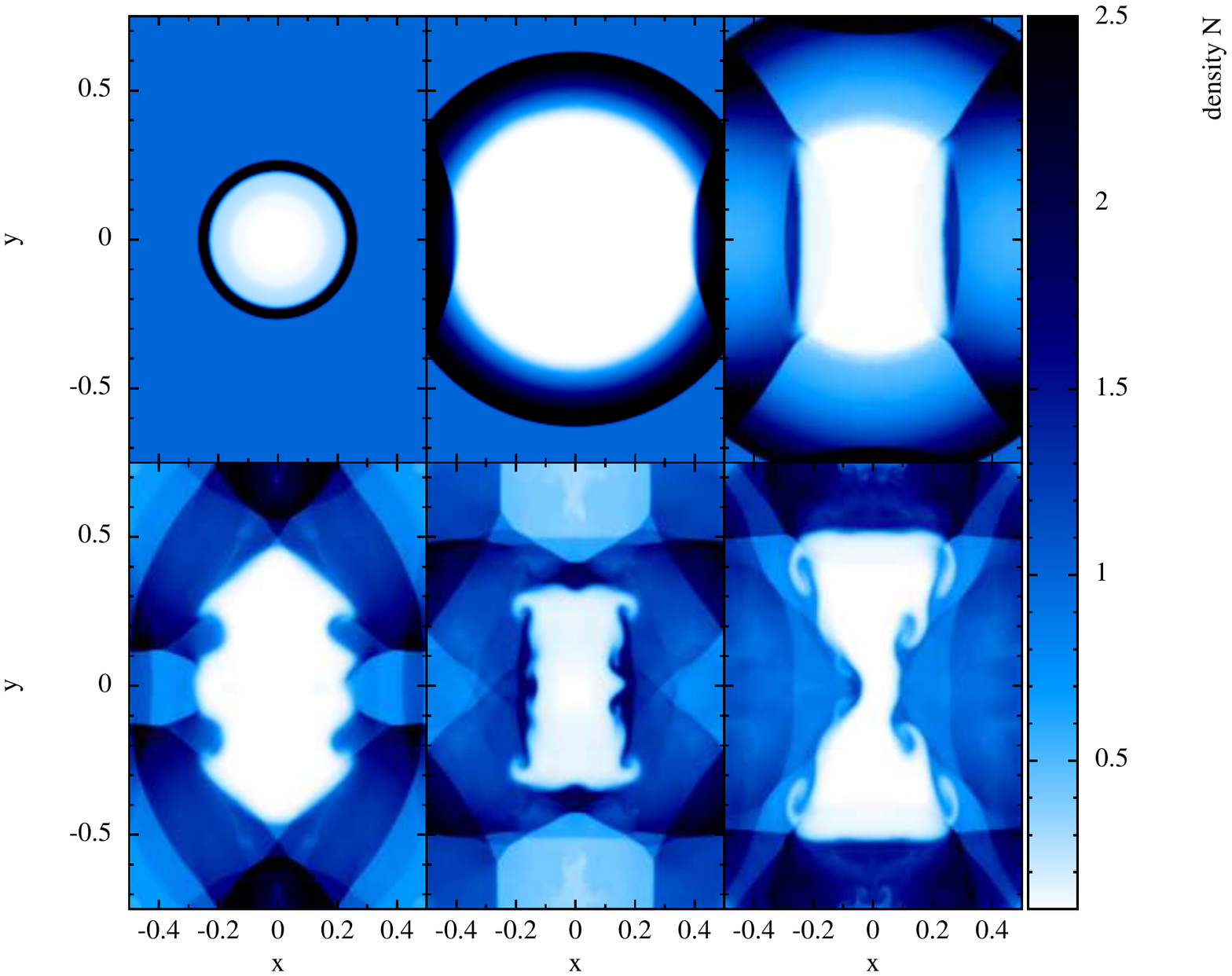}
   \caption{Blast in a box: shown are snapshots at $t= 0.2, 0.7, 1.0, 2.0, 3.0, 4.0$
            (left to right, top to bottom, $\mathcal{F}_1$) of the computing frame baryon number density.
             Note the mushroom-like structures appearing in the central and upper/lower parts of panels 5 and 6.}
   \label{fig:blast_in_box_1}
\end{figure*}
%-----------------------------------------------------------------------
The results (density) for the $\mathcal{F}_1$ formulation at $t=$ 0.2, 0.7, 1.0, 2.0, 3.0 and 4.0 are shown in 
Fig.~\ref{fig:blast_in_box_1}. Note the perfectly spherically symmetric expansion of the overpressured bubble
despite the lack of exact particle symmetry with respect to the explosion center. Despite the use of an exact hexagonal
lattice with its sixfold symmetry for the initial particle configuration no ``particle pile up'' is observed along those
directions. This is one of the appreciated qualities
of the Wendland kernel, some of the other kernels are explored below. The shocks are reflected back and forth
from the boundaries and produce a number of Richtmyer-Meshkov ``mushrooms'' (high density moving into
the central, low-density region; also near $x=0$ at the upper and lower boundary), see the last panel of Fig.~\ref{fig:blast_in_box_1}.
Note that they are hard to resolve since --on the one hand-- the density there is lowest here and we are using a 
very large kernel support ($\eta= 2.2$) so that the resolution length in this region is rather large.
% some comments
Since our initial particle distribution is not symmetric with respect to the reflecting boundaries, 
the central Richtmyer-Meshkov instabilities are not expected to show a symmetry with respect
to the coordinate axis. This is different from simulations by fixed-mesh codes where the grid is usually
aligned with the reflecting boundaries and the symmetry in the instabilities can be considered as a quality
measure of the simulation.\\
% working of dissipation switches
To illustrate the working of the dissipation switches in such a geometrically complicated situation we show
in Fig.~\ref{fig:BiB_diss_switch} snapshots at $t=0.5$ of the simulation shown in Fig.~\ref{fig:blast_in_box_1}.
As can be seen from the lower left panel, the shock trigger really only switches on at the shock location, 
where it produces a razor-sharp circle of high $K$-values. The noise trigger produces dissipation
where particles are arranging themselves after the shock as passed. \\
%-----------------------------------------------------------------------
\begin{figure*} 
\includegraphics[width=20cm,angle=0]{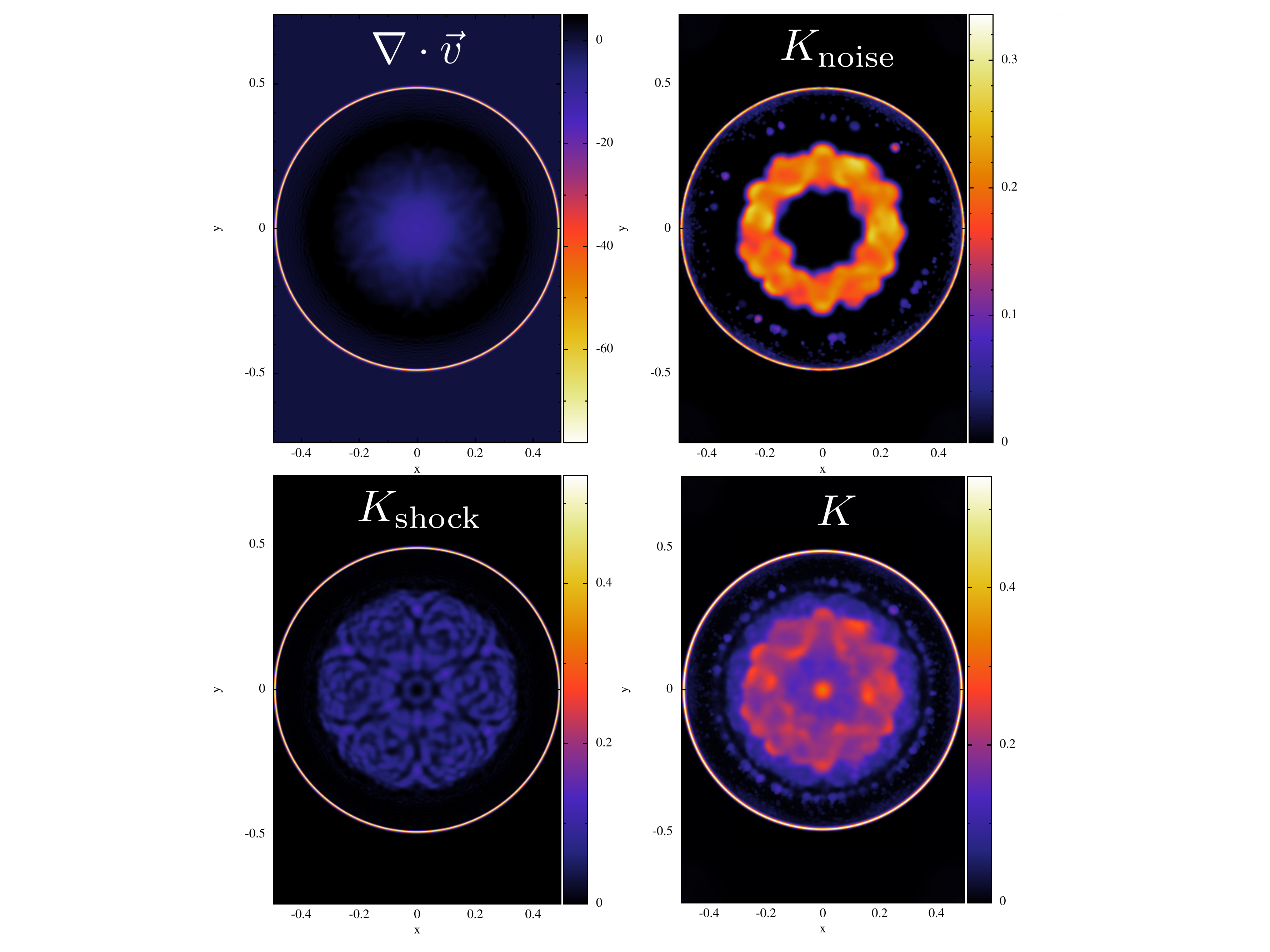}
   \caption{Illustration of the dissipation switches at the ``blast in a box'' problem (at $t=0.5$).
                  Shown are snapshots of $\nabla \cdot \vec{v}$, upper left, the instantaneous value of $K_{\rm noise}$, added to
                  calm down particles when they are noisy, see Eq.~(\ref{eq:K_noise}), upper right,
                  the instantaneous value of $K_{\rm shock}$, see Eq.~(\ref{eq:K_shock}), lower left, and the currently used value of $K$.}
   \label{fig:BiB_diss_switch}
\end{figure*}
%-----------------------------------------------------------------------
We also briefly compare the four SPH formulations for the case where only 200K particles are
used, see Fig.~\ref{fig:BiB_F1_to_F4}.   in the low-density central region. 
Vague anticipations of the Richtmyer-Meshkov instabilities appear in $\mathcal{F}_1$ to $\mathcal{F}_3$, 
but not in $\mathcal{F}_4$. The Richtmyer-Meshkov instabilities for $\mathcal{F}_2$ seem actually 
slightly more pronounced than those of $\mathcal{F}_1$. We attribute this to the the  larger noise level for the case
of kernel-gradient formulations that helps triggering the instabilities.
The mushroom-like structure near $x=0$ at the upper and lower boundaries are most developed for 
$\mathcal{F}_1$. As outlined above, however, the major deficiency here is simply  the lack of resolution.\\
%-----------------------------------------------------------------------
\begin{figure*} 
\includegraphics[width=14cm,angle=-90]{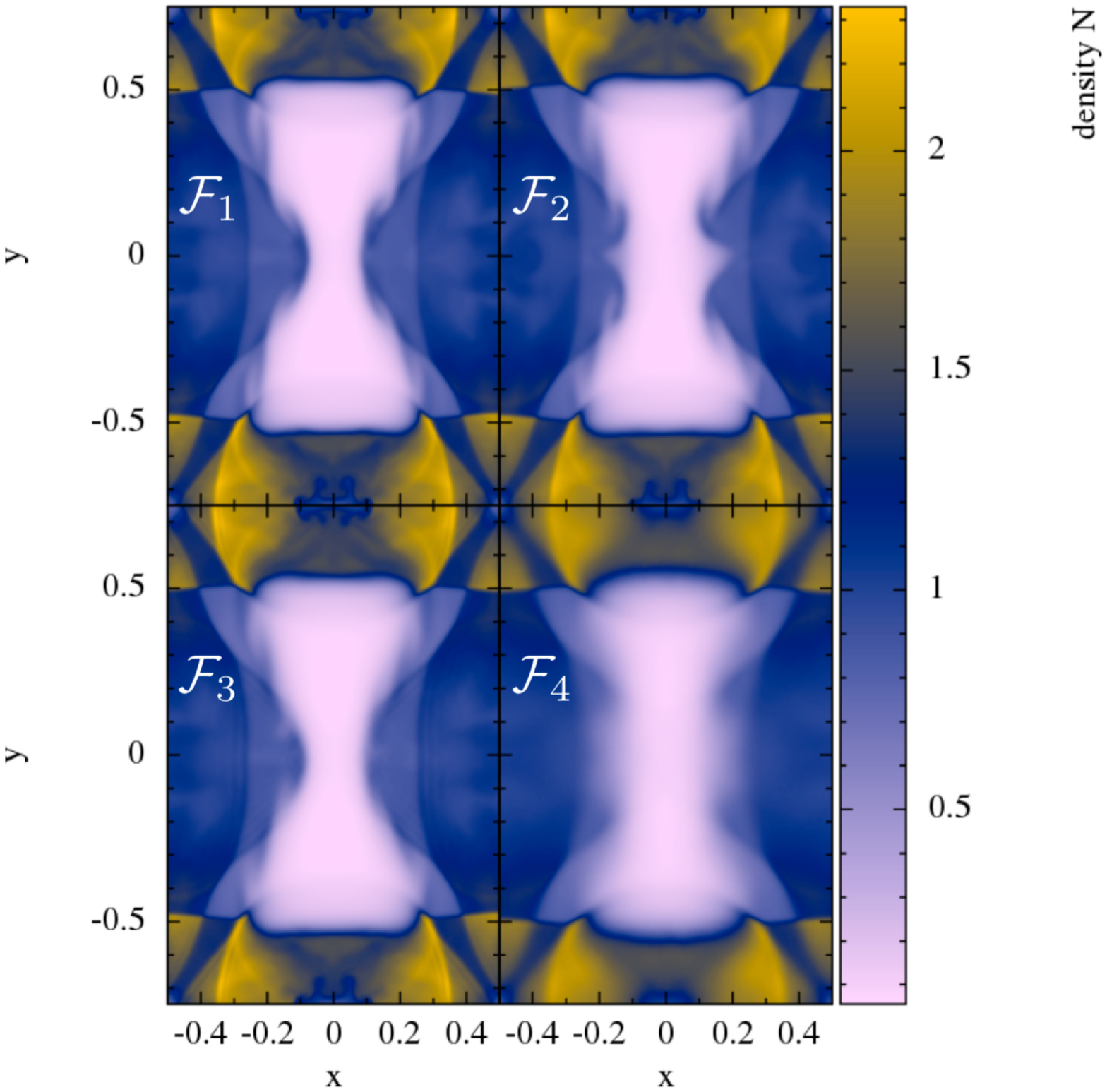}
\vspace*{0cm}
   \caption{Comparison of the different formulations ($\mathcal{F}_1$, upper left, to
                 $\mathcal{F}_4$, lower right) for a low-resolution version (200K) of the ``blast-in-box''
                 test.}
   \label{fig:BiB_F1_to_F4}
\end{figure*}
%-----------------------------------------------------------------------
% Gradient comparison
We use this to compare the new gradient prescription with the standard, kernel-gradient approach. We use 200K particles
and compare $\mathcal{F}_1$ and  $\mathcal{F}_2$. In Fig.~\ref{fig:F1_F2_blast_in_box} we show
density snapshots at t= 0.3. At this stage the pattern should be perfectly  spherically symmetric, however,
in the $\mathcal{F}_2$ case the sixfold symmetry of the underlying hexagonal lattice becomes visible while
$\mathcal{F}_1$ shows practically perfect symmetry. Note, however, that a special colour scheme
was chosen to make the noticeable but still moderate differences visible.\\
%-----------------------------------------------------------------------
\begin{figure*} 
\includegraphics[width=14cm,angle=0]{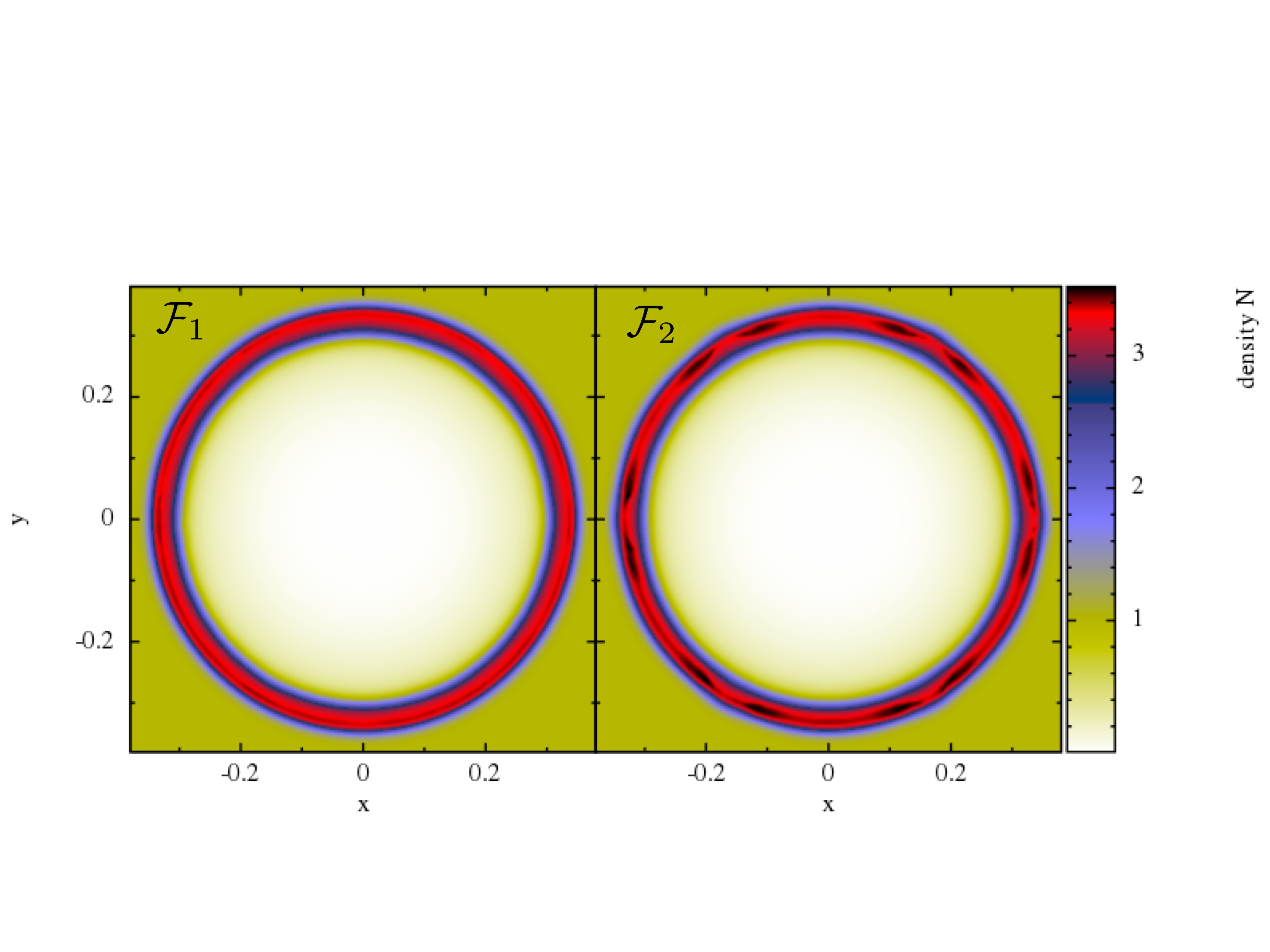}
\vspace*{0cm}
   \caption{Comparison between the IA-gradient ($\mathcal{F}_1$, left) and the standard kernel gradient 
                 ($\mathcal{F}_2$, right) for a low-resolution version (200K) of the ``blast-in-box''
                 test. The IA-gradient shows essentially perfect symmetry while the kernel-gradient
                 shows signs of sixfold symmetry of the underlying hexagonal particle lattice.
                 Note, however, that the differences are small and a particular colour scheme has been 
                 chosen to them well visible. }
   \label{fig:F1_F2_blast_in_box}
\end{figure*}
%-----------------------------------------------------------------------
% Kernel comparison
We perform  another test to compare the performance of  different kernels under challenging conditions.
In particular, we are interested in the level of noise and the question whether the grid symmetries are reflected
in aggregated quantities such as the density. To this end we performed this test with the $\mathcal{F}_1$ formulation, 
but once we use the CS kernel ($\eta=1.2$), once the $M_6$ kernel ($\eta=1.6$), once $W_{\rm H,9}$ ($\eta=2.2$)
and once the Wendland kernel $W_{3,3}$ ($\eta=2.2$). For this test 79K particles were used. The results
at $t= 0.45$, just before the shock hits the first set of walls, is shown in Fig.~\ref{fig:blast_in_box_kernel_comparison}. 
The first row shows the density $N$ and the second and third one show $\nabla \cdot \vec{v}$ which is 
a sensitive indicator for the presence of noise. The kernel functions are noted in the panels. Clearly, after the 
passage of the shock the particles have to re-arrange themselves into a new configuration, so some particle motion 
and therefore a non-zero value of $\nabla \cdot \vec{v}$ is expected. The cubic spline kernel produces the noisiest 
results, followed by $M_6$ and the higher order kernels. The Wendland kernel performs slightly better than 
$W_{\rm H,9}$, especially close to the explosion center where it produces the most symmetric results of all kernels. 
It is also the only kernel that does not show any sign of the sixfold symmetry 
of the original hexagonal lattice configuration. Consistent with our previous tests, the Wendland 
kernel produces the cleanest and least noisy results.
%-----------------------------------------------------------------------
\begin{figure*} 
   \centerline{\includegraphics[width=22cm,angle=0]{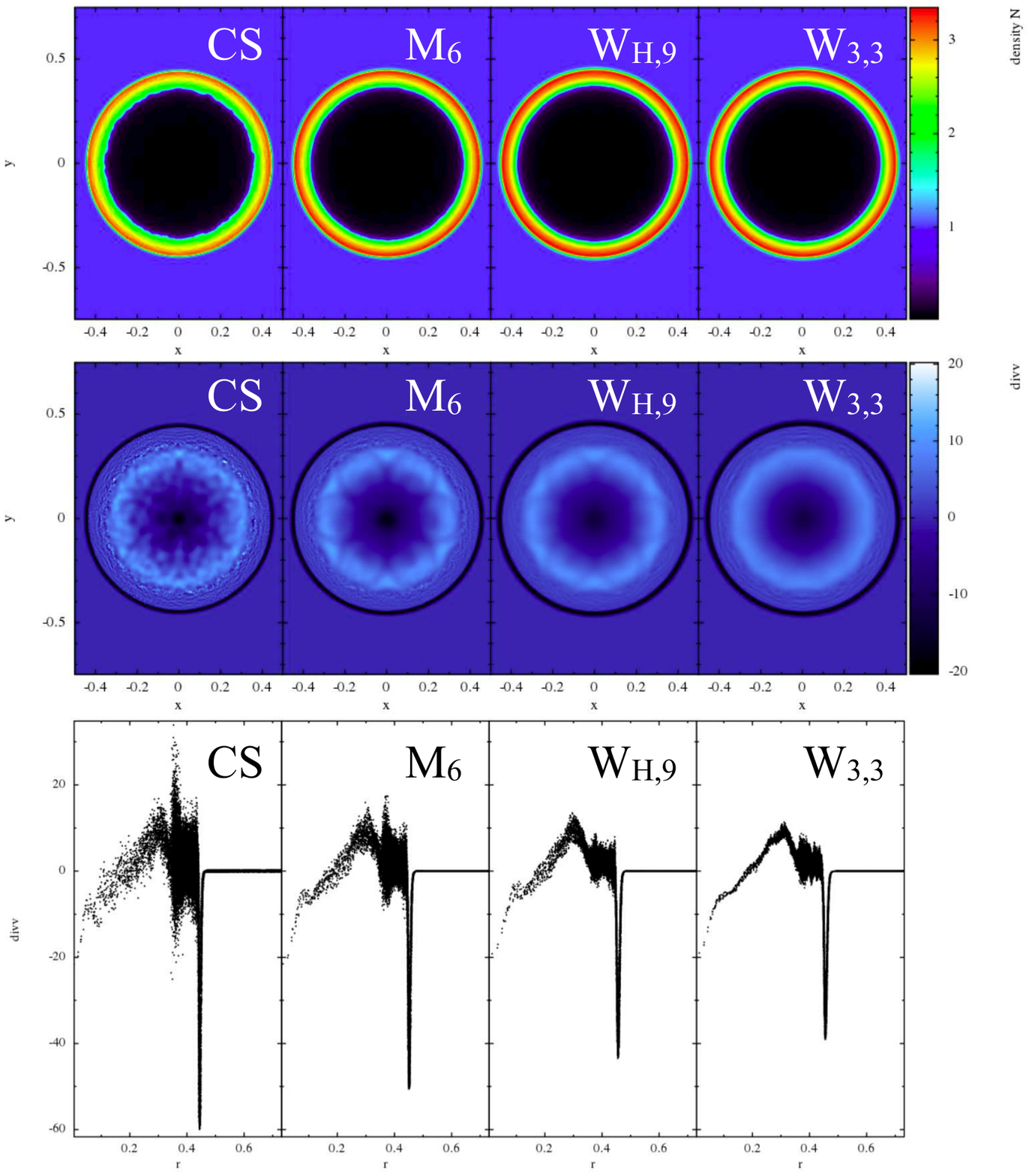}} 
   \caption{Importance of the kernel choice in the ``blast-in-a-box'' test.
                 All tests are performed with the $\mathcal{F}_1$ formulation, apart from the kernel choice.
                 The upper row shows the computing frame baryon number density and the other 
                 rows show $\nabla \cdot \vec{v}$ as a indicator for the presence of noise, one color-coded in the
                 $XY$-plane (middle) and once as function of the distance from the explosion center (bottom).
                 The first column shows the result for the cubic spline kernel ($\eta=1.2$), the second the 
                 $M_6$ kernel ($\eta=1.6$) and column three and four show the results for $W_{\rm H,9}$ and
                 the Wendland kernel (both with $\eta=2.2$). }
   \label{fig:blast_in_box_kernel_comparison}
\end{figure*}
%-----------------------------------------------------------------------

\section{Summary}
\label{sec:summary}

In this paper we have explored the impact of various SPH discretization choices. Motivated by
future relativistic applications, we have developed a 2D special-relativistic SPH code, called SPHINCS\_SR,
that allows to explore a large number  of different choices.
All the tests of this paper --whether in the Newtonian or special-relativistic limit-- 
have been performed with this new code. Part of the motivation of this paper was
to show that modern SPH formulations perform very well even in challenging tests 
where more traditional SPH approaches fail badly.\\
The effects of the improvements are the following.
\bi
 \i {\em Gradients}\\
% gradients
The first measure concerns the calculations of gradients. We have explored here in particular a
prescription that starts from an integral-based function representation and requires the 
(analytical) inversion of a small matrix. By neglecting 
appropriate terms, one can recover the same desirable anti-symmetry property in particle indices as the 
standard SPH kernel gradient (for radial kernel functions). Thus, numerical conservation 
can be ensured in a similar manner as in standard SPH. Such gradient prescriptions had been 
explored earlier in an astrophysical context by \cite{garcia_senz12} and \cite{cabezon12a}
and, in an engineering context, by \cite{jiang14}. As we show in Sec.~\ref{sec:gradient_accuracy}, the gradient 
accuracy can be dramatically increased ($\sim$ 10 orders of magnitude!) if the particle distribution 
is regular, see Fig.~\ref{fig:acc_different_gradients}. The new prescription also yields much more 
accurate results under less idealized circumstances. The SPH formulations with the new gradient
prescriptions are able to resolve smaller details and they result in less velocity noise
 (compare $\mathcal{F}_1$ and $\mathcal{F}_3$), see, for example, Fig.~\ref{fig:Gresho_formulation} 
and \ref{fig:Sod_F1_to_F4}. Moreover, the gradient prescription turns out to be
very beneficial in resolving fluid instabilities, see Sec.\ref{sec:instabilities}.\\
% kernels
 \i {\em Kernel choice}\\
Apart from the commonly used, ``standard'' SPH kernel M$_4$ and the M$_6$ kernel, we have explored
examples of peaked kernels and in particular some high-order members of a recently suggested kernel 
family \citep{cabezon08} and a high-order  Wendland kernel \citep{wendland95}. 
We found that the most commonly used SPH kernel M$_4$ actually performs rather poorly,
see for example Figs. \ref{fig:dens_accuracy}, \ref{fig:grad_accuracy}, \ref{fig:Gresho_kernel} and \ref{fig:blast_in_box_kernel_comparison}.
The M$_6$ kernel is better, but the results still can be substantially improved  by employing higher order kernels.
The overall best performance was found for the Wendland kernel. It allows for only very little velocity noise,
even in highly dynamical situations, see Figs. \ref{fig:noise_box}, \ref{fig:Gresho_kernel} and 
\ref{fig:blast_in_box_kernel_comparison}. The explored peaked kernels performed very poorly 
in practically every respect, even for very large kernel supports.\\
% volume element
\i {\em Volume elements}\\
Motivated by recent suggestions of \cite{saitoh13} and \cite{hopkins13}, we have  generalized our previous
relativistic SPH formulation to a class of more general volume elements, see Eq.~(\ref{eq:gen_vol_element}),
that differ from the traditional choice $m/\rho$ in the Newtonian and $\nu/N$ in the special-relativistic
case. We have in particular explored the case where
the weighting quantity is a power of the pressure, $X_b= P_b^k$, $k=0.05$. This formulation removes spurious
surface tension effects, see Fig.~\ref{fig:surface_tension}, and it performs in all of the tests at least as good, but 
sometimes substantially better than the standard choice $X_b= \nu_b$ (corresponding to the usual SPH 
density sum). In fact, for the standard SPH choice $X_b= \nu_b$  together with low dissipation we have seen in a number of
tests ``lattice ringing effects'', see for example, Figs. \ref{fig:KH_num_trigger}, \ref{fig:compar_blast_bubble_II_v2},
\ref{fig:compar_blast_bubble_I_v2} and \ref{fig:BiB_F1_to_F4}. Other choices for the weight $X$ are
certainly possible and should be explored in future studies. \\
\i {\em Dissipation triggers}\\
% dissipation triggers
We have also designed new triggers to decide where dissipation should be applied. 
Our general strategy is ``react immediately, decay fast'': if the triggers indicate a desired value 
that is higher than the currently used value, the dissipation parameter is raised immediately to 
the indicated value \citep{cullen10} and subsequently it decays exponentially 
on a very short time scale. We trigger on both shocks and velocity noise. Our shock trigger is
based on the temporal change of the velocity divergence, very similar to the approach of
\cite{cullen10}. We also trigger on the occurrence of velocity noise, 
see Eqs.~(\ref{eq:N_trigg_1}) and (\ref{eq:N_trigg_2}). The first of these noise triggers only releases very little 
dissipation, based on fluctuations in the sign of $\nabla \cdot \vec{v}$. This small amount of extra dissipation 
substantially improves the convergence rate in the Gresho-Chan test, see Fig.~\ref{fig:Gresho_convergence},
but for a number of tests probably very good results would be obtained even if this trigger was ignored.
The second trigger hardly ever switches on, but when it does so, it efficiently damps possibly remaining 
post-shock oscillations/noise.  The addition of such noise triggers allows to safely choose a very short decay time
for the dissipation parameter, since possibly appearing noise is efficently taken care of.
An illustration of the functioning of our dissipation triggers is shown in Figs.~\ref{fig:2D_Sod_initial_distrib} 
and \ref{fig:BiB_diss_switch}.
In summary, our treatment takes at each time step for each particle a decision 
on the required dissipation value. This leads to a very local dissipation and essentially removes 
unwanted effects while providing accurate and robust solutions, even in strong shocks, see 
Sec.~\ref{sec:riemann_probs}.
\ei
% Different formulations
To disentangle the different effects in benchmark tests, we use four different SPH formulations, in the paper
referred to as $\mathcal{F}_1$ to $\mathcal{F}_4$, which are explained in detail at the beginning of 
Sec.~\ref{sec:tests}. The first one, $\mathcal{F}_1$, contains all suggested improvements, while $\mathcal{F}_4$
uses every time the worst choices (CS-kernel, direct kernel derivatives, standard SPH volume element and
constant, large dissipation parameters), choices that are actually not too far from what is implemented
in a number of frequently used SPH codes. The $\mathcal{F}_1$ formulation delivers excellent results, even in those
tests where the ``standard choices'' fail completely. For example, $\mathcal{F}_4$ does not converge
to the correct solution in the Gresho-Chan vortex, consistent with earlier findings of \cite{springel10a}, while
$\mathcal{F}_1$ converges in this test close to linearly. Another example are the weakly triggered or untriggered 
fluid instabilities, see Secs.~\ref{sec:KH} and \ref{sec:RT}, where $\mathcal{F}_4$ hardly shows any evolution
at all while all other formulations show a healthy growth of the instabilities.\\
We have found $\mathcal{F}_1$ to be a major improvement over commonly made choices, whether in Newtonian 
or relativistic tests. So far, no efforts have been undertaken
to optimize any of the SPH formulations in terms of computational speed and in this form $\mathcal{F}_1$ 
takes approximately twice as much time
as $\mathcal{F}_4$, mainly due to the required matrix inversion and the substantially larger neighbor number
($\eta= 2.2$ rather than 1.2; see Eq.~(\ref{eq:h_vol})). While we think that the results more than justify the additional
computational effort even in the purely hydrodynamic case, the inclusion of other physics ingredients such as self-gravity 
may actually make the extra effort practically negligible.\\
In recent years, a number of projects have been carried out to compare numerical hydrodynamics 
methods that are commonly  used in astrophysics 
\citep{price10,creasey11,bauer12,scannapieco12,torrey12,sijacki12,nelson13,hubber13,bird13,hayward13}.
At least to some extent these investigations were triggered by Lagrangian Voronoi-tesselation codes 
having become available both for Newtonian \citep{springel10b} and special-relativistic hydrodynamics 
\citep{duffell11}. Where shortcomings of SPH were identified, they were  attributed to excessive dissipation,
velocity noise and gradient accuracy. As demonstrated in this study, all these issues can be substantially 
improved by the suggested measures. Comparisons of the suggested $\mathcal{F}_1$ SPH formulation
(or its Newtonian equivalents) with other methods are left to future studies.

\section*{Acknowledgements}
It is a pleasure to acknowledge inspiring discussions with Daniel Price and Joe Monaghan during
a sabbatical stay at Monash University. Also the hospitality of the University of Queensland in 
Brisbane and Monash University, Clayton, Vic 3800, Australia is gratefully acknowledged. The 
stay in Australia was supported by the DFG by a grant to initiate and intensify bilateral 
collaboration.
This work has further been supported by the Deutsche Forschungsgemeinschaft (DFG) under 
grant number RO-3399/5-1, by the Swedish Research Council (VR) under grant 621-2012-4870
and by the CompStar network, COST Action MP1304. 
This work has profited from visits to Oxford which were supported by a DAAD grant 
``Projektbezogener Personenaustausch mit Gro{\ss}britannien'' under grant number 
313-ARC-XXIII-Ik and from visits to La Scuola Internazionale Superiore di Studi Avanzati (SISSA), 
Trieste, Italy. It is a pleasure to acknowledge in particular the hospitality of John Miller 
(Oxford, Trieste).  S.R. also gratefully acknowledges the hospitality of Andrew MacFadyen 
(New York University) and of Enrico Ramirez-Ruiz (UC Santa Cruz)  and their home institutions 
where part of this work was carried out. The stay in Santa Cruz was generously supported by 
the David and Lucile Packard Foundation. Many of the figures of this article were 
produced with the visualization software SPLASH \citep{price07a}. The author also thanks 
Marius Dan and Franco Vazza for their careful reading of the manuscript.
It is a pleasure to acknowledge detailed conversations with Walter Dehnen and to thank him
for his insightful comments and a number of useful suggestions.

\bibliographystyle{mn2e}
\bibliography{astro_SKR.bib}
\end{document}